\documentclass[12pt,reqno]{amsart}
\usepackage{amssymb,multicol,amscd}
\usepackage[mathscr]{eucal}
\newcommand{\ufields}[1]{U^{#1}(x)} 
\newcommand{\umodule}[1]{G^{#1}(\ufields{})} 
\newcommand{\uconnection}[2]{\omega_{#1}^{#2}(x)} 
\newcommand{\ualg}{\mathfrak{h}} 
\newcommand{\ustrconst}[2]{G^{#1}_{#2}} 

\newcommand{\linfields}[1]{U_1^{#1}(x)} 
\newcommand{\linconnection}[2]{\omega_{0#1}^{#2}(x)} 
\newcommand{\linstrconst}[2]{f^{#1}_{#2}} 
\newcommand{\linbasis}[1]{e_{#1}} 
\newcommand{\fields}[2]{\bundsec{#1}^{#2}} 
\newcommand{\linrepres}[3]{t_{#1}{}^{#2}{}_{#3}} 
\newcommand{\gparameter}[1]{\epsilon^{#1}(x)} 
\newcommand{\gparameterp}[1]{\varepsilon^{#1}(x)} 
\newcommand{\glparameter}[1]{\epsilon_0^{#1}(x)} 

\newcommand{\symalg}{\mathfrak{g}_\module} 
\newcommand{\symconnection}[2]{w_{#1}^{#2}(x)} 
\newcommand{\symbasis}[1]{E_{#1}} 
\newcommand{\symrepres}[3]{T_{#1}{}^{#2}{}_{#3}} 
\newcommand{\symstrconst}[2]{h^{#1}_{#2}} 

\newcommand{\Vermaex}[1]{\mathfrak{V}_{#1}} 
\newcommand{\vacVermaex}[1]{\ketv{#1}} 
\newcommand{\bundsectex}[1]{\ketv{\Phi_{#1}(x)}} 
\newcommand{\bundsectlex}[2]{\ketv{\Phi_{#1}{}_{,#2}(x)}} 
\newcommand{\subVermaex}[1]{\mathfrak{P}_{#1}} 
\newcommand{\irrVermaex}[1]{\mathfrak{I}_{#1}} 

\newcommand{\alg}{\mathfrak{f}} 
\newcommand{\roots}{\Pi} 
\newcommand{\subroots}{\bar\Pi} 
\newcommand{\subalg}{{\smm a}_{\subroots}} 
\newcommand{\Cartan}{{\smm h}} 
\newcommand{\psubalg}{{\smm p}_{\subroots}} 
\newcommand{\Levi}{{\smm l}_{\subroots}} 
\newcommand{\rad}{{\smm r}_{\subroots}} 
\newcommand{\dmn}{{{M}}} 
\newcommand{\module}{{\smm M}} 
\newcommand{\filtration}[1]{{\module_{(#1)}}} 
\newcommand{\gradation}[1]{{\module_{[#1]}}} 
\newcommand{\Minkowski}{{\oR^\dmn}} 
\newcommand{\bundle}{{\cB}} 
\newcommand{\subbundle}{{\cb}} 
\newcommand{\sect}[1]{\Gamma(#1)} 
\newcommand{\genconnection}[2]{\omega_{#1}{}^{#2}(x)} 
\newcommand{\diff}{\mathsf{D}}  
\newcommand{\bundsec}[1]{\ketv{\Phi^{#1}(x)}} 
\newcommand{\subbundsec}[1]{\ketv{\phi^{#1}(x)}} 
\newcommand{\forms}{\Xi} 
\newcommand{\moduleforms}{\smm F} 
\newcommand{\moduleformslp}[2]{\moduleforms_{[#1]}^{#2}} 
\newcommand{\moduleformsp}[1]{\moduleforms^{#1}} 
\newcommand{\cohomology}[1]{H^{#1}(\rad,\module)} 
\newcommand{\cohomologylp}[2]{H_{[#1]}^{#2}(\rad,\module)} 
\newcommand{\equations}{\mathsf{R}} 
\newcommand{\extdiff}{{\rm d}} 
\newcommand{\sgminus}{\sigma_-} 
\newcommand{\closed}[1]{\mathfrak{c}^{#1}} 
\newcommand{\exact}[1]{\mathfrak{e}^{#1}} 
\newcommand{\kphi}[1]{\ketv{\varphi_{[#1]}(x)}}             
\newcommand{\kchi}[1]{\ketv{\chi_{[#1]}(x)}}                
\newcommand{\tphi}[1]{\ketv{\tilde\varphi_{[#1]}(x)}}       
\newcommand{\tpsi}[1]{\ketv{\tilde\psi_{[#1]}(x)}}          
\newcommand{\irrmodule}{\mathfrak{I}} 
\newcommand{\extmodule}{\mathfrak{E}} 
\newcommand{\comodule}{\mathfrak{M}^\natural} 

\newcommand{\commut}[2]{\left[#1,\,#2\right]}  
\renewcommand{\d}[1]{\frac{\partial}{\partial #1}} 
\newcommand{\ld}[1]{\stackrel{\leftarrow}{\frac{\partial}{\partial #1}}} 
\def\ketv#1{\mathchoice{{\left|{#1}\right\rangle}}%
       {|{#1}\rangle}{|{#1}\rangle}{|{#1}\rangle}}
\def\brav#1{\mathchoice{{\left\langle{#1}\right|}}%
       {\langle{#1}|}{\langle{#1}|}{\langle{#1}|}}
\newcommand{\half}{\frac{1}{2}} 
\newcommand{\one}{{1\kern-4pt1}} 
\newcommand{\ptl}{\partial} 

\newcommand{\exactsq}[3]{0\longrightarrow {#3}\longrightarrow
           {#2}\longrightarrow {#1}\longrightarrow 0}
\newcommand{\subplus}{\lefteqn{\subset}+}
\newcommand{\Img}{\mathop{\mathsf{Im}}}

\def\cA{\mathcal{A}} \def\cB{\mathcal{B}} 
\def\cD{\mathcal{D}} \def\cE{\mathcal{E}} \def\cF{\mathcal{F}}
 \def\cH{\mathcal{H}} 
 \def\cK{\mathcal{K}} \def\cL{\mathcal{L}}
  
\def\cP{\mathcal{P}}  
 \def\cT{\mathcal{T}}

 \newcommand{\oC}{\mathbb{C}}
\newcommand{\oN}{\mathbb{N}} \newcommand{\oZ}{\mathbb{Z}}
 \newcommand{\oR}{\mathbb{R}}

\newcommand{\ga}{\alpha}
\newcommand{\gb}{\beta}
\newcommand{\gga}{\gamma}   \newcommand{\Gga}{\Gamma}
\newcommand{\gd}{\delta}    \newcommand{\Gd}{\Delta}
\newcommand{\gep}{\epsilon}

\newcommand{\gl}{\lambda}   \newcommand{\Gl}{\Lambda}
\newcommand{\go}{\omega}    \newcommand{\Go}{\Omega}
\newcommand{\gs}{\sigma}

\newtheorem{Thm}{Theorem}[section]

\newtheorem{Lemma}[Thm]{Lemma}

\newtheorem{Dfn}[Thm]{Definition}
{
\newtheorem{Rem}[Thm]{Remark}
}

\theoremstyle{definition}

\newenvironment{prf}{%
\noindent{\sc Proof.}}%
{\noindent{\hfill\mbox{\rule{.5em}{.5em}}\,}\par\medskip}

\newcommand{\so}{\mathfrak{o}} 
\newcommand{\iso}{\mathfrak{iso}} 
\newcommand{\translations}{\mathfrak{t}} 
\newcommand{\socohomology}[2]{H^{#1}(\translations(\dmn),{#2})} 
\newcommand{\vacVerma}[1]{\mathfrak{N}_{(#1)}} 
\newcommand{\vacbasis}[2]{\ketv{(#1)}^{#2}} 
\newcommand{\Verma}[1]{\mathfrak{V}_{(#1)}} 
\newcommand{\Vermal}[2]{\mathfrak{V}_{(#1)#2}} 
\newcommand{\Vermabasis}[2]{\ketv{#1}^{#2}} 
\newcommand{\singmodule}[1]{\mathfrak{S}_{(#1)}} 
\newcommand{\singbasis}[1]{\ketv{s}^{#1}} 
\newcommand{\subVerma}[2]{\mathfrak{P}_{(#1),(#2)}} 
\newcommand{\msubVerma}[1]{\mathfrak{P}_{(#1)}} 
\newcommand{\quotVerma}[1]{\mathfrak{O}_{(#1)}} 
\newcommand{\quotVermaex}[1]{\mathfrak{Q}_{#1}}
\newcommand{\irrVerma}[1]{\mathfrak{J}_{(#1)}} 
\newcommand{\coVerma}[1]{\mathfrak{V^\natural}_{(#1)}} 
\newcommand{\coVermabasis}[2]{{}_{#2}\brav{#1}} 
\newcommand{\covacmodule}[1]{\mathfrak{N^\natural}_{(#1)}} 
\newcommand{\covacbasis}[2]{{}_{#2}\brav{(#1)}} 
\newcommand{\coirrVerma}[1]{\mathfrak{J^\natural}_{(#1)}} 
\newcommand{\coVermal}[2]{\mathfrak{V}^\natural_{(#1)#2}} 
\newcommand{\sqVerma}[2]{\mathfrak{V}_{(#1)_{#2}}} 
\newcommand{\sqvacVerma}[2]{\mathfrak{N}_{(#1)_{#2}}} 
\newcommand{\sqtvacVerma}[2]{\tilde{\mathfrak{N}}_{(#1)_{#2}}} 
\newcommand{\sqtirrVerma}[2]{\tilde{\mathfrak{J}}_{(#1)_{#2}}} 
\newcommand{\sqquotVerma}[2]{\mathfrak{O}_{(#1)_{#2}}} 
\newcommand{\sqirrVerma}[2]{\mathfrak{J}_{(#1)_{#2}}} 

\newcommand{\oneline}{
   \unitlength=1pt
   \begin{picture}(16,10)
   \put(-3.4,2.6){\line(1,0){23}}
   \end{picture}
}

\newcommand{\doubleline}{
   \unitlength=1pt
   \begin{picture}(20,10)
   \put(-4,3.8){\line(1,0){29}}
   {
   }
   \put(-4,2.3){\line(1,0){29}}
   \end{picture}
}

\makeatletter
  \@addtoreset{equation}{section}
\makeatother

\def\tilde{\widetilde}

\newcommand{\nn}{\nonumber}

\def\smm{\mathfrak}
\newcommand{\cb}{{\smm b}}       

\newcommand{\soW}{{\smm{V}}} 

\newcommand{\lattice}{{\mathbf{Q}}}
\newcommand{\weights}{{\mathbb{L}}}

\newcommand{\sof}{\mathfrak{f}}

\newcommand{\ket}[1]{\mathchoice{%
     {\left|{#1}\right\rangle}}{|{#1}\rangle}{|{#1}\rangle}{|{#1}\rangle}}

\def\bar{\overline}

\newcommand{\drop}[1]{}

\def\tensor{\otimes}

\renewcommand{\atop}[2]{\genfrac{}{}{0pt}{}{#1}{#2}}

\begin{document}

\makeatletter
\renewcommand{\@oddhead}{\sc\hfill
Unfolded form of linear conformal equations in $\dmn$-dimensions and
\dots\hfill\rm\footnotesize\thepage}
\makeatother
\hfuzz=1pt \vfuzz=1.6pt \addtolength{\baselineskip}{4pt}

\title{Unfolded form of conformal equations in $\dmn$ dimensions
       \\[2mm]
      and~$\so(\dmn+2)$-modules}

\author[Shaynkman]{
 O.~V.~Shaynkman$\,{}^{1}$
   }

\author[Tipunin]{I.~Yu.~Tipunin$\,{}^{2}$ }

\author[Vasiliev]{M.~A.~Vasiliev$\,{}^{3}$
}
\maketitle
\vspace{-20pt}
\begin{center}
{\small\it I.E.Tamm Theory Department, Lebedev Physics Institute, Leninski
prospect 53,\\ 119991, Moscow, Russia}
\end{center}
\begin{abstract}
A constructive procedure is proposed for formulation of linear
differential equations invariant under global symmetry
transformations forming a semi-simple Lie algebra~$\alg$. Under
certain conditions $\alg$-invariant systems of differential
equations are shown to be associated with $\alg$-modules that are
integrable with respect to some parabolic subalgebra of $\alg$.
The suggested construction is motivated by the unfolded
formulation of dynamical equations developed in the higher spin
gauge theory and provides a starting point for generalization to
the nonlinear case. It is applied to the conformal algebra
$\so(\dmn,2)$ to classify  all linear conformally invariant
differential equations in Minkowski space. Numerous examples of conformal
equations are discussed from this perspective.
\end{abstract}

\footnotetext[1]{\scriptsize{\tt shayn@lpi.ru}}
\footnotetext[2]{\scriptsize{\tt tipunin@lpi.ru}}
\footnotetext[3]{\scriptsize{\tt vasiliev@lpi.ru}}

{\footnotesize\tableofcontents}

\section{Background and Introduction
\label{Sct.:_Unfolded_Formulation_of_Dynamical_Equations}} In this
paper we apply a method of the analysis of dynamical systems
called {\it unfolded formulation} to classify all  conformally
invariant linear differential equations in any space-time dimension $\dmn>2$.
This method, suggested originally for the analysis of
higher spin dynamical systems \cite{Ann}-\cite{VD},
proved to be useful for the
analysis of problems of deformation quantization \cite{Frrr,Fed}.

Unfolded formulation of a system of
partial differential equations in a space-time with coordinates
$x^m$ $(m=0,\ldots \dmn-1)$ consists of its reformulation in
the first-order form with respect to all coordinates.
As such, it is a generalization of the first-order form
of ordinary ({\it i.e} $\dmn=1$) differential equations $\dot{q}^i = G^i (q)$.
More precisely, unfolded equations have the form

\begin{equation}
\label{unfolded_system} \extdiff\ufields{\Go}=\umodule{\Go}\,.
\end{equation}
Here $\extdiff=\xi^m \d{x^m}$
is the exterior differential\footnote{ Throughout this paper we
use the notation $\xi^m$ for the basis  1-forms conventionally
denoted $\extdiff x^m$.}. $\ufields{\Go}$ denotes a set of
variables being differential forms (i.e., polynomials in
$\xi^m$). The condition
\begin{equation}\label{unfolded_consistancy_conditions}
\umodule{\Go} \wedge \frac{\gd \umodule{\Gl}} {\gd \ufields{\Go}}
= 0
\end{equation}
is imposed on $\umodule{\Gl}$ to guarantee that the system is
formally consistent. (It is assumed that only wedge products of
differential forms appear in (\ref{unfolded_system}) and
(\ref{unfolded_consistancy_conditions}), {\it i.e.} $\umodule{\Go}$ is a
polynomial of $\ufields{\Go}$ containing no derivatives in
$\xi^m$ and $x^m$.)

In the case of ordinary differential equations
the variables $q^i(t)$ taken at any $t=t_0$ provide
a full set of initial data. For an $\dmn>1$
unfolded  field-theoretical system the knowledge of the fields
$U(x)$ at any $x^m=x^m_0$ also reconstructs $U(x)$
in some neighborhood of $x^m_0$. Therefore, to unfold a
field-theoretical system with infinitely many degrees of freedom
it is necessary to introduce
infinitely many auxiliary fields. The latter  identify with all
derivatives of dynamical fields ({\it i.e.} with infinitely many
generalized momenta).

Unfolded formulation, which is available for any dynamical system,
has a number of properties proved to be useful for the analysis of
various aspects of linear and nonlinear dynamics
(see \cite{Sakh}
for a recent review).
The property  of the unfolded formulation which is of particular
importance for the analysis of this paper
is that it makes  symmetries of a model manifest.
In particular, unfolded formulation of any dynamical system possessing
one or another linearly realized global symmetry $g$ is formulated
in terms of some $g$-module. This simple observation makes it trivial
to list  unfolded dynamical systems of a given symmetry. The nontrivial
part of the problem is to single out nontrivial dynamical systems in this
list that result from unfolding of certain differential equations.
(Note that, generally,  unfolded equations may describe an infinite set of
constraints with no differential equations among them.) As we show in
this paper, nontrivial $g$-invariant
differential equations are associated with the unfolded equations
based on  $g$-modules resulting  from factorization of generalized Verma $g$-modules
over singular submodules.
 Our  scheme is quite general and can be applied to the analysis of various
dynamical systems. In this paper we   apply  this analysis to classification
of conformally invariant   linear differential equations.

Let us now analyze relevant properties of unfolded equations more carefully.
 Due to (\ref{unfolded_consistancy_conditions})
the system (\ref{unfolded_system}) is invariant under the gauge
transformations
\begin{equation}
\label{gauge_transformations} \delta
\ufields{\Go}=\extdiff\gparameter{\Go}+\gparameter{\Gl}\wedge
\frac{\gd \umodule{\Go}}{\gd \ufields{\Gl}}\,,
\end{equation}
where the gauge parameters $\gparameter{\Go}$ are arbitrary
functions of the coordinates $x^m$. Let $\go^A(x)=\xi^m\uconnection{m}{A}$
be the set of 1-forms in $\ufields{\Go}$. The requirement that the
restriction $\umodule{A} \Big |_{\uconnection{}{}}
=-\ustrconst{A}{BC} \uconnection{}{B}\wedge\uconnection{}{C}$ to
the sector of 1-forms $\uconnection{}{A}$ is compatible with
(\ref{unfolded_consistancy_conditions}) implies that
$\ustrconst{A}{BC}$ satisfy (super)Jacobi identities thus being
structure coefficients of some Lie (super)algebra\footnote{To
introduce superalgebraic structure it is enough to let the 1-forms
$\uconnection{}{A}$, which correspond to the even(odd) elements of
superalgebra $\ualg$,  be Grassmann even(odd).}~${\ualg}$.  As a
result, the restriction of equation (\ref{unfolded_system}) to the
sector of 1-forms amounts to the flatness condition on
$\uconnection{}{A}$. In higher spin theories, ${\ualg}$ is some
infinite dimensional higher spin symmetry algebra
\cite{HSalg}-\cite{SS}, \cite{BHS,VD,east}, which contains one or
another finite dimensional space--time symmetry subalgebra $\alg$.
For example, $\alg=\so(n,2)$ appears either as anti-de Sitter
$(n=\dmn-1)$  or as conformal $(n=\dmn)$ algebra in $\dmn$
dimensions.

Let $\linconnection{}{\Go}$ be a fixed 1-form taking values
in~$\alg$, i.e. $\linconnection{}{}= \linconnection{}{i}
\linbasis{i}$, where $\linbasis{i}$ is a basis in $\alg$. Equation
(\ref{unfolded_system}) for $\ufields{\Go}=\linconnection{}{\Go}$
is equivalent to the zero curvature condition
\begin{equation}\label{zero_curvature_equation}
\extdiff \linconnection{}{i}+\linconnection{}{j} \wedge
\linconnection{}{k} \linstrconst{i}{jk}=0\,,
\end{equation}
where $\linstrconst{i}{jk}$ are structure coefficients of $\alg$.
For $\alg$ isomorphic to Poincar\'e algebra, $\linconnection{}{i}$
is usually identified with the flat space gravitational field with
co-frame and Lorentz connection corresponding to generators of
translations $\cP_n$ and Lorentz rotations $\cL_{mn}$,
respectively. The components of the co-frame part of the
connection are required to form a non-degenerate $\dmn\times\dmn$
matrix in which case we call connection non-degenerate. For
example, Minkowski space--time in Cartesian coordinates is
described by zero Lorentz connection and co-frame $\xi^n \cP_n$ so
that the components of the co-frame 1-form $e_n{}^m = \delta_n^m$
form a non-degenerate matrix. The freedom in the choice of a
non-degenerate $\linconnection{}{i}$  encodes the coordinate
choice ambiguity.

One can analyse equation (\ref{unfolded_system}) perturbatively by
setting
\begin{equation}
\ufields{\Go} =\linconnection{}{\Go}+\linfields{\Go}\,,
\end{equation}
where $\linfields{\Go}$ describes first-order fields
(fluctuations), while $\linconnection{}{\Go}$ is zero-order. Let
$\fields{p}{\gl}$ be the subset of $p$-forms contained in
$\linfields{\Gl}$ (we use Dirac ket notation for the future
convenience). The linearized part of equations
(\ref{unfolded_system}) associated with the $p$-forms reduces to
some equations of the form
\begin{equation}\label{covariant_constancy_equation}
\diff \fields{p}{\gl}=0\,,
\end{equation}
with
\begin{equation}\label{covariant_derivative}
\diff\fields{p}{\eta}=(\extdiff\gd^\eta_\lambda+\linconnection{}{i}
\linrepres{i}{\eta}{\gl})\fields{p}{\gl}\,.
\end{equation}
The identity (\ref{unfolded_consistancy_conditions}) implies that
the matrices $\linrepres{i}{\eta}{\gl}$ form a representation of
$\alg$ (i.e. $\commut{\linrepres{j}{}{}}{\linrepres{k}{}{}}
=2\linstrconst{i}{jk}\linrepres{i}{}{}$). Let $\module$ be the
$\alg$-module  associated with $\fields{p}{\gl}$, i.e.
$\fields{p}{\gl}$ be a section of the trivial bundle
$\bundle=\module\times \Minkowski$ with the fiber $\module$ and
the $\dmn$-dimensional Minkowski base space $\Minkowski$. The
covariant derivative $\diff$ (\ref{covariant_derivative}) in
$\bundle$ is flat,
\begin{equation}\label{covariant_derivative_flat}
\diff\diff=0
\end{equation}
as a consequence of (\ref{zero_curvature_equation}).

Let the associative algebra $A_\module$ be the quotient of the
universal enveloping algebra of $\alg$ over the ideal
$Ann(\module)$ that annihilates the representation $\module$, i.e.
$A_\module =U(\alg)/Ann(\module)$. Let $\symbasis{I}$ be a basis
of $A_\module$ and $\symrepres{I}{\eta}{\gl}$ be the
representation of $A_\module$ induced from the representation
$\linrepres{i}{\eta}{\gl}$. If $\linconnection{}{i}\linbasis{i}$
satisfying equation (\ref{zero_curvature_equation}) is (locally)
represented in a pure gauge form
\begin{equation}
\linconnection{}{i}\linbasis{i}=g(x)\extdiff g^{-1}(x)\,
\end{equation}
with an invertible element $g(x)=g^I(x)\symbasis{I}\in A_\module$,
the generic local solution of equation
(\ref{covariant_constancy_equation}) gets the form
\begin{equation}
\label{solution_for_fields}
\fields{p}{\eta}=g^I(x)\symrepres{I}{\eta}{\gl}
\ketv{\Phi^p(x_0)}^\gl\,.
\end{equation}
We see that $\ketv{\Phi^p(x_0)}^\gl$ plays a role of  initial data
for equation (\ref{covariant_constancy_equation}), fixing
$\fields{p}{\eta}\Big |_{x\in\varepsilon(x_0)}$ in a neighborhood
$\varepsilon(x_0)$ of a point $x_0$ such that $g(x_0)=1$. As a result,
solutions of equation (\ref{covariant_constancy_equation}) are
parameterized by elements of the $\alg$-module $\module$. If
the $\alg$-module $\module$ is finite dimensional, we will call
the corresponding equation (\ref{covariant_constancy_equation})
topological because it describes at most  ${\rm dim} (\module)$
degrees of freedom.

The system (\ref{zero_curvature_equation}),
(\ref{covariant_constancy_equation}) is invariant under the gauge
transformations (\ref{gauge_transformations})
\begin{align}
\label{connection_gauge_transformations} &\gd
\linconnection{}{i}=\extdiff\gparameter{i}-
2\gparameter{j}\linconnection{}{k}\linstrconst{i}{jk}\,,\\
\label{fields_gauge_transformations} &\gd
\fields{p}{\eta}=\extdiff |\gparameterp{}\rangle^\eta-(-1)^p
\linconnection{}{i}\linrepres{i}{\eta}{\gl}|\gparameterp{}\rangle^\gl
-\gparameter{i}\linrepres{i}{\eta}{\gl} \fields{p}{\gl}\,,
\end{align}
where the $(p-1)$-form $|\gparameterp{}\rangle^\eta$ and 0-form
$\gparameter{i}$ are infinitesimal gauge symmetry parameters.
(Note that if $p=0$ then $|\gparameterp{}\rangle^\eta\equiv 0$.)
Any fixed solution $\linconnection{}{i}$ of equation
(\ref{zero_curvature_equation}) (called vacuum solution) breaks
the local $\alg$ (super)symmetry associated with $\gparameter{i}$
to its stability subalgebra with the infinitesimal parameter
$\glparameter{i}$ satisfying equation
\begin{equation}\label{global_symmerty_parameter}
\extdiff\glparameter{i}-2\glparameter{j}\linconnection{}{k}
\linstrconst{i}{jk}=0\,.
\end{equation}
This equation is consistent due to the zero curvature equation
(\ref{zero_curvature_equation}), and its generic (local) solution
is parameterized by the values of $\gep_0^i(x_0)$,
\begin{equation}
\label{solution_for_symmetry_parameter}
\glparameter{i}\linbasis{i}=\gep_0^i(x_0)g(x)\linbasis{i}g^{-1}(x)\,.
\end{equation}
The leftover global symmetry
\begin{align}
&\gd \fields{p}{\eta}=\gep_0^i(x_0)\Big(g(x)
\linrepres{i}{}{}g^{-1}(x)\Big)^\eta{}_\gl\fields{p}{\gl}\,,\qquad
&\gd \linconnection{}{i}=0\,
\end{align}
with the  symmetry parameters $\gep_0^i(x_0)$ forms the Lie
(super)algebra $\alg$. {}From the Poincar\'e lemma it follows that
the gauge symmetries (\ref{fields_gauge_transformations}) of
$\fields{p}{\eta}$ associated with the parameters
$|\gparameterp{}\rangle^\eta$, which are  $p-1 >0$ forms, do not
give rise to  additional global symmetries of
(\ref{zero_curvature_equation}) and
(\ref{covariant_constancy_equation}) in the topologically trivial
situation.

In fact, equations (\ref{zero_curvature_equation}) and
(\ref{covariant_constancy_equation}) have a larger symmetry
$\symalg\supset \alg$ manifest. Let $\symalg$ be the Lie
(super)algebra built from $A_\module$ via (super)commutators. One
can  extend~(\ref{zero_curvature_equation})
and~(\ref{covariant_constancy_equation}) to
\begin{align}
\label{sup_zero_curvature_equation} &\extdiff
\symconnection{}{I}+\symconnection{}{J}
\symconnection{}{K}\symstrconst{I}{JK}=0\,,\\
\label{sup_covariant_constancy_equation}
&\diff\fields{p}{\eta}=(\extdiff\gd^\eta_\gl+\symconnection{}{I}
\symrepres{I}{\eta}{\gl})\fields{p}{\gl}=0\,,
\end{align}
where  $\xi^m\symconnection{m}{I}$ are the gauge fields of
$\symalg$, and $\symstrconst{I}{JK}$ are the structure
coefficients of $\symalg$. The system
(\ref{sup_zero_curvature_equation}),
(\ref{sup_covariant_constancy_equation}) is consistent in the
sense of (\ref{unfolded_consistancy_conditions}) and has global
symmetry   $\symalg$ for any  $\symconnection{}{I}$, which solves
(\ref{sup_zero_curvature_equation})\,. Since $\alg$ is canonically
embedded into $\symalg$, setting $\symconnection{}{I}
\symbasis{I}=\linconnection{}{i} \linbasis{i}$ one recovers the
system (\ref{zero_curvature_equation}),
(\ref{covariant_constancy_equation}) thus proving invariance of
the system (\ref{zero_curvature_equation}),
(\ref{covariant_constancy_equation}) under the infinite
dimensional global symmetry  $\symalg$. Infinite dimensional
symmetries of this class appear in the field-theoretical models as
higher spin symmetries.

This approach is universal: any system of $\alg$-invariant linear
differential equations can be reformulated in the form
(\ref{zero_curvature_equation}),
(\ref{covariant_constancy_equation}) by introducing auxiliary
variables associated with the appropriate (usually
infinite dimensional) $\alg$-module $\module$ \cite{unfolded}
(also see examples below). As a result, classification of
$\alg$-invariant linear systems of differential equations is
equivalent to  classification of $\alg$-modules $\module$ of an
appropriate class. More precisely, let $\alg$,
$\psubalg\subset\alg$ and $\module$ be, respectively, some
semi-simple Lie algebra, its parabolic subalgebra and
$\alg$-module integrable with respect to $\psubalg$ (for necessary
definitions see section \ref{Sct.:_General_Construction}). We show
that, for a non-degenerate flat connection 1-form $\go^i_0 (x)$,
the covariant constancy equation
(\ref{covariant_constancy_equation}) on a $p$-form
$\bundsec{p}^\gl$ taking values in $\module$ encodes an
$\alg$-invariant system of differential equations
$\equations_\module\subbundsec{p}^\gl=0$ on a $p$-form
$\subbundsec{p}^\gl$ from the $p$-th cohomology $\cohomology{p}$
of the radical $\rad\subset\psubalg$ with coefficients
in~$\module$. For Abelian radical $\rad$ we prove that each
differential operator from $\equations_\module$ corresponds to an
element of $\cohomology{p+1}$ and vice versa. We introduce
classification of $\alg$-invariant systems of equations
$\equations_{\module}$ by reducibility of $\alg$-modules
$\module$. $\alg$-invariant systems that correspond to (reducible)
irreducible $\alg$-modules $\module$ are called $\hbox{\rm(non-)}$primitive.
Non-primitive systems contain nontrivial subsystems and can be
described as extensions of the primitive ones.

This general construction is applied to classification of linear
homogeneous conformally invariant equations on
$\subbundsec{0}\in\cohomology{0}$, where we set
$\alg=\so(\dmn,2)$\footnote{In fact we consider only complex case.
Thus $\so(\dmn,2)\sim\so(\dmn+2)$.}, $\rad=\translations(\dmn)$
(the algebra of translations) and
$\psubalg=\iso(\dmn)\oplus\so(2)$ (i.e., the direct sum of
Poincar\'e algebra and the algebra of dilatations). Conformally
invariant equations are determined by $\socohomology{1}{\module}$.
Examples of primitive equations include Klein--Gordon and Dirac
equations and their conformal generalizations to higher
(spinor-)tensor fields, conformal equations on $p$-forms and, in
particular, (anti)selfduality equations. Examples of non-primitive
equations correspond to reducible $\module$ and include $\dmn=4$
electrodynamics with and without external current and its higher
spin generalization to higher tensors in the flat space of any
even dimension. Note that our construction allows us to write
these systems both in gauge invariant  and in  gauge fixed form.
In the latter case we automatically obtain conformally invariant
gauge conditions. A number of examples of conformal systems are
considered in sections \ref{Sct.:_Examples of the Simplest
Conformal Systems} and \ref{Sct.:_Conformal_equations}.

To find $\socohomology{1}{\irrmodule}$ with coefficients in an
irreducible integrable with respect to $\iso(\dmn)\oplus\so(2)$
conformal module $\irrmodule$ we consider a generalized Verma
module $\mathfrak{V}$ of $\so(\dmn+2)$  such that $\irrmodule$ is
its irreducible quotient. We calculate
$\socohomology{1}{\irrmodule}$ for any $\irrmodule$. As an
$\iso(\dmn)\oplus\so(2)$-module, $\socohomology{1}{\irrmodule}$ is
shown to be isomorphic to the space of certain systems of singular
and subsingular vectors in $\mathfrak{V}$. As a result, the form
of a primitive system of conformal differential equations
$\equations_{\irrmodule}$ encoded by the covariant constancy
equation (\ref{covariant_constancy_equation}) is completely
determined by these systems of singular and subsingular vectors in
$\mathfrak{V}$. Since any reducible integrable with respect to
$\iso(\dmn)\oplus\so(2)$ module $\module$ is an
extension of some irreducible modules ${\irrmodule}$,
$\socohomology{1}{\module}$ can be easily calculated in terms of
$\socohomology{1}{\irrmodule}$, thus allowing classification of
all possible conformal differential equations.

Practical calculating of $\cohomology{p}$ may be difficult for a
general pair $\psubalg\subset\alg$ because the structure of
generalized Verma modules is not known in the general case. In the
relatively simple case where $\psubalg=\iso(\dmn)\oplus\so(2)$ and
$\alg=\so(\dmn+2)$ we calculate the structure of
generalized  Verma modules using the results of \cite{Vogan,[BB]}.
This allows us to calculate $\socohomology{p}{\module}$ for any
integrable with respect to $\iso(\dmn)\oplus\so(2)$
module $\module$.

Let us note that our approach has significant parallels with
important earlier works. In particular, the relation between
conformally quasi-invariant\footnote{\label{quasiinvariant} An
operator $g$ is called $\sof$-quasi-invariant for a Lie algebra
$\sof$ if for any $f\in\sof$ there exists an operator $h$ such
that $[g,f]=hg$.} differential operators and singular vectors in
the generalized Verma modules of the conformal algebra was
originally pointed out in~\cite{Kos} for a particular case. For
any semi-simple Lie algebra $\sof$ and some its parabolic
subalgebra $\psubalg$, a correspondence between
homogeneous $\sof$-(quasi)invariant linear differential operators
acting on a finite set of $\psubalg$-covariant fields and jet
bundle $\psubalg$-homomorphisms was studied in \cite{ER}.
Namely, let
the Lie groups ${\cA}$ and
${\cP}\subset \cA$ correspond to the Lie algebras $\alg$ and
$\psubalg$, respectively,
and $\cE$ and $\cF$ be homogeneous
vector bundles with the base $\cA / \cP$ and, respectively, the
fibers ${\smm E}$ and ${\smm F}$ being some finite dimensional
$\psubalg$-modules. $J^k\cE$ is the $k^{\rm th}$ associated jet
bundle of $\cE$. By taking the projective limit
\begin{equation} J^\infty\cE\rightarrow\cdots\rightarrow
J^{k+1}\cE\rightarrow J^k\cE\rightarrow \cdots\rightarrow
J^1\cE\rightarrow\cE
\end{equation}
one finds \cite{ER} that there exists a
class of $\alg$-(quasi)invariant linear
differential operators corresponding to $\alg$-homo\-mor\-phisms
$J^\infty\cE\rightarrow J^\infty\cF$. To establish relation with
our approach one observes that the $\alg$-module dual to the
module $J^\infty\cE$ identifies with the generalized Verma module
induced from the $\psubalg$-module ${\smm E}$, i.e. $
\soW=(J^\infty\cE)^\natural, $ where~$(J^\infty\cE)^\natural$ is
the contragredient module to~$J^\infty\cE$. The image of the
highest--weight subspace of~$(J^\infty\cF)^\natural$
in~$(J^\infty\cE)^\natural$ under the dual mapping
$(J^\infty\cF)^\natural\rightarrow (J^\infty\cE)^\natural$ is
spanned by singular vectors. We expect that $\Minkowski$ in our
construction corresponds to the big cell of~$\cA / \cP$ and the
sections of the bundle~$\soW\times\Minkowski$
satisfying~(\ref{covariant_constancy_equation}) along with
appropriate boundary conditions coincide with sections of the
bundle~$J^\infty\cE$ over~$\cA / \cP$.

The  approach developed in this paper allows one to classify  all
$\alg$-invariant homogeneous differential equations on a finite
number of fields that form finite dimensional modules of a
parabolic subalgebra $\psubalg\subset\alg$ with the Abelian
radical $\rad\subset\psubalg$. Equations of this class are
referred to as $\alg_{\psubalg}$-invariant equations in the rest
of this paper. In particular we give the full list of conformally
invariant equations in Minkowski space. In the case of  even
space--time dimension this list is broader than that of  \cite{ER}
because of taking into account the equations resulting from
subsingular vectors.

Apart from giving a universal tool for classification of various
$\alg$-invariant linear equations, the unfolded formulation is
particularly useful for the study of their nonlinear
deformations~\cite{Ann}. Once some set of linear equations is
formulated in the unfolded form (\ref{zero_curvature_equation})
(\ref{covariant_constancy_equation}), the problem is to check if
there exists a nonlinear unfolded system (\ref{unfolded_system}),
which gives rise to the linear equations in question in the free
field limit. In particular, nonlinear dynamics of higher spin
gauge fields in various dimensions was formulated this way in
\cite{more,VD}. This paper is the first step towards realization
of the full scale program of the study of nonlinear deformations
of $\alg$-invariant equations. In fact, the analysis of this paper
clarifies some ways towards nonlinear deformation. In particular,
one can consider extensions of the modules $\module$ associated
with the free fields of the model by the ``current" modules
contained in the tensor products of $\module$.

Let us note that the unfolded equations (\ref{unfolded_system})
can be thought of as a particular $L_\infty$ algebra \cite{A-inf,A-inf_}
(and references therein). The specific property of the
system (\ref{unfolded_system}), extensively used in the analysis
of higher spin models \cite{Ann,more,VD}, is that it is invariant
under diffeomorphisms and, therefore, is ideally suited for the
description of theories which contain gravity. It is important to
note that in this case a nonlinear deformation within the system
(\ref{unfolded_system}) may deform the $\alg$-symmetry
transformations by some field-dependent terms originating from
(\ref{gauge_transformations}), that may complicate the description
of this class of deformations within the manifestly
$\alg$-symmetric schemes. For example this happens when gravity
or (conformal gravity) is described in this formalism with the
Weyl tensor 0-form interpreted as a particular dynamical field of
the system, added to the right hand side of
(\ref{sup_zero_curvature_equation}) \cite{Ann,tri}. Note that such
a deformation is inevitable in any theory of gravitation because
no global symmetry $\alg$ is expected away from a particular
$\alg$-symmetric vacuum.  Within unfolded formulation deformations of this
class also admit a natural module extension interpretation.

The content of the rest of the paper is as follows. In section
\ref{Sct.:_Examples of the Simplest Conformal Systems} we consider
unfolded formulation of some simple conformal systems. In
particular, conformal scalar is considered in section
\ref{Sct.:_Conformal_Scalar}, conformal spinor is considered in
section \ref{Sct.:_Conformal_Spinor}, conformal $p$-forms are
considered in section \ref{Sct.:_Conformal_p-Forms} and $\dmn=4$
electrodynamics is considered in section
\ref{Sct.:_M4_Electrodynamics}. The general construction, which
allows us to classify $\alg_{\psubalg}$-invariant linear differential
equations for any semi-simple Lie algebra $\alg$ and $\psubalg\subset\alg$ with
Abelian radical $\rad$ is given in
section \ref{Sct.:_General_Construction}. In section
\ref{Sct.:_Conformal_Systems_of_equations} we apply this
construction to the conformal algebra $\so(\dmn,2)$. Irreducible
finite-dimensional representations of the Lorentz algebra are
considered in section
\ref{Sct.:_Irreducible_Tensors_and_Spin-tensors}. Conformal
modules (in particular generalized Verma modules and
contragredient to generalized Verma modules) are discussed in
sections \ref{Sct.:_Generalized_Verma_Modules} and
\ref{Sct.:_Contragredient_Modules}, respectively. In section
\ref{Sct.:_Structure of so(M+2) generalized Verma modules} we
collect relevant facts about submodule structure of conformal
generalized Verma modules for the cases of odd (section
\ref{Sct.:_M=2q+1}) and even (section \ref{Sct.:_M=2q})
space--time dimensions. Cohomology with coefficients in
irreducible conformal modules is calculated in section
\ref{Sct.:_Cohomology of irreducible o(M+2)-modules}.
Examples of calculating cohomology with coefficients in reducible
conformal modules are given in section \ref{Sct.:_Examples of
calculating cohomology for reducible conformal modules}. In
section \ref{Sct.:_Conformal_equations} we formulate an algorithm
that permits us to obtain explicit form of any conformal equation
thus completing the analysis of conformally invariant equations.
Conformal generalizations of the Klein--Gordon and the Dirac
equations to the fields with block--type (rectangular) Young
symmetries are given in section
\ref{Sct.:_Conformal_Klein-Gordon_and_Dirac_-like_equations_for_a_block}.
Generalization of $\dmn=4$ equations for massless higher spin
fields to a broad class of tensor fields in the flat space of
arbitrary even dimension is given in section
\ref{Sct.:_Conformal_higher_spins_in_even_dimensions}.
Fradkin--Tseytlin conformal higher spin equations in even
dimensions are considered in section
\ref{Sct.:_F_T_conformal_higher_spins_in_even_dimensions}. In
section \ref{Sct.:_Conclusions} we conclude our results. In
Appendix A we sketch the analysis of submodule structure of
generalized Verma  modules for odd
and even
dimensions. Corresponding homomorphism
diagrams are given in Appendix~\ref{app:hom_diag}.

\section{The Simplest Conformal Systems
\label{Sct.:_Examples of the Simplest Conformal Systems}}
The nonzero commutation relations of the conformal algebra
$\so(\dmn,2)$ are
\begin{alignat}{2}\label{conformal_algebra_commutators}
   &\commut{\cL^{mn}}{\cL^{rs}}=\eta^{mr}\cL^{ns}+\eta^{ms}
   \cL^{rn}-\eta^{nr}\cL^{ms}-\eta^{ns}\cL^{rm},\kern-80pt\notag\\
   &\commut{\cL^{mn}}{\cP^s}=\eta^{ms}\cP^n-\eta^{ns}\cP^m,
   &&\commut{\cL^{mn}}{\cK^s}=\eta^{ms}\cK^n-\eta^{ns}\cK^m,\\
   &\commut{\cD}{\cP^n}=-\cP^n, &&\commut{\cD}{\cK^n}=\cK^n,\notag\\
   &\commut{\cP^n}{\cK^m}=2\eta^{nm}\cD+2\cL^{nm},\notag
\end{alignat}
where $\eta^{mn}$ is an invariant metric of the Lorentz algebra
$\so(\dmn-1,1)$ and $\cL^{nm}$, $\cP^n$, $\cK^n$, and $\cD$ are
generators of $\so(\dmn-1,1)$ Lorentz rotations, translations,
special conformal transformations and dilatation, respectively.
Minkowski metric  $\eta^{mn}$ and its inverse $\eta_{mn}$ are
used to raise and lower Lorentz indices.

Let $\bundsec{}{}= e_\eta\fields{}{\eta}$
be a 0-form section of the trivial
bundle $\Minkowski\times\module$.
Here $\module$ is
some $\so(\dmn,2)$-module. In most examples in this section we
consider the case with an irreducible module $\module \sim
\irrVermaex{\Gd}$ where $\irrVermaex{\Gd}$ is a
 quotient of
the generalized Verma module $\Vermaex{\Gd}$ freely generated by
$\cK^n$ from a vacuum Lorentz representation $\vacVermaex{\Gd}^A$
having a definite conformal weight $\Gd\in \oC$
\begin{equation}
\cD\vacVermaex{\Gd}^A=\Gd\vacVermaex{\Gd}^A
\end{equation}
and annihilated by $\cP^n$
\begin{equation}
\cP^n\vacVermaex{\Gd}^A=0\,.
\end{equation}

To describe  Minkowski space in Cartesian coordinates,
we choose the flat connection
\begin{equation}
\diff=\xi^n(\ptl_n+\cP_n)\,.
\end{equation}

\subsection{Conformal Scalar\label{Sct.:_Conformal_Scalar}}
In order to describe a conformal scalar field let us consider
the generalized Verma module $\Vermaex{\Gd,0}$ induced from the
trivial Lorentz representation with the basis vector $\vacVermaex{\Gd,0}$
satisfying $\cL^{nm}\vacVermaex{\Gd,0}=0$.
The generic element of $\Vermaex{\Gd,0}$ is
\begin{equation}\label{scalar_generalized_Verma_module}
\sum_{l=0} \frac{1}{l!} C_{n_1\ldots n_l} \cK^{n_1}\ldots
\cK^{n_l}\vacVermaex{\Gd,0}\,,
\end{equation}
where $C_{n_1\ldots n_l}\in\oC$ are totally symmetric tensor
coefficients.

Let $\bundsectex{\Gd,0}$ be a section of the trivial bundle
$\Minkowski\times\Vermaex{\Gd,0}$, i.e.,
\begin{equation}
\bundsectex{\Gd,0}=\sum_{l=0} \frac{1}{l!} C_{n_1\ldots n_l}(x)
\cK^{n_1}\ldots \cK^{n_l}\vacVermaex{\Gd,0}\,,
\end{equation}
where $C_{n_1\ldots n_l}(x)$ are some  functions on
$\Minkowski$. The covariant constancy
condition~(\ref{covariant_constancy_equation}) for the field
$\bundsectex{\Gd,0}$
\begin{equation}\label{covariant_scalar}
\diff\bundsectex{\Gd,0}=0
\end{equation}
is equivalent to the infinite system of equations
\begin{equation}\label{covariant_scalar_'}
\ptl_n\bundsectlex{\Gd,0}{l-1}+\cP_n\bundsectlex{\Gd,0}{l}=0\,,\qquad
l\geq 1\,,
\end{equation}
where
\begin{equation}
\bundsectlex{\Gd,0}{l}=\frac{1}{l!}C_{n_1\ldots
n_l}(x)\cK^{n_1}\ldots\cK^{n_l} \vacVermaex{\Gd,0}\,.
\end{equation}
With the definition
\begin{equation}
\ptl_n\vacVermaex{\Gd,0}=0\,,
\end{equation}
(\ref{covariant_scalar_'}) amounts to the system of equations
\begin{equation}\label{covariant_scalar_l}
\ptl_n C_{m_1\ldots m_{l-1}}(x)+2(\Gd+l-1) C_{nm_1\ldots
m_{l-1}}(x)-(l-1)\eta_{n(m_1}C_k{}^k{}_{m_2\ldots m_{l-1})}(x)=0
\end{equation}
for $l\geq 1$, where parentheses imply symmetrization over the
indices denoted by the same letter, i.e.,
\begin{equation}\label{parentheses_meaning}
\eta_{n(m_1}C_k{}^k{}_{m_2\ldots
m_{l-1})}=\frac{1}{l-1}(\underbrace{\eta_{nm_1}C_k{}^k{}_{m_2\ldots
m_{l-1}}+\eta_{nm_2}C_k{}^k{}_{m_1m_3\ldots m_{l-1}}+\ldots}_{l-1
\scriptsize \hbox{ terms}})\,.
\end{equation}

For $\Gd\not\in\half\oZ$ (\ref{covariant_scalar_l}) expresses all
tensors $C_{m_1\ldots m_{l}}(x)$ via the derivatives of $C(x)$
imposing no differential conditions on the latter. For
half-integer $\Gd$ the situation is more interesting. For
example, for $\Gd=\half\dmn-1$ system (\ref{covariant_scalar_l})
imposes the Klein--Gordon equation on $C(x)$ and expresses all
higher rank tensors in terms of the higher derivatives of~$C(x)$
and~$C^{mn}(x)\eta_{mn}$. Indeed, the first two equations in
(\ref{covariant_scalar_l}) are
\begin{align}
\label{covariant_scalar_1}
&\ptl_n C(x)+2\Gd C_n(x)=0\,,\\
\label{covariant_scalar_2} &\ptl_n
C_m(x)+2(\Gd+1)C_{nm}(x)-\eta_{nm}C_k{}^k(x)=0\,.
\end{align}
Contracting (\ref{covariant_scalar_2}) with $\eta^{nm}$ and
substituting $C_n(x)$ from (\ref{covariant_scalar_1}) we obtain
\begin{equation}\label{covariant_scalar_1_2}
-\frac{1}{2\Gd}\square C(x)+(2\Delta+2-\dmn)C_k{}^k(x)=0\,.
\end{equation}
Thus, for $\Gd=\frac{1}{2}\dmn-1$, $\Gd\neq 0$ (i.e., $\dmn\neq
2$) (\ref{covariant_scalar_1_2}) is equivalent to the
Klein--Gordon equation for $C(x)$
\begin{equation}\label{Klein_Gordon_equation}
\square C(x)=0\,.
\end{equation}

Algebraically, the situation is as follows. Whenever $\Gd$ is not
half-integer $\cP_n\bundsectlex{\Gd,0}{l}\neq 0$ for any
$\bundsectlex{\Gd,0}{l}$ with $l \ge 1$ and the module
$\Vermaex{\Gd,0}$ is irreducible. This means that it is possible
to solve the chain (\ref{covariant_scalar_l}) by expressing each
$\bundsectlex{\Gd,0}{l}$ via derivatives of
$\bundsectlex{\Gd,0}{l-1}$ for~($l \ge 1$). Abusing notations,
$\bundsectlex{\Gd,0}{l}=-(\cP^{-1})^n\ptl_n
\bundsectlex{\Gd,0}{l-1}\,,\quad l\geq1\,.$ For $\Gd=\half\dmn-1$,
the module $\Vermaex{\Gd,0}$ is reducible because the identity
\begin{equation}\label{singular_module_for_scalar}
\cP_n\singbasis{}=  0\,,\qquad \singbasis{}=\cK_m \cK^m
\vacVermaex{\Gd,0}
\end{equation}
implies that $\singbasis{}$ is a singular vector,
 i.e.{} it is a vacuum vector of the
submodule~$\subVermaex{\Gd,0}\subset\Vermaex{\Gd,0}$ generated
from $\singbasis{}$ by $\cK^n$. Effectively, the algebraic
condition (\ref{singular_module_for_scalar}) imposes the
Klein--Gordon equation on
$\bundsectlex{\Gd,0}{0}=C(x)\vacVermaex{\Gd,0}$. The same time,
since the coefficient in front of $C_n^n\cK_m\cK^m
\vacVermaex{\Gd,0}\in\bundsectlex{\Gd,0}{2}$ in equation
(\ref{covariant_scalar_l}) with $l=2$ vanishes, $C_n^n(x)$ cannot
be expressed in terms of derivatives of $\bundsectex{\Gd,0}$, thus
becoming an independent field. Setting $C_n^n(x)=0$ is equivalent
to restriction of $\Minkowski\times\Vermaex{\Gd,0}$ to the bundle
$\oR^{\dmn}\times\irrVermaex{\Gd,0}$ with the irreducible fiber
$\irrVermaex{\Gd,0}=\Vermaex{\Gd,0}/\subVermaex{\Gd,0}$. As a
result, the conformally invariant
equation~(\ref{Klein_Gordon_equation})
corresponds to the irreducible
$\so(\dmn,2)$-module~$\irrVermaex{\Gd,0}$, thus being primitive.

More generally, the  generalized Verma module $\Vermaex{\Gd,0}$ is
reducible for  $\Gd=\half\dmn-n$. Starting
from $\Vermaex{\frac{\dmn}{2}-n,0}$ one obtains the conformal equation
$\square^n C(x)=0$ associated with
$\irrVermaex{\frac{\dmn}{2}-n}=\Vermaex{\frac{\dmn}{2}-n}/
\subVermaex{\frac{\dmn}{2}-n}$.

\subsection{Conformal Spinor\label{Sct.:_Conformal_Spinor}}
Massless Dirac equation admits an analogous reformulation. Let
the module $\Vermaex{\Gd,1/2}$ be generated by $\cK^n$ from the
spinor module of the $\so(\dmn-1,1)$ subalgebra with the basis
elements $\vacVermaex{\Gd,1/2}^\ga$ ($\ga=1,\ldots
,2^{{[}\dmn/2]}$ is the spinor index)
\begin{equation}
\cL^{nm}\vacVermaex{\Gd,1/2}^\ga=\frac{1}{4}(\gga^m\gga^n-\gga^n\gga^m)^\ga{}_\gb
\vacVermaex{\Gd,1/2}^\gb\,.
\end{equation}
Here $\gga^n{}^\ga{}_\gb$ are gamma matrices
\begin{equation}\label{gamma_matrices}
\gga^{n\gga}{}_\gb\gga^{m\ga}{}_\gga+
\gga^{m\gga}{}_\gb\gga^{n\ga}{}_\gga=
(\gga^n\gga^m+\gga^m\gga^n)^\ga{}_\gb=2\eta^{nm}\gd^\ga_\gb\,.
\end{equation}
The covariant constancy condition
(\ref{covariant_constancy_equation}) imposed on the field
\begin{equation}
\bundsectex{\Gd,1/2}=\sum_{l=0} \frac{1}{l!} C_{m_1 \ldots
m_{l},\ga}(x) \cK^{m_1}\ldots \cK^{m_l}\vacVermaex{\Gd,1/2}^\ga\,,
\end{equation}
(i.e.{} on the section of the
bundle~$\Minkowski\times\Vermaex{\Gd,1/2}$) is equivalent to the
system of equations
\begin{align} \label{covariant_spinor_l}
&\ptl_n C_{m_1 \ldots m_{l-1},\ga}(x)+2(\Gd+l-1) C_{nm_1\ldots
m_{l-1},\ga}(x)-(l-1)\eta_{n(m_1}C_{k}{}^k{}_{m_2 \ldots m_{l-1}),\ga}(x)+\nn\\
&{}+\half(\gga^q\gga_n-\gga_n\gga^q)^\gb{}_\ga C_{qm_1
\ldots m_{l-1},\gb}(x)=0\,,\qquad l\geq 1\,.
\end{align}
Whenever $\Gd$ is not half-integer, the system
(\ref{covariant_spinor_l}) just expresses all higher rank
spinor--tensors in terms of higher derivatives of $C_\ga(x)$. For
example, from (\ref{covariant_spinor_l}) it follows that ($l=1$)
\begin{equation}\label{covariant_spinor_1}
\gga^{n\ga}{}_\gb\Big(\ptl_n
C_\ga(x)+(2\Gd-\dmn+1)C_{n,\ga}(x)\Big)=0\,.
\end{equation}

For $\Gd=(\dmn-1)/2$ the coefficient in front of $C_{n,\ga}(x)$
vanishes and we arrive at the massless Dirac equation for
$C_\ga(x)$
\begin{equation}\label{Dirac_equation}
\gga^{n\ga}{}_\gb\ptl_n C_\ga(x)=0\,.
\end{equation}
Other equations of the system (\ref{covariant_spinor_l}) with
$\Gd=(\dmn-1)/2$ express higher rank spinor--tensors in terms of
higher derivatives of~$C_\ga(x)$ and~$\gga^{n\ga}{}_\gb
C_{n,\ga}(x)$.

Algebraically, the situation is analogous to the case of the
Klein--Gordon equation. For $\Gd=(\dmn-1)/2$ the module
$\Vermaex{\Gd,1/2}$ is reducible. It contains the submodule
$\subVermaex{(\dmn-1)/2,1/2}\subset\Vermaex{(\dmn-1)/2,1/2}$
generated by $\cK_n$ from the singular vectors
\begin{equation}
\singbasis{\ga}=\gga_m{}^\ga{}_{\gb}\cK^m
\vacVermaex{(\dmn-1)/2,1/2}^\gb\,.
\end{equation}
Setting $\gga^{n\ga}{}_\gb C_{n,\ga}(x)=0$ is equivalent to the
restriction to the subbundle
$\Minkowski\times\irrVermaex{(\dmn-1)/2,1/2}$, where the
irreducible module
$\irrVermaex{(\dmn-1)/2,1/2}=\Vermaex{(\dmn-1)/2,1/2}/
\subVermaex{(\dmn-1)/2,1/2}$ corresponds to the primitive
conformal equation~(\ref{Dirac_equation}).

\subsection{Conformal p-Forms\label{Sct.:_Conformal_p-Forms}}
Consider a trivial bundle $\Minkowski\times\Vermaex{\Gd,p}$, where
the module $\Vermaex{\Gd,p}$ is induced from the rank $p$
$(p\le\dmn)$ totally antisymmetric tensor module of
$\so(\dmn-1,1)$ with the basis $\vacVermaex{\Gd,p}^{k_1\ldots
k_p}$
\begin{equation}
\cL_{nm}\vacVermaex{\Gd,p}^{k_1\ldots k_p}=
p\gd_n^{{[}k_1}\vacVermaex{\Gd,p}_m{}^{k_2\ldots k_p]}-
p\gd_m^{{[}k_1}\vacVermaex{\Gd,p}_n{}^{k_2\ldots k_p]}\,.
\end{equation}
Here square brackets imply antisymmetrization over indices
denoted by the same letter
\begin{equation}\label{square_braces_meaning}
\gd_n^{{[}k_1}\vacVermaex{\Gd}_m{}^{k_2\ldots k_p]}=
\frac{1}{p}(\underbrace{\gd_n^{k_1}\vacVermaex{\Gd}_m{}^{k_2\ldots
k_p}- \gd_n^{k_2}\vacVermaex{\Gd}_m{}^{k_1 k_3\ldots
k_p}+\ldots}_{p \scriptsize\hbox{ terms}})\,.
\end{equation}
Consider a section $\bundsectex{\Gd,p}$ of the bundle
$\Minkowski\times\Vermaex{\Gd,p}$
\begin{equation}
\bundsectex{\Gd,p}=\sum_{l=0}\frac{1}{l!}C_{m_1\ldots
m_l;k_1\ldots k_p}(x)
\cK^{m_1}\ldots\cK^{m_l}\vacVermaex{\Gd,p}^{k_1\ldots k_p}\,,
\end{equation}
where the tensor $C_{m_1 \dots m_l;k_1\ldots k_p}(x)$ is totally
symmetric in the indices $m$ and totally antisymmetric in the
indices $k$. (The  semicolon separates the groups of totally
symmetric and antisymmetric indices).

Equation (\ref{covariant_constancy_equation}) for the field
$\bundsectex{\Gd,p}$ amounts to
\begin{align}\label{covariant_forms_l}
\ptl_n C_{m_1 \ldots m_{l-1};k_1 \ldots k_p}(x)&+
2(\Gd+l-1)C_{nm_1\ldots m_{l-1};k_1 \ldots k_p}(x)-
(l-1)\eta_{n(m_1} C_q{}^q{}_{m_2 \ldots m_{l-1});k_1 \ldots k_p}(x)+\nn\\
{}&+2pC_{m_1 \ldots m_{l-1}{[}k_1;n k_2\ldots k_p]}(x)-
2p\eta_{n{[}k_1}C_{m_1\ldots m_{l-1}q;}{}^q{}_{k_2\ldots
k_p]}(x)=0\,, \qquad l\geq 1\,.
\end{align}
The differential equations imposed by the system (\ref{covariant_forms_l})
depend on the conformal weight $\Gd$.

\begin{enumerate}
\item $\Gd\not\in\half\oZ$.\\
(\ref{covariant_forms_l})  imposes no differential restrictions,
just expressing all higher rank tensor fields in terms of
derivatives of the field $C_{;k_1\ldots k_p}(x)$.
\item $\dmn$ is
odd, $\Gd=p=0\,,1\,,\ldots,\dmn-1$ or $\dmn$ is even,
$\Gd=p=0,\frac{\dmn}{2}+1,\frac{\dmn}{2}+2,\ldots,\dmn-1$.\\
In this case (\ref{covariant_forms_l}) imposes the closedness
condition on the $\Gd$-form $C_{;k_1\ldots k_\Gd}(x)$
\begin{equation}\label{closery_condition}
\ptl_{{[}k_{\Gd+1}}C_{;k_1\ldots k_\Gd]}(x)=0
\end{equation}
and expresses all higher rank tensor fields in terms of
derivatives of $C_{;k_1\ldots k_\Gd}(x)$ and
$C_{[k_{\Gd+1};k_1\ldots k_\Gd]}(x)$. Actually, consider
(\ref{covariant_forms_l}) at $l=1$. We have
\begin{equation}\label{covariant_forms_1}
\ptl_n C_{;k_1 \ldots k_p}(x)+ 2\Gd C_{n;k_1 \ldots
k_p}(x)+2pC_{{[}k_1;n k_2\ldots k_p]}(x)-
2p\eta_{n{[}k_1}C_{q;}{}^q{}_{k_2\ldots k_p]}(x)=0\,.
\end{equation}
Total antisymmetrization of indices in
(\ref{covariant_forms_1}) gives
\begin{equation}\label{covariant_forms_1_antisymmetrization}
\ptl_{{[}k_{p+1}}C_{;k_1\ldots
k_p]}(x)+2(\Gd-p)C_{{[}k_{p+1};k_1\ldots k_p]}(x)=0\,,
\end{equation}
For $\Gd=p$ we obtain (\ref{closery_condition}).
\item $\dmn$ is
odd, $\Gd=\dmn-p=0,1,\ldots,\dmn-1$ or $\dmn$ is even,
$\Gd=\dmn-p=0,\frac{\dmn}{2}+1,\frac{\dmn}{2}+2,\ldots,\dmn-1$.\\
In this case (\ref{covariant_forms_l}) imposes the dual form of
equation (\ref{closery_condition}) implying that the polyvector
$C^{;k_1\ldots k_{\dmn-\Gd}}(x)$ conserves
\begin{equation}\label{conservation_condition}
\ptl_nC^{;nk_2\ldots k_{\dmn-\Gd}}(x)=0\,.
\end{equation}
Also (\ref{covariant_forms_l}) expresses all higher rank tensor
fields in terms of derivatives of the fields $C^{;k_1\ldots
k_{\dmn-\Gd}}(x)$ and $C_q{}^{;qk_2\ldots k_{\dmn-\Gd}}(x)$.
Indeed, contracting indices in (\ref{covariant_forms_1}) with
$\eta^{nk_1}$, one obtains (\ref{conservation_condition}) from
\begin{equation}\label{covariant_forms_1_contruction}
\ptl^nC_{;n k_2 \ldots k_p}(x)+2(\Gd+p-\dmn)C^n{}_{;nk_2\ldots
k_p}(x)=0\,.
\end{equation}

\item $\dmn$ is even, $\Gd=p=1,2,\ldots,\frac{\dmn}{2}-1$.\\
In this case, (\ref{covariant_forms_l}) imposes on $C_{;k_1\ldots
k_\Gd}(x)$ equation (\ref{closery_condition}) along with equation
\begin{equation}\label{subsingular_forms_closeness}
\square^{\dmn/2-\Gd}\ptl^nC_{;nk_2\ldots k_\Gd}(x)=0
\end{equation}
and expresses all higher rank tensor fields in terms of
derivatives of the fields $C_{;k_1\ldots k_\Gd}(x)$,
$C_{[k_{\Gd+1};k_1\ldots k_\Gd]}(x)$, and
$C^{n_1\ldots n_{\dmn/2-\Gd}}{}_{n_1\ldots n_{\dmn/2-\Gd}}{}^q{}_{;qk_2\ldots
k_{\Gd}}(x)$.
\item $\dmn$ is even, $\Gd=\dmn-p=1,2,\ldots,\frac{\dmn}{2}-1$.\\
Now (\ref{covariant_forms_l}) imposes on $C_{;k_1\ldots
k_{\dmn-\Gd}}(x)$ equation (\ref{conservation_condition}) along
with
\begin{equation}\label{subsingular_forms_conservation}
\square^{\dmn/2-\Gd}\ptl_{[k_{\dmn-\Gd+1}}C_{;k_1\ldots k_{\dmn-\Gd}]}(x)=0
\end{equation}
and expresses all higher rank tensor fields in terms of
derivatives of the fields $C_{;k_1\ldots k_{\dmn-\Gd}}(x)$,
$C^{n_1\ldots n_{\dmn/2-\Gd}}{}_{n_1\ldots
n_{\dmn/2-\Gd}[k_{\dmn-\Gd+1};k_1\ldots k_{\dmn-\Gd}]}(x)$, and
$C^q{}_{;qk_2\ldots k_{\dmn-\Gd}}(x)$. Note that system
(\ref{conservation_condition}),
(\ref{subsingular_forms_conservation}) is dual to system
(\ref{closery_condition}), (\ref{subsingular_forms_closeness}).

\item $\dmn$
is even, $\Gd=p=\frac{\dmn}{2}$.\\ In this case, the vacuum
vectors $\ketv{\dmn/2,\dmn/2}^{k_1\ldots k_{\dmn/2}}$ form a
reducible $\so(\dmn,2)$-module. The irreducible
parts are singled out by the
additional (anti)selfduality conditions
\begin{equation}\label{anti_selfduality_condition}
\ketv{\dmn/2,\dmn/2}^{k_1\ldots
k_{\dmn/2}}_\pm=\pm\frac{i^{\dmn^2/4}}{(\dmn/2)!} \gep_{p_1\ldots
p_{\dmn/2}}{}^{k_1\ldots k_{\dmn/2}}
\ketv{\dmn/2,\dmn/2}^{p_1\ldots p_{\dmn/2}}_\pm\,,
\end{equation}
which in the complex case  can be imposed for any even space--time
dimension. Equation (\ref{covariant_forms_l}) imposes primitive
equation
\begin{equation}
\label{antiselfduality_equation_1}
\ptl^nC_{;nk_2\ldots k_{\dmn/2}}(x)=0\,
\end{equation}
on the (anti)selfdual field $C_{;k_1\ldots k_{\dmn/2}}(x)$
\begin{equation}\label{antiselfduality_equation_2}
C_{;k_1\ldots k_{\dmn/2}}(x)=\pm\frac{i^{\dmn^2/4}}{(\dmn/2)!}
\gep^{p_1\ldots p_{\dmn/2}}{}_{k_1\ldots k_{\dmn/2}} C_{;p_1\ldots
p_{\dmn/2}}(x)
\end{equation}
and expresses all higher rank tensor fields in terms of
derivatives of the fields $C_{;k_1\ldots k_{\dmn/2}}(x)$ and
$C^q{}_{;qk_2\ldots k_{\dmn/2}}(x)$.
\end{enumerate}

Vanishing coefficients in front of higher tensors in
(\ref{covariant_forms_1_antisymmetrization}) and
(\ref{covariant_forms_1_contruction}) imply
the appearance of the singular vectors
\begin{align}
\label{singular_vector_for_closure_condition}
&\singbasis{k_1\ldots k_{\Gd+1}}=
\cK^{{[}k_1}\vacVermaex{\Gd,\Gd}^{k_2\ldots k_{\Gd+1}]}\,,\\
\label{singular_vector_for_conservation_condition} &\singbasis{k_1
\ldots k_{\dmn-\Gd-1}}= \cK_n\vacVermaex{\Gd,\dmn-\Gd}^{n
k_1\ldots k_{\dmn-\Gd-1}}\,
\end{align}
in $\Vermaex{\Gd,p}$ for $\Gd=p=0,\ldots,\dmn-1$ and
$\Gd=\dmn-p=0,\ldots,\dmn-1$, respectively. These singular vectors
induce proper submodules
$\subVermaex{\Gd,\Gd}\subset\Vermaex{\Gd,\Gd}$ and
$\subVermaex{\Gd,\dmn-\Gd}\subset\Vermaex{\Gd,\dmn-\Gd}$. In the
cases 2 and~3 the  quotients
$\quotVermaex{\Gd,\Gd}=\Vermaex{\Gd,\Gd}/\subVermaex{\Gd,\Gd}$ and
$\quotVermaex{\Gd,\dmn-\Gd}=\Vermaex{\Gd,\dmn-\Gd}/
\subVermaex{\Gd,\dmn-\Gd}$ are irreducible and, therefore,
equations (\ref{closery_condition}) and
(\ref{conservation_condition}) are primitive. In the cases~4 and~5
the modules $\quotVermaex{\Gd,\Gd}$  and
$\quotVermaex{\Gd,\dmn-\Gd}$  are reducible. They contain
submodules $\subVermaex{\Gd,\Gd}'\subset\quotVermaex{\Gd,\Gd}$ and
$\subVermaex{\Gd,\dmn-\Gd}'\subset\quotVermaex{\Gd,\dmn-\Gd}$
generated from the subsingular vectors
\begin{align}
&\ketv{s'}^{k_1\ldots k_{\Gd-1}}=(\cK_n\cK^n)^{\dmn/2-\Gd}
\cK_m\ketv{\Gd,\Gd}^{mk_1\ldots k_{\Gd-1}}\,,\\
&\ketv{s'}^{k_1\ldots k_{\dmn-\Gd+1}}=(\cK_n\cK^n)^{\dmn/2-\Gd}
\cK^{[k_1}\ketv{\Gd,\Gd}^{k_2\ldots k_{\dmn-\Gd+1}]}\,,
\end{align}
respectively. The quotients
$\quotVermaex{\Gd,\Gd}'=\quotVermaex{\Gd,\Gd}/\subVermaex{\Gd,\Gd}'$
and $\quotVermaex{\Gd,\dmn-\Gd}'
=\quotVermaex{\Gd,\dmn-\Gd}/\subVermaex{\Gd,\dmn-\Gd}'$ are
irreducible and systems (\ref{closery_condition}),
(\ref{subsingular_forms_closeness}) and
(\ref{conservation_condition}),
(\ref{subsingular_forms_conservation}) are primitive. Note that
in the cases~4 and~5 the systems (\ref{closery_condition}) and
(\ref{conservation_condition}) alone are also conformally
invariant but non-primitive.

In the case 6 the singular vector
(\ref{singular_vector_for_closure_condition}) coincide (modulo a
sign) with the singular vector
(\ref{singular_vector_for_conservation_condition}). This vector
contained in both generalized Verma modules
$\Vermaex{\dmn/2,\dmn/2+}$ and $\Vermaex{\dmn/2,\dmn/2-}$
generated from the selfdual and the antiselfdual vacuum Lorentz
representations correspondingly. The quotients
$\quotVermaex{\dmn/2,\dmn/2\pm}=\Vermaex{\dmn/2,\dmn/2\pm}/\subVermaex{\dmn/2,\dmn/2}$
are irreducible and, therefore, system
(\ref{antiselfduality_equation_1}),
(\ref{antiselfduality_equation_2}) is primitive.

\subsection{$\dmn=4$ Electrodynamics\label{Sct.:_M4_Electrodynamics}}

Primitive conformally invariant equations constructed with the use
of irreducible conformal modules are the simplest ones in the
sense that it is impossible to impose any  stronger conformally
invariant equations that admit nontrivial solutions. As follows
from the examples 4 and 5 in section
\ref{Sct.:_Conformal_p-Forms}, non-primitive equations not
necessarily reduce to a set of independent primitive subsystems.

A somewhat trivial example of a non-primitive system is provided
by the case 6 in section \ref{Sct.:_Conformal_p-Forms} with the
relaxed (anti)selfduality condition
(\ref{anti_selfduality_condition}). Namely, consider the module
$\Vermaex{\dmn/2,\dmn/2}$ induced from the reducible vacuum
$\ketv{\dmn/2,\dmn/2}^{k_1,\ldots,k_{\dmn/2}}$. It contains both
singular vectors (\ref{singular_vector_for_closure_condition}) and
(\ref{singular_vector_for_conservation_condition}). Thus the
equation (\ref{covariant_forms_l}) imposes the system
(\ref{closery_condition}), (\ref{conservation_condition}) on the
field $C_{;k_1\ldots k_{\dmn/2}}(x)$ and expresses all higher
rank tensor fields in terms of derivatives of the fields
$C_{;k_1\ldots k_{\dmn/2}}(x)$, $C_{[k_{\dmn/2+1};k_1\ldots
k_{\dmn/2]}}(x)$, and $C^q{}_{;qk_2\ldots k_{\dmn/2}}(x)$. This
system is non-primitive because it reduces to the combination of
the independent subsystems for selfdual and anti-selfdual parts.
For $\dmn =4$ it coincides with the free Maxwell equations
formulated in terms of field strengths.

A less trivial important example of a nontrivial
non-primitive system, which allows us to illustrate the idea of
the general construction is provided by the potential formulation
of the $\dmn=4$ electrodynamics. Consider the $\dmn=4$ irreducible
module $\irrVermaex{A}=\quotVermaex{1,1}/\subVermaex{1,1}'$, (see
explanation to the case 4 at the end of section
\ref{Sct.:_Conformal_p-Forms}). The covariant constancy condition
(\ref{covariant_constancy_equation}) for the section
\begin{equation}
\bundsectex{A}=\sum_{l=0} \frac{1}{l!} A_{m_1\ldots m_l;k} (x)
\cK^{m_1}\ldots \cK^{m_l}\ketv{A}^{k}\,,\qquad m,k=1,\ldots ,4
\end{equation}
encodes the following differential equations on $A_{;k}(x)$:
\begin{align}\label{M4_electrodynamics_A_close}
\ptl_{[ n} A_{;k ]}(x)=0\,, \\
\square\ptl^kA_{;k}(x)=0\,.\label{Maxwell_gauge}
\end{align}
Let  us extend the irreducible module $\irrVermaex{A}$ to a module
$\extmodule_{A,F}$ by ``gluing" the module
$\mathfrak{K}_F=\mathfrak{Q}_{2,2+}\oplus\mathfrak{Q}_{2,2-}$ (see
explanation to the case 6 at the end of section
\ref{Sct.:_Conformal_p-Forms}) to $\irrVermaex{A}$ as follows. The
module $\extmodule_{A,F}$ is generated from the vacuum vectors
$\ketv{A}^k$ and $\ketv{F}^{k_1k_2}$ of the modules
$\Vermaex{A}=\Vermaex{1,1}$ and $\Vermaex{F}=\Vermaex{2,2}$,
respectively, with the following additional relations imposed
\begin{align}
\cK^{[n}\ketv{A}^{k]}=0\,,&&
\cK_m \cK^m \cK_k \ketv{A}^k=0\,,\label{relation_example_IA}\\
\cK^{[n}\ketv{F}^{k_1k_2]}=0\,,&&
\cK_n\ketv{F}^{nk}=0\,,\label{relation_example_IF}\\
\cP^n\ketv{F}^{k_1k_2}=-\eta^{n[k_1}\ketv{A}^{k_2]}\,.
\label{relation_example_IA_IF}
\end{align}
Here the conditions (\ref{relation_example_IA}) and
(\ref{relation_example_IF}) single out $\irrVermaex{A}$ and
$\mathfrak{K}_F$ from the generalized Verma modules $\Vermaex{A}$
and $\Vermaex{F}$, respectively. The condition
(\ref{relation_example_IA_IF}) ``glues" the modules
$\irrVermaex{A}$ and $\mathfrak{K}_F$ into $\extmodule_{A,F}$.

Consider the section
\begin{equation}
\bundsectex{A,F}=\sum_{l=0} \frac{1}{l!} A_{m_1\ldots m_l;k} (x)
\cK^{m_1}\ldots \cK^{m_l}\ketv{A}^k+ \sum_{l=0} \frac{1}{l!}\,
F_{m_1\ldots m_l;k_1k_2} (x) \cK^{m_1}\ldots
\cK^{m_l}\ketv{F}^{k_1k_2}
\end{equation}
of the bundle $\oR^4\times\extmodule_{A,F}$. The covariant
constancy condition $\diff\bundsectex{A,F}=0$ amounts to the
infinite differential system
\begin{eqnarray}
& \ptl_n A_{m_1 \ldots m_{l-1};k}(x)+ 2lA_{nm_1 \ldots
m_{l-1};k}(x)-
(l-1)\eta_{n(m_1} A_q{}^q{}_{m_2 \ldots m_{l-1});k}(x)+\nn\\
&{}+2A_{m_1 \ldots m_{l-1}k;n}(x)- 2\eta_{nk}A_{m_1 \ldots
m_{l-1}q;}{}^q(x)- F_{m_1\ldots m_{l-1};nk}(x)=0\,,
\label{M4_electrodynamics_l_1}
\end{eqnarray}
\begin{eqnarray}
& \ptl_n F_{m_1 \ldots m_{l-1};k_1k_2}(x)+ 2(l+1)F_{nm_1 \ldots
m_{l-1};k_1k_2}(x)-
(l-1)\eta_{n(m_1}F_q{}^q{}_{m_2 \ldots m_{l-1});k_1k_2}(x)+\nn\\
&{}+4F_{m_1 \ldots m_{l-1}[k_1;nk_2]}(x)- 4\eta_{n[k_1}F_{m_1
\ldots  m_{l-1}q;}{}^q{}_{k_2]}(x)=0
\label{M4_electrodynamics_l_2}
\end{eqnarray}
for $l=1,2,\ldots$. The subsystem (\ref{M4_electrodynamics_l_2})
coincides with the system (\ref{covariant_forms_l}) for $\dmn=4$
and $\Gd=p=2$. It expresses all higher components $F_{m_1 \ldots
m_l;k_1k_2}(x)$ via the higher derivatives of the
field~$F_{;k_1k_2}(x)$ (note that components $F^q{}_{;qk_2}(x)$
and $F_{[k_3;k_1k_2]}(x)$ are set to zero in the bundle
$\oR^4\times\extmodule_{A,F}$ due to the relation
(\ref{relation_example_IF})) and imposes Maxwell equations on the
field strength 2-form $F_{;k_1k_2}(x)$
\begin{eqnarray}
\ptl_{[n}F_{;k_1k_2]}(x)=0\label{Maxwell_1}\,,\\
\ptl^n F_{;nk}(x)=0\label{Maxwell_2}\,.
\end{eqnarray}

The subsystem (\ref{M4_electrodynamics_l_1}) is a deformation of
the system (\ref{covariant_forms_l}) for $\irrVermaex{A}$ by the
additional terms containing the fields $F_{m_1 \ldots
m_l;k_1k_2}(x)$  resulting from  the ``gluing" condition
(\ref{relation_example_IA_IF}) which links the vacuums
$\ketv{A}^k$ and $\ketv{F}^{k_1k_2}$. The system
(\ref{M4_electrodynamics_l_1})  expresses all higher fields
$A_{m_1\ldots m_l;k}(x)$ ($l \geq 1$) via the higher derivatives
of~$A_{;k}(x)$ (in $\oR^4\times\extmodule_{A,F}$ components
$A_{[k_2;k_1]}(x)=0$ and $A^n{}_n{}^q{}_{;q}(x)=0$ due to
(\ref{relation_example_IA})) and also imposes the differential
equation (\ref{Maxwell_gauge}) on $A_{;k}(x)$ and the constraint
\begin{equation}
\ptl_{[k_1}A_{;k_2]}(x)=F_{;k_1k_2}(x) \label{Maxwell_potential}
\end{equation}
on $F_{;k_1k_2}(x)$. The constraint (\ref{Maxwell_potential})
replaces  the closedness condition
(\ref{M4_electrodynamics_A_close}) for the potential 1-form
$A_{;k}(x)$. The point is that the singular vector
$\ketv{s}^{k_1k_2}=\cK^{[k_1}\ketv{1,1}^{k_2]}$ from the module
$\Vermaex{A}$ responsible for (\ref{M4_electrodynamics_A_close})
is ``glued" in the module $\extmodule_{A,F}$ by the field
$F_{;k_1k_2}(x)$ in (\ref{relation_example_IA_IF}). As a result,
the field $F_{;k_1k_2}(x)$ replaces
 zero on the right hand side of
(\ref{M4_electrodynamics_A_close}) giving rise to the constraint
(\ref{Maxwell_potential}), which identifies
 $A_{;k}(x)$ with the
potential for the field strength $F_{;k_1k_2}(x)$.

Thus the infinite system (\ref{M4_electrodynamics_l_1}) and
(\ref{M4_electrodynamics_l_2}) provides the potential formulation
of $\dmn=4$ electrodynamics (\ref{Maxwell_1}), (\ref{Maxwell_2})
and (\ref{Maxwell_potential}) along with infinitely many
constraints on the auxiliary fields $A_{m_1 \ldots m_l;k}(x)$ and
$F_{m_1 \dots m_l;k_1k_2}(x)$ for $l\geq 1$. Equation
(\ref{Maxwell_gauge}) is the conformally invariant gauge
condition, considered originally in \cite{Mayer,Bayen_Flato}. The
system (\ref{Maxwell_1}), (\ref{Maxwell_2}),
(\ref{Maxwell_potential}) and (\ref{Maxwell_gauge}) is
non-primitive. Its primitive reduction results from the condition
$F_{;k_1k_2}(x)=0$.

The module $\extmodule_{A,F}$ can be further extended by the
module $\irrVermaex{J}=\Vermaex{3,1}/\subVermaex{3,1}$ (see
explanation to the case 3 at the end of section
\ref{Sct.:_Conformal_p-Forms}) to a module $\extmodule_{A,F,J}$ as
follows. $\extmodule_{A,F,J}$ is generated from the totally
antisymmetric vacua $\ketv{A}^k$, $\ketv{F}^{k_1k_2}$
and~$\ketv{J}^k$ with the properties (\ref{relation_example_IA}),
(\ref{relation_example_IF}), (\ref{relation_example_IA_IF}) along
with
\begin{align}
&\cK_k\ketv{J}^k=0\,,\label{relation_example_IJ}\\
&\cP^n\ketv{J}^k=-\frac{2}{3}\ketv{F}^{nk}\,.
\label{relation_example_IA_IF_IJ}
\end{align}
The covariant constancy condition for the section
\begin{eqnarray}
&&\bundsectex{A,F,J}=\sum_{l=0} \frac{1}{l!} A_{m_1\ldots m_l;k}
(x)\cK^{m_1}\ldots \cK^{m_l}\ketv{A}^k+ \\
&&{}+ \sum_{l=0}\frac{1}{l!}\, F_{m_1\ldots m_l;k_1k_2} (x)
\cK^{m_1}\ldots\cK^{m_l}\ketv{F}^{k_1k_2}+ \sum_{l=0}\frac{1}{l!}
J_{m_1\ldots m_l;k} (x) \cK^{m_1}\ldots \cK^{m_l}\ketv{J}^k
\nonumber
\end{eqnarray}
of the trivial bundle $\oR^4\times\extmodule_{A,F,J}$ contains
several parts.  The first one is the system
(\ref{M4_electrodynamics_l_1}), which gives rise to equations
(\ref{Maxwell_potential}), (\ref{Maxwell_gauge}).  The second one
is the system for the fields~$J_{m_1\ldots m_l;k}(x)$ of the form
(\ref{covariant_forms_l}) with $\dmn=4$ and $\Delta=\dmn-p=3$.
This system encodes equation
\begin{equation}\label{Maxwell_current_conservation}
\ptl^k J_{;k}(x)=0
\end{equation}
on the field~$J_{;k}(x)$ and expresses all the higher
fields~$J_{m_1\ldots m_l;k}(x)$ ($l\geq 1$) in terms of  higher
derivatives of~$J_{;k}(x)$ (in $\oR^4\times\extmodule_{A,F,J}$
component $J^q{}_{;q}(x)=0$ due to
(\ref{relation_example_IJ})). The third part reads
\begin{eqnarray}
& \ptl_n F_{m_1 \ldots m_{l-1};k_1k_2}(x)+ 2(l+1)F_{nm_1 \ldots
m_{l-1};k_1k_2}(x)-
(l-1)\eta_{n(m_1} F_q{}^q{}_{m_2 \ldots m_{l-1});k_1k_2}(x)+\nn\\
&{}+4F_{m_1 \ldots m_{l-1}[k_1;nk_2]}(x)- 4\eta_{n[k_1}F_{m_1
\ldots m_{l-1}q;}{}^q{}_{k_2]}(x)-\frac{2}{3}\eta_{n[k_1}
J_{m_1\ldots m_{l-1};k_2]}(x)=0 \label{M4_electrodynamics_l_3}
\end{eqnarray}
for $l=1,2,\ldots$. It is a deformation  of the system
(\ref{M4_electrodynamics_l_2})  with the additional terms
containing $J_{m_1 \ldots m_l;k}(x)$, which result from the
``gluing" condition (\ref{relation_example_IA_IF_IJ}). This system
encodes the Bianchi identities  (\ref{Maxwell_1}) along with
the second pair of Maxwell equations with external current
\begin{equation}\label{Maxwell_current}
\ptl^n F_{;nk}(x)=J_{;k}(x)
\end{equation}
and expresses $F_{m_1 \ldots m_l;k_1k_2}(x)$ for $l \geq 1$ via
the derivatives of $F_{;k_1k_2}(x)$. Thus the covariant constancy
condition (\ref{covariant_constancy_equation}) for the bundle
$\oR^4\times \extmodule_{A,F,J}$ encodes the non-primitive system
of differential equations (\ref{Maxwell_1}),
(\ref{Maxwell_potential}), (\ref{Maxwell_gauge}),
(\ref{Maxwell_current}) and (\ref{Maxwell_current_conservation}).
Note that analogous differential system was derived in
\cite{P_S_T} in terms of a 5-potential that transforms according
to a non-decomposable representation of $SU(2,2)$ (see also
\cite{Fr_P} and references therein).

This system admits two interpretations. The first one with
$J_{m_1\ldots m_l;k}(x)$ treated as independent fields restricted
only by equations (\ref{covariant_constancy_equation}) is that it
provides the off-mass-shell  version of the Maxwell
electrodynamics, which accounts for all differential consequences
of the Bianchi identities. Another interpretation comes out when
the field~$J_{;k}(x)$ is a nonlinear combination of some other
``matter'' fields. In that case, equations
(\ref{covariant_constancy_equation}) should be treated as Maxwell
equations describing electromagnetic interactions of the matter
fields. Clearly, for this to be possible it is necessary to
single out the module $\irrVermaex{J}$ from the tensor product of
some other ``matter modules'' that leads to a nonlinear system
describing electromagnetic interactions of matter fields from
which the current $J_{;k}(x)$ is built.  The equation
(\ref{Maxwell_current_conservation}) imposes the conservation
condition on this current.

Finally let us note that to have a gauge invariant form of the
Maxwell equations (i.e. to relax the gauge condition
(\ref{Maxwell_gauge})) one has to  consider the further extension
$\extmodule_{A,F,J,G}$ of the module $\extmodule_{A,F,J}$ with the
module $\irrVermaex{G}=\Vermaex{4,0}$. The module
$\extmodule_{A,F,J,G}$ is defined by the relations
(\ref{relation_example_IA}), (\ref{relation_example_IF}),
(\ref{relation_example_IA_IF}), (\ref{relation_example_IJ}),
(\ref{relation_example_IA_IF_IJ}) along with
\begin{equation}
\cP^n\ketv{G}=-\frac{1}{16}\cK^m\cK_m\ketv{A}^n\,,
\end{equation}
where $\ketv{G}$ is the vacuum of the module $\irrVermaex{G}$.
Consider a section
\begin{align}
\bundsectex{A,F,J,G}&=\sum_{l=0} \frac{1}{l!} A_{m_1\ldots m_l;k}
(x)\cK^{m_1}\ldots \cK^{m_l}\ketv{A}^k+ \sum_{l=0}\frac{1}{l!}\,
F_{m_1\ldots m_l;k_1k_2} (x)
\cK^{m_1}\ldots\cK^{m_l}\ketv{F}^{k_1k_2}+\nn\\
&{}+\sum_{l=0}\frac{1}{l!} J_{m_1\ldots m_l;k} (x) \cK^{m_1}\ldots
\cK^{m_l}\ketv{J}^k+ \sum_{l=0}\frac{1}{l!} G_{m_1\ldots m_l} (x)
\cK^{m_1}\ldots\cK^{m_l}\ketv{G}\label{AFJG-section}
\end{align}
of the bundle $\oR^4\times\extmodule_{A,F,J,G}$. The consequences
of the covariant constancy condition imposed
on~(\ref{AFJG-section}) are analogous to those for the
section~$\bundsectex{A,F,J}$ but with subsystem
(\ref{M4_electrodynamics_l_1}) replaced with
\begin{align}
\ptl_n A_{m_1 \ldots m_{l-1};k}(x)+ 2lA_{nm_1 \ldots
m_{l-1};k}(x)- (l-1)\eta_{n(m_1} A_q{}^q{}_{m_2 \ldots
m_{l-1});k}(x)+&\nn\\
{}+2A_{m_1 \ldots m_{l-1}k;n}(x)-2\eta_{nk}A_{m_1\ldots
m_{l-1}q;}{}^q(x)- F_{m_1\ldots m_{l-1};nk}(x)-&\nn\\
{}-\frac{1}{16}(l-1)(l-2)\eta_{nk}\eta_{(m_1m_2}G_{m_3\ldots
m_{l-1})}=0&\,,
\label{M4_electrodynamics_l_1g}
\end{align}
and additional subsystem of the form (\ref{covariant_forms_l})
with $\dmn=4$, $\Delta=\dmn-p=4$ for the fields~$G_{m_1\ldots
m_l}(x)$. $G$-dependent terms in (\ref{M4_electrodynamics_l_1g})
modify equation (\ref{Maxwell_gauge}) to
\begin{equation}\label{Maxwell_gauge_invariant}
\square\ptl^kA_{;k}(x)=G(x)\,.
\end{equation}
Subsystem for the fields~$G_{m_1\ldots m_l}(x)$ expresses higher
components of $G_{m_1\ldots m_l}(x)$ ($l\geq 1$) in terms of
derivatives of $G(x)$.

In section \ref{Sct.:_Conformal_higher_spins_in_even_dimensions}
we consider a generalization of this construction  to a case of
an almost arbitrary tensor structure of the field strength in any
even space--time dimension $\dmn > 2$.

\section{General Construction}\label{Sct.:_General_Construction}
Let $\alg$ be a complex semi-simple\footnote{In fact the following
consideration remains essentially the same for any Kac--Moody
algebra.} Lie algebra with simple roots
$\roots=(\alpha_0,\alpha_1,\dots,\alpha_q)$. Then~$\alg$ is
generated by elements~$\cH_i$, $\cE_i$ and~$\cF_i$, $0\leq i\leq
q$ with the relations
\begin{align}
   &\commut{\cH_i}{\cE_j}=A_{ij}\cE_j\,,&
   &\commut{\cH_i}{\cF_j}=-A_{ij}\cF_j\,,\\
   &\commut{\cE_i}{\cF_j}=\delta_{ij}\cH_j\,,\\
   &\bigl({\rm ad}\cE_i\bigr)^{1-A_{ij}}\cE_j=0\,,&
   &\bigl({\rm ad}\cF_i\bigr)^{1-A_{ij}}\cF_j=0\,,\qquad i\neq
   j\,,
\end{align}
where no summation over repeated indices is assumed and
\begin{equation}
   A_{ij}=\alpha_j(\cH_i)\,,\qquad A_{i,j\neq i}\leq 0\,,
   \qquad A_{ii}=2
\end{equation}
is the Cartan matrix. The transformation $\tau$
\begin{equation}\label{Chevalley_involution}
\tau(\cE_i)=\cF_i\,,\quad \tau(\cF_i)=\cE_i\,,\quad
\tau(\cH_i)=\cH_i
\end{equation}
generates the involutive antilinear antiautomorphism of $\alg$
called the Chevalley involution.

Choose a subset of the set of simple roots
$\subroots\subset\roots$. Let $\subalg\subset\alg$ denote the
semi-simple subalgebra generated by elements $\cE_i$, $\cF_i$,
$\cH_i$ such that $\alpha_i\in\bar \Pi$. $\Cartan_{\subroots}$ is
the Cartan subalgebra of~$\subalg$. Let~$\psubalg$ be the
parabolic subalgebra with respect to~$\subroots$, i.e.~$\psubalg$
is generated by~$\cH_i$, $\cE_i$ with~$0\leq i\leq q$ and~$\cF_i$
corresponding to simple roots in~$\subroots$.
Evidently,~$\subalg\subset\psubalg\subset\alg$ for
any~$\subroots$. The parabolic subalgebra~$\psubalg$ admits the
Levi--Maltsev decomposition~$\psubalg=\Levi\subplus\rad$, where
$\Levi=\Cartan_{\roots\backslash\subroots}\subplus\subalg$ is the
Levi factor of~$\psubalg$ and~$\rad$ is the radical of~$\psubalg$.
The linear space~$\alg$ can thus be decomposed into the direct
sum~$\alg=\subalg\oplus\Cartan_{\roots\backslash\subroots}
\oplus\rad\oplus\alg / \psubalg$. Let us choose a basis
$\Big(\cL_\beta,\cD_I,\cP_a,\cK_a \Big )$ of~$\alg$ such that the
elements~$\cL_\beta$, $\cD_I$, $\cP_a$ and~$\cK_a$ form some bases
in~$\subalg$, $\Cartan_{\roots \backslash \subroots}$, $\rad$
and~$\alg / \psubalg$, respectively. Note that the
involution~$\tau$ maps~$\rad$ to~$\alg / \psubalg$ and vice versa.
Therefore, both for $\cP_a$ and for $\cK_a$ the index $a$ takes
values~$a=0,\ldots \dmn-1$, where~$\dmn={\rm dim}(\rad)={\rm
dim}(\alg/\psubalg)$.

Note that the commutation relations of~$\alg$ in the
basis~$\Big(\cL_\beta,\cD_I,\cP_a,\cK_a \Big )$ have the following
structure
\begin{align}\label{commutation_relations_for_f}
\commut{\cL}{\cL}&\sim\cL\,, & \commut{\cP}{\cP}&\sim\cP\,, &
\commut{\cK}{\cK}&\sim\cK\,,\nn\\
\commut{\cD}{\cL}&\sim\cL\,, & \commut{\cL}{\cP}&\sim\cP\,, &
\commut{\cL}{\cK}&\sim\cK\,,\\
\commut{\cP}{\cK}&\sim\cL+\cD+\cP+\cK\,, &
\commut{\cD}{\cP}&\sim\cP\,, & \commut{\cD}{\cK}&\sim\cK\,,\nn\\
\commut{\cD}{\cD}&=0\,,\nn
\end{align}
where $\cL_\beta$, $\cD_I$, $\cP_a$, $\cK_a$ are operators of
generalized Lorentz transformations,  dilatations, translations
 and special conformal
transformations, respectively.

Let $\module$ be some (usually infinite dimensional) $\alg$-module
with the following properties. $\module$ decomposes into the
direct sum of irreducible finite dimensional modules of~$\Levi$.
The action of the Cartan subalgebra~$\Cartan_\Pi\subset\alg$ is
diagonalizable in~$\module$. The action of the radical~$\rad$ is
locally nilpotent in~$\module$, i.e.{} $\module$ admits a
filtration by $\Levi$-modules
\begin{eqnarray}\label{filtration}
&\filtration{0}\subset\filtration{1}\subset\cdots
\subset\filtration{f}\subset\cdots\subset\module\nn\\
&\module=\bigcup_{f=0}^{\infty}\filtration{f}\,, \end{eqnarray}
where a $\Levi$-module $\filtration{f}$ is such that
\begin{equation}\label{definition_of_filtration}
(\rad)^{f+1}\filtration{f}\equiv 0\,, \end{equation} i.e.
 a product of any $f+1$
elements from~$\rad$ annihilates any vector from~$\filtration{f}$.

The filtration (\ref{filtration}) gives rise to the grading on
$\module$
\begin{equation}\label{gradation}
\module=\bigoplus_{l=0}^{\infty} \gradation{l}\,.
\end{equation}
Here $\gradation{0}=\filtration{0}$ and $\gradation{l}$ ($l\geq
1$) is the preimage of the quotient morphism
\begin{equation}
q:\qquad\filtration{l}\rightarrow\filtration{l}/\filtration{l-1}
\end{equation}
$\gradation{l}=q^{-1}\Big(\filtration{l}/\filtration{l-1}\Big)$,
where $q^{-1}$ is a homomorphism of $\Levi$ modules satisfying
$qq^{-1}=1$.  $q^{-1}$ is fixed uniquely provided that
$\filtration{l-1}$ does not contain $\Levi$-irreducible
submodules isomorphic to some of the $\Levi$-irreducible
submodules of $\filtration{l}/\filtration{l-1}$. Otherwise, to
fix the arbitrariness in $q^{-1}$, an appropriate additional
prescription is needed. We demand every $\gradation{l}$, which is
called level $l$ submodule of $\module$, to form a finite
dimensional module of $\Levi$.  An element $r\in\rad$ decreases
the grading
\begin{equation}
r :\qquad \gradation{l}\rightarrow\gradation{l-n(r)}\,,
\end{equation}
where $n(r)\geq 1 $ is an integer.
Note that if $\rad$ is Abelian then $n(r)=1$ for any $r\in\rad$.

Let ~$\forms$ be the Grassmann algebra on $\xi^n$,
$n=0,1,\ldots,\dmn-1$, $\xi^n \xi^m = - \xi^m \xi^n$  and $\xi^n$
are identified with space--time basis 1-forms. Consider the tensor
product $\moduleforms=\module\otimes\forms$. $\moduleforms$ is
bi-graded by the level of $\module$ (\ref{gradation}) and by the
exterior form degree of~$\forms$
\begin{equation}\label{forms_gradation}
\moduleforms=\bigoplus_{p=0}^{\dmn}\bigoplus_{l=0}^{\infty}
\moduleformslp{l}{p}=\bigoplus_{p=0}^{\dmn}\moduleformsp{p}\,,
\end{equation}
where $\moduleformslp{l}{p}$ is the space of $p$-forms taking
values in $\gradation{l}$.  $\moduleformsp{p}$ is the space of
$p$-forms taking values in the whole module $\module$.

Consider the trivial vector bundle
$\bundle=\Minkowski\times\moduleforms$ over $\Minkowski$
\begin{equation}
   \begin{array}{ccc}
    \moduleforms&\longrightarrow&\bundle\\
     {}&{}&\downarrow\\
     {}&{}&\Minkowski
   \end{array}
\end{equation}
with the fiber $\moduleforms$.  Let $\sect{\bundle}$ denote the
space of sections of $\bundle$. We define the covariant
derivative in $\bundle$
\begin{equation}
\label{general_differrential}
\diff=\xi^n\ptl_n+\xi^{n}\genconnection{n}{\gb}\cL_\beta
+\xi^{n}\genconnection{n}{a}\cP_{a}+\xi^{n}\genconnection{n}{I}\cD_{I}\,,
\end{equation}
where $x^n,$ $n=0,1,\ldots,\dmn-1$ are the space--time coordinates
in~$\Minkowski$. The connection 1-forms $\genconnection{n}{\gb}$,
$\genconnection{n}{a}$ and $\genconnection{n}{I}$ are chosen to
satisfy the zero curvature equation
(\ref{covariant_derivative_flat}). We require
$\genconnection{n}{a}$ to be non-degenerate
\begin{equation}\label{general_frame_non-degenerate}
{\rm det}\left |\genconnection{n}{a}\right | \neq 0\,.
\end{equation}

In the rest of this paper we focus on the case of Abelian $\rad$,
\begin{equation}
\commut{\cP_a}{\cP_b}=0\,.
\end{equation}
In this case (\ref{covariant_derivative_flat})
and~(\ref{general_frame_non-degenerate}) admit the simple solution
\begin{equation}
\diff=\xi^n\ptl_n+\xi^n\delta^a_n\cP_a\,,
\end{equation}
with $\genconnection{n}{\ga}=\genconnection{n}{I}=0$
and~$\genconnection{n}{a}=\delta_n^a$, where $\delta_n^a$ is
identified with the flat space co-frame in Cartesian coordinates.
Choosing different solutions of (\ref{covariant_derivative_flat})
allows one to analyse the problem in any other coordinates. Having
fixed  the flat frame in the form of Kronecker delta, in what
follows we will not distinguish between the base and the fiber
indices.

Let us introduce the exterior differential
\begin{equation}\label{d}
\extdiff=\xi^n\ptl_n\;:\qquad
\moduleformslp{l}{p}\rightarrow\moduleformslp{l}{p+1}
\end{equation}
and the operator
\begin{equation}\label{gs}
\sgminus=\xi^n\cP_n\;:\qquad
\moduleformslp{l}{p}\rightarrow\moduleformslp{l-1}{p+1}\,.
\end{equation}
We have
\begin{equation}\label{differential_for_Abelian_radical}
\diff=\extdiff+\sgminus\,.
\end{equation}
{}From (\ref{covariant_derivative_flat}), (\ref{d}) and (\ref{gs})
it follows that the operators $\extdiff$ and $\sgminus$ are
nilpotent and anticommutative
\begin{align}
&\extdiff\extdiff=0\,,\qquad \sgminus\sgminus=0\,,\nn\\
\label{properties_of_sigma_minus}
&\extdiff\sgminus+\sgminus\extdiff=0\,.
\end{align}

Let $\closed{}\subset\moduleforms$ and
 $\exact{}\subset\closed{}\subset\moduleforms$
be the spaces of $\sgminus$-closed and $\sgminus$-exact
forms, respectively,
\begin{equation}
  \sgminus\closed{}=0\,, \qquad
\exact{}=\sgminus\moduleforms\,.
\end{equation}
The cohomology~$\cohomology{}$ of~$\rad$ is the
quotient~$\closed{}/\exact{}$. Let $p$ be the quotient mapping
\begin{equation}
p:\qquad\closed{}\rightarrow\cohomology{}\,.
\end{equation}
This mapping is a $\Levi$-homomorphism.
We define the mapping
\begin{equation}
p^{-1}:\qquad\cohomology{}\rightarrow\closed{}
\end{equation}
such that $pp^{-1}=1$ and $p^{-1}$ is a $\Levi$-homomorphism.
These requirements fix $p^{-1}$ uniquely provided that $\exact{}$
does not contain $\Levi$-irreducible submodules isomorphic to some
of the $\Levi$-irreducible submodules of $\closed{}/\exact{}$.
Otherwise, to fix the arbitrariness in $p^{-1}$, an appropriate
additional prescription is needed. The space $\moduleforms$
decomposes into the direct sum of $\Levi$-modules
\begin{equation}\label{moduleforms_decomposition}
\moduleforms=H\oplus \exact{}\oplus F\,.
\end{equation}
Here $H$ denotes $p^{-1}(\cohomology{})$, $\exact{}$ complements
$H$ to $\closed{}$ and $F$ complements $\closed{}$ to
$\moduleforms$. The gradings~(\ref{forms_gradation}) of
$\moduleforms$ induces the gradings of $H$, $\exact{}$ and $F$.
Let $H^p_{[l]}$, $\exact{p}_{[l]}$ and $F^p_{[l]}$ denote
corresponding homogeneous subspaces. Note that
$H^0=\closed{0}=\moduleformslp{0}{0}$ and thus $p^{-1}$ is
identical in the sector of 0-forms.

Introduce the subbundle $\subbundle=\Minkowski\times H$ of the
bundle $\bundle$
\begin{equation}
   \begin{array}{ccc}
    H&\longrightarrow&\subbundle\\
     {}&{}&\downarrow\\
     {}&{}&\Minkowski
   \end{array}
\end{equation}
with the fiber $H\subset\moduleforms$. Let $\sect{\subbundle}$
denote the space of sections of~$\subbundle$. Let a $p$-form
$\subbundsec{p}\in\sect{\subbundle}$ be a section of
$\subbundle$. Now we are in a position to formulate
$\alg_{\psubalg}$-invariant differential equations on $\subbundsec{p}$ as the
conditions for~$\subbundsec{p}$ to admit a lift to a $p$-form
$\bundsec{p}\in\sect{\bundle}$ such that
\begin{equation}\label{Con.:_lift}
  \begin{split}
\diff\bundsec{p}=0\,,\\
\kern55pt\left.\bundsec{p}\right|_\subbundle=\subbundsec{p}\,.\kern-30pt
  \end{split}
\end{equation}
Here $\left.\bundsec{p}\right|_\subbundle$ is the projection of
$\moduleforms$ to $H$ in the decomposition
(\ref{moduleforms_decomposition}). Call a section
$\bundsec{p}\in\sect{\bundle}$ $\diff$-horizontal if
$\diff\bundsec{p}=0$. Call a section
$\bundsec{p}\in\sect{\bundle}$ $\diff$-horizontal lift of
$\subbundsec{p}\in\sect{\subbundle}$ if it
satisfies~\eqref{Con.:_lift}. Taking into account
(\ref{covariant_derivative_flat}), the equation
$\diff\bundsec{p}=0\, $ is invariant under the gauge
transformation
\begin{equation}
\label{gauge}
\delta \bundsec{p} = \diff| \epsilon^{p-1} (x)\rangle\,,
\end{equation}
where $\epsilon^{p-1} \in\sect{\bundle}$ is an arbitrary
$p-1$-form. Note that for $p\geq 2$ (\ref{gauge}) is invariant
under the second order gauge transformation
\begin{equation}
\label{2_gauge} \delta |\gep^{p-1}(x)\rangle=
\diff|\chi^{p-2}(x)\rangle\,,
\end{equation}
where $|\chi^{p-2}(x)\rangle$ is an arbitrary $p-2$-form. For
$p\geq 3$ (\ref{2_gauge}) is invariant under the third order gauge
transformation and so on.

We will distinguish between $T$ (trivial), $D$ (differential) and
$A$ (algebraic)  classes of gauge transformations with the gauge
parameters $| \epsilon_{T}^{p-1}
(x)\rangle=|\psi_{T}^{p-1}(x)\rangle+\diff |\chi^{p-2}_T
(x)\rangle$, $| \epsilon_{D}^{p-1}
(x)\rangle=|\psi_{D}^{p-1}(x)\rangle+\diff
|\chi^{p-2}_D(x)\rangle$ and $| \epsilon_{A}^{p-1}
(x)\rangle=|\psi_{A}^{p-1}(x)\rangle+\diff
|\chi^{p-2}_A(x)\rangle$, respectively, with some $p-1$-forms
$|\psi_{T}^{p-1}(x)\rangle\in\exact{}$,
$|\psi_{D}^{p-1}(x)\rangle\in H$, $|\psi_{A}^{p-1}(x)\rangle\in
F{}$. The ambiguity in the  second-order gauge parameters
$|\chi^{p-2}_T (x)\rangle$, $|\chi^{p-2}_D (x)\rangle$ and
$|\chi^{p-2}_A (x)\rangle$ manifests the fact that the
decomposition into the $T$, $D$, and $A$  gauge transformations is
not unique. One can see, in particular, that any
$T$-transformation reduces to a linear combination of some
$A$-transformation and $D$-transformation and can therefore be
discarded. Indeed, let $|\epsilon_{T[l]}^{p-1}
(x)\rangle=\gs_-|\chi_{T[l+1]}^{p-2}(x)\rangle$ be a level-$l$
$T$-transformation parameter. Taking into account
(\ref{properties_of_sigma_minus}) one gets
\begin{equation}\label{T_gauge}
\gd_T\bundsec{p}=
\extdiff|\epsilon_{T[l]}^{p-1}(x)\rangle=
-\gs_-\extdiff|\chi_{T[l+1]}^{p-2}(x)\rangle=
-\diff \extdiff|\chi_{T[l+1]}^{p-2}(x)\rangle\,.
\end{equation}
Decompose $-\extdiff|\chi_{T[l+1]}^{p-2}(x)\rangle$ into
a combination  of level $l+1$ $D$, $A$ and $T$ gauge parameters.
If the resulting level $l+1$ $T$-parameter is nonzero
one applies the same procedure, and so on.

The roles of the $D$ and $A$ gauge transformations are as
follows. The variation of $\bundsec{p}$ under $D$-transformations
is purely differential
\begin{equation}\label{D_gauge}
\gd_D\bundsec{p}=\extdiff|\gep_D^{p-1}(x)\rangle\,.
\end{equation}
$D$-transformations generalize the gradient transformations in
electrodynamics and linearized diffeomorphisms in gravity.
$A$-transformations are gauge transformations of the form
\begin{equation}\label{A_gauge}
\gd_A\bundsec{p}=
\extdiff|\gep_A^{p-1}(x)\rangle+\gs_-|\gep_A^{p-1}(x)\rangle
\end{equation}
with a nonzero second term. These are analogous to the linearized
local Lorentz transformations in gravity.

Now, following to \cite{conf3}, we prove that the existence of a $\diff$-horizontal lift
(see  \eqref{Con.:_lift}) is governed by $\cohomology{p+1}$.
\begin{Thm}\label{Theorem_1}
$\!$\\[-12pt]
\begin{enumerate}
\item Let $\subbundsec{p}\in\sect{\subbundle}$ and let there exist
$\bundsec{p}_1$ and $\bundsec{p}_2\in\sect{\bundle}$ that are
$\diff$-horizontal lifts of $\subbundsec{p}$. Then
$\bundsec{p}_1-\bundsec{p}_2=\gd_A|\chi^{p-1}(x)\rangle$ for some
$|\chi^{p-1}(x)\rangle\in\sect{\bundle}$ (see \eqref{A_gauge}).
\item The two statements are equivalent
  \begin{enumerate}
  \item any section $\subbundsec{p}\in\sect{\subbundle}$ has a $\diff$-horizontal
           lift to a $\bundsec{p}\in\sect{\bundle}$
  \item $\cohomology{p+1}=0$.
  \end{enumerate}
\item If $\cohomology{p+1}\neq0$, there exists a system of differential
equations
\begin{equation}\label{Con.:_equations}
\equations\subbundsec{p}=0
\end{equation}
such that any solution of \eqref{Con.:_equations} admits a $\diff$-horizontal
lift to a $\bundsec{p}\in\sect{\bundle}$ and all $\subbundsec{p}\in\sect{\subbundle}$
admitting such a lift satisfy \eqref{Con.:_equations}.
\end{enumerate}
\end{Thm}
\begin{prf}
Let us look for a lift $\bundsec{p}$ in the form
\begin{equation}\label{lift_decomposition}
\bundsec{p}=\kphi{0}+
\kphi{1}+\kphi{2}+\cdots\,,
\end{equation}
where $\kphi{l}\in\moduleformslp{l}{p}$. The condition
$\left.\bundsec{p}\right|_\subbundle=\subbundsec{p}$ fixes the
first term in this decomposition
$\left.\kphi{0}\right|_\subbundle=\subbundsec{p}\cap\moduleformslp{0}{p}$
modulo a $\sgminus$-exact form
$\kphi{0}{}_\exact{}\in\exact{p}_{[0]}$. The freedom in
$\kphi{0}{}_\exact{}\in\exact{p}_{[0]}$ is a consequence of
$A$-gauge symmetry, i.e. $\kphi{0}$ is reconstructed modulo an
$A$-gauge part (which, of course, also contributes to $\kphi{1}_H$).

Suppose that $\cohomology{p+1}$ is trivial, i.e.
$\closed{p+1}=\exact{p+1}$. To reconstruct $\bundsec{p}$ we use
the following step-by-step procedure. The zero level part of
(\ref{Con.:_lift}) reads
\begin{align}
&\extdiff \kphi{0}  +\sgminus\kphi{1}=0\,,\label{level_0}\\
&\left.\kphi{1}\right|_\subbundle=
\subbundsec{p}\cap\moduleformslp{1}{p}\,.\label{boundary_1}
\end{align}
Since  $\kphi{0}$ has the lowest grading, it is $\sgminus$-closed.
$\extdiff  \kphi{0}$ is also $\sgminus$-closed because
$\extdiff\sgminus+\sgminus\extdiff=0$. Since $\cohomology{p+1}$ is
trivial, $\extdiff  \kphi{0}$ is $\sgminus$-exact
\begin{equation}
\extdiff  \kphi{0} =\sgminus\kchi{1}\,
\end{equation}
for some $\kchi{1}$. Setting $\kphi{1}=-\kchi{1}$ we solve the
equation (\ref{level_0}) modulo an arbitrary $\gs_-$-closed form
$\kphi{1}_\closed{}\in\closed{p}_{[1]}$. The condition
(\ref{boundary_1}) fixes $\kphi{1}_\closed{}$ modulo an arbitrary
$\gs_-$-exact form $\kphi{1}{}_\exact{}\in\exact{p}_{[1]}$, which
parameterizes the level 1 restriction of some $A$-gauge part with
level 2 gauge parameter. As a result,
$\kphi{1}\in\moduleformslp{1}{p}$ is expressed via the first
derivatives of $\subbundsec{p}\cap\moduleformslp{0}{p}$ and via
$\subbundsec{p}\cap\moduleformslp{1}{p}$ modulo an arbitrary
$A$-gauge part.  The first level part of (\ref{Con.:_lift})
\begin{align}
&\extdiff\kphi{1} +\sgminus\kphi{2}=0\,,\label{level_1}\\
&\left.\kphi{2}\right|_\subbundle=
\subbundsec{p}\cap\moduleformslp{2}{p}\label{boundary_2}
\end{align}
is considered analogously. $\extdiff\kphi{1}$ is $\sgminus$-closed
because $\sgminus\extdiff\kphi{1}=
-\extdiff\sgminus\kphi{1}=\extdiff ^2\kphi{0}=0$. Introducing
$\kchi{2}\in\moduleformslp{2}{p}$ such that
$\extdiff\kphi{1}=\sgminus\kchi{2}$ and setting
$\kphi{2}=-\kchi{2}$ we solve equation (\ref{level_1}) modulo an
arbitrary $\sgminus$-closed form
$\kphi{2}{}_\closed{}\in\closed{p}_{[2]}$. The condition
(\ref{boundary_2}) fixes $\kphi{2}{}_\closed{}$ modulo an
arbitrary $\sgminus$-exact form
$\kphi{2}{}_\exact{}\in\exact{p}_{[2]}$, which parameterizes the
level 2 restriction of some $A$-gauge part with level 3 gauge
parameter. As a result $\kphi{2}$ is expressed via the second
derivatives of $\subbundsec{p}\cap\moduleformslp{0}{p}$, via the
first derivatives of $\subbundsec{p}\cap\moduleformslp{1}{p}$ and
via the $\subbundsec{p}\cap\moduleformslp{2}{p}$ modulo some
$A$-gauge terms. Repetition of this procedure reconstructs the
lift $\bundsec{p}$ in the form (\ref{lift_decomposition}) with
$\kphi{l}$ expressed in terms of derivatives of $\subbundsec{p}$
modulo an $A$-gauge part.

Suppose now  that $\cohomology{p+1}$ is nontrivial. Then it
decomposes into a sum of some definite grade nonzero subspaces
\begin{equation}
\cohomology{p+1}=\cohomologylp{l_1}{p+1}\oplus
\cohomologylp{l_2}{p+1}\oplus\cdots\,,
\end{equation}
where $0\leq l_1<l_2<\cdots$. Carrying out the first $l_1$ steps
of the described  procedure we solve~(\ref{Con.:_lift})
up to the $l_1-1$-th level, expressing all
$\kphi{l}$ with $1\leq l\leq l_1$ via derivatives of
$\subbundsec{p}$ modulo some $A$-gauge part. The level $l_1$ sector
of~(\ref{Con.:_lift}) reads
\begin{align}\label{level_l1}
&\extdiff\kphi{l_1}+\sgminus\kphi{l_1+1}=0\\
&\left.\kphi{l_1+1}\right|_\subbundle=
\subbundsec{p}\cap\moduleformslp{l_1+1}{p}\,.\label{boundary_l1+1}
\end{align}
{}From the  level $l_1-1$ sector of~(\ref{Con.:_lift}) it follows
that the $p+1$-form $\extdiff\kphi{l_1}$ is $\sgminus$-closed.
Equation (\ref{level_l1}) imposes however a stronger condition
that $\extdiff\kphi{l_1}$ is $\sgminus$-exact thus requiring
those combinations of $\extdiff\kphi{l_1}$ that belong to the
cohomology class $\cohomology{p+1}$ to vanish. This imposes some
differential equations on $\subbundsec{p}$ of orders not higher
than $l_1$
\begin{equation}\label{equation_l1}
\equations_{[l_1]}\subbundsec{p}=0\,.
\end{equation}
In addition, equation (\ref{level_l1}), (\ref{boundary_l1+1})
expresses $\kphi{l_1+1}$ via derivatives of $\subbundsec{p}$
modulo an arbitrary $A$-gauge part
$\kphi{l_1+1}{}_\exact{}\in\exact{p}_{[l_1+1]}$.

Solving further (\ref{Con.:_lift}) level by
level we fix $\kphi{l_1+1},\ldots,\kphi{l_2}$ modulo an arbitrary
$A$-gauge part. At the level $l_2$, equations
\begin{align}\label{level_l2}
&\extdiff\kphi{l_2}+\sgminus\kphi{l_2+1}=0\\
&\left.\kphi{l_2+1}\right|_\subbundle=\subbundsec{p}\cap\moduleformslp{l_2+1}{p}\,
\label{boundaru_l2+1}
\end{align}
 fix $\kphi{l_2+1}$ in terms of derivatives of $\subbundsec{p}$
modulo an $A$-gauge part
$\kphi{l_2+1}_\exact{}\in\exact{p}_{[l_2+1]}$ and impose some
additional differential equations of  orders not higher than
$l_2$
\begin{equation}\label{equation_l2}
\equations_{[l_2]}\subbundsec{p}=0.
\end{equation}

Repetition of this procedure reconstructs modulo an $A$-gauge
part a lift $\bundsec{p}$
in the form (\ref{lift_decomposition}) for $\subbundsec{p}$
satisfying the system of differential equations
\begin{align}\label{system_of_equations}
\equations_{[l_1]}&\subbundsec{p}=0\,,\nn\\
\equations_{[l_2]}&\subbundsec{p}=0\,,\\
&\cdots\nn
\end{align}

To show that the system (\ref{system_of_equations}) is necessarily
nontrivial if $\cohomology{p+1}$ is nonzero, let us construct a
section $\subbundsec{p}\in\sect{\subbundle}$ that does not
satisfy~(\ref{system_of_equations}). Let us choose some
$\tpsi{l_1}\in H^{p+1}_{[l_1]}$ such that $\extdiff\tpsi{l_1}=0$
(for example one can choose $\tpsi{l_1}\in H^{p+1}_{[l_1]}$ to be
$x$-independent). Then $\tpsi{l_1}=\extdiff\tphi{l_1}$ for some
$\tphi{l_1}$. Decompose
$\tphi{l_1}=\tphi{l_1}{}_H+\tphi{l_1}{}_{\exact{}}+\tphi{l_1}{}_F$
in accordance with~(\ref{moduleforms_decomposition}). Consider now
the $l_1-1$-th level part of equation (\ref{Con.:_lift})
\begin{equation}\label{level_l1-1}
\extdiff\tphi{l_1-1}+\sgminus\tphi{l_1}=0\,.
\end{equation}
Because $\sgminus\tphi{l_1}$ is $\extdiff$-closed
($\extdiff\sgminus\tphi{l_1}=-\sgminus\extdiff\tphi{l_1}=0$) we
can solve it for
$\tphi{l_1-1}=\tphi{l_1-1}{}_H+\tphi{l_1-1}{}_{\exact{}}+\tphi{l_1-1}{}_F$.
Repeating this ``inverse'' procedure we find
$\tphi{l_1}{}_H,\ldots,\tphi{0}{}_H$, arriving at the field
$\subbundsec{p}=\tphi{0}{}_H+\cdots+\tphi{l_1}{}_H$ that solves
(\ref{Con.:_lift}) for the levels $0,1\ldots
l_1 -1$ but satisfies the modified equation (\ref{Con.:_lift})
with the nonzero right hand side proportional to $\tpsi{l_1}\in
H^{p+1}_{[l_1]}$ at the level $l_1$, thus
violating~(\ref{system_of_equations}).
\end{prf}
\begin{Rem}\rm
If there exists a $\diff$-horizontal lift of 0-form
$\subbundsec{0}$ to a 0-form $\bundsec{0}\in\sect{\bundle}$ then
it is unique.
\end{Rem}
\begin{prf}
Gauge symmetries (\ref{gauge}) trivialize in the sector of
0-forms.
\end{prf}
\begin{Rem}\rm
Consider the subbundle $\subbundle'=\Minkowski\times H\oplus
\exact{}$ of the bundle $\bundle$
\begin{equation}
   \begin{array}{ccc}
    H\oplus \exact{}&\longrightarrow&\subbundle'\\
     {}&{}&\downarrow\\
     {}&{}&\Minkowski
   \end{array}
\end{equation}
with the fiber $H\oplus\exact{}=\closed{}\subset\moduleforms$. If
there exists a $\diff$-horizontal lift of $\subbundsec{p}$ to a
$p$-form $\bundsec{p}\in\sect{\bundle}$ then it is unique.
\end{Rem}
\begin{prf}
Restriction to ${\subbundle'}$ fixes some  $A$-gauge.
\end{prf}
\begin{Rem}\label{Thm.:_Remark_3_4}\rm
Theorem \ref{Theorem_1} allows the following interpretation.
Given equation \eqref{Con.:_lift}, a section $\bundsec{p}$
decomposes into
$\bundsec{p}=\bundsec{p}_H+\bundsec{p}_\exact{}+\bundsec{p}_F$.
The subsection $\bundsec{p}_H$ describes dynamical fields subject
to some differential equations \eqref{Con.:_equations}. Solutions
of these differential equations are moduli of solutions of the
equation \eqref{Con.:_lift}. The part $\bundsec{p}_F$ describes
(usually infinite) set of fields expressed by the equation
\eqref{Con.:_lift} via derivatives of the dynamical fields. The
fields of this class are called auxiliary fields and the equations
that express them are called constraints. The $A$-gauge symmetry
\eqref{A_gauge} (generalized local Lorentz symmetry) allows one
to get rid of $\sigma_-$-exact terms $\bundsec{p}_\exact{}$. The
$D$-gauge symmetry \eqref{D_gauge} with the parameters in
$\cohomology{p-1}$ acts on the dynamical fields $\bundsec{p}_H$
and is the gauge symmetry of equations \eqref{Con.:_equations}.
\end{Rem}
\begin{Rem}\rm
According to \eqref{solution_for_fields} solutions of
\eqref{Con.:_lift} are parameterized by the values of
$\left.\ketv{\Phi^p(x)}\right|_{x\in\gep(x_0)}$ at a
neighborhood $\gep(x_0)$ of any point $x_0$. This is because the
equation \eqref{Con.:_lift} expresses all higher level ($l\geq
1$) components of $\bundsec{p}$ via higher derivatives
of~$\subbundsec{p}$. As a result, the fields $\subbundsec{p}$ can
be expressed modulo gauge symmetries in terms of
$\left.\ketv{\Phi^p(x)}\right|_{x\in\gep(x_0)}$ by virtue of the
Taylor expansion.
\end{Rem}

In the rest of this paper we mostly confine ourselves to the
sector of 0-forms, which turns out to be reach enough to
reformulate any $\alg_{\psubalg}$-invariant linear differential system in the
unfolded form by virtue of introducing appropriate auxiliary
fields. In other words, for any $\alg_{\psubalg}$-invariant linear
differential system $\equations\subbundsec{0}=0$, there exists
some $\alg$-module $\module_\equations$, which gives rise
to~$\equations\subbundsec{0}=0$ by virtue of the procedure
described above\footnote{Note that any equation
$\equations\subbundsec{p}=0$ can be rewritten in terms of 0-forms
by converting indices of forms into tangent indices with the aid
of the frame field. The formulation in terms of higher forms may
be useful however for the analysis of nonlinear dynamics and will
be discussed elsewhere.}. Thus the problem of  listing all linear
$\alg_{\psubalg}$-invariant differential systems is equivalent to the problem
of  calculating the cohomology $\cohomology{0}$ and
$\cohomology{1}$ for any $\alg$-module $\module$. An important
subclass of such systems is formed by those associated with
irreducible $\module$.
\begin{Dfn}
A system of $\alg_{\psubalg}$-invariant linear differential equations
\begin{equation}\label{Dfn.:_primitive_system}
\equations\subbundsec{0}=0
\end{equation}
is called primitive if the $\alg$-module $\module_\equations$
corresponding to  \eqref{Dfn.:_primitive_system}  as in
Theorem~\ref{Theorem_1} is irreducible.
\end{Dfn}
Reducible modules can be treated as extensions of the irreducible
ones. Let $\irrmodule_1$ and $\irrmodule_2$ be some irreducible
$\alg$-modules. Consider a module $\module$ defined by the exact
sequence
\begin{equation}\label{ext_I1_I2}
0\longrightarrow\irrmodule_1\longrightarrow
\module\longrightarrow
\irrmodule_2\longrightarrow 0\,.
\end{equation}
A trivial possibility is $\module=\irrmodule_1\oplus\irrmodule_2$.
The non-primitive system corresponding to
$\irrmodule_1\oplus\irrmodule_2$ decomposes into
 two independent primitive subsystems
\begin{align}
\label{system_corresponding_I1}
&\equations_{\irrmodule_1}\ketv{\phi^0_{\irrmodule_1}(x)}=0\,,\\
\label{system_corresponding_I2}
&\equations_{\irrmodule_2}\ketv{\phi^0_{\irrmodule_2}(x)}=0\,,
\end{align}
where $\equations_{\irrmodule_1}$,
$\ketv{\phi^0_{\irrmodule_1}(x)}$ and $\equations_{\irrmodule_2}$,
$\ketv{\phi^0_{\irrmodule_1}(x)}$ correspond to $\irrmodule_1$ and
$\irrmodule_2$, respectively. For some particular irreducible
$\irrmodule_1$ and $\irrmodule_2$, a module
$\module=\extmodule_{\irrmodule_1,\irrmodule_2}$ non-isomorphic
to~$\irrmodule_1\oplus\irrmodule_2$ may also exist however. The
non-primitive system corresponding to
$\extmodule_{\irrmodule_1,\irrmodule_2}$
\begin{equation}\label{system_corresponding_ext_I1_I2}
\equations_{\extmodule_{\irrmodule_1,\irrmodule_2}}
\ketv{\phi^0_{\extmodule_{\irrmodule_1,\irrmodule_2}}(x)}=0
\end{equation}
contains the system (\ref{system_corresponding_I2}) for the
dynamical fields $\ketv{\phi^0_{\irrmodule_2}(x)}$ associated with
$\module=\irrmodule_2$. The system~(\ref{system_corresponding_I1})
results from~(\ref{system_corresponding_ext_I1_I2})
at~$\ketv{\phi^0_{\irrmodule_2}(x)}=0$, which means that the space
of solutions of the non-primitive system
(\ref{system_corresponding_ext_I1_I2}) contains the invariant
subspace of solutions of the
system~(\ref{system_corresponding_I1}). In other words, the
equations that contain  $\extdiff\ketv{\phi^0_{\irrmodule_2}(x)}$
are   $\ketv{\phi^0_{\irrmodule_1}(x)}$ independent, while those,
that contain  $\extdiff\ketv{\phi^0_{\irrmodule_1}(x)}$, contain
some terms with  $\ketv{\phi^0_{\irrmodule_2}(x)}$.

Further extensions of the types
\begin{equation}\label{ext_I3_I1_I2}
\exactsq{\extmodule_{\irrmodule_1,\irrmodule_2}}
{\module'}{\irrmodule_3}
\end{equation}
or
\begin{equation}\label{ext_I1_I2_I3}
\exactsq{\irrmodule_3}{\module''}
{\extmodule_{\irrmodule_1,\irrmodule_2}}
\end{equation}
with  indecomposable modules~$\module'$ and~$\module''$  can also
be considered. As a result,  all possible $\alg_{\psubalg}$-invariant linear
differential equations can be classified in terms of extensions of
the primitive equations. Some examples of nontrivial extensions
are considered in section \ref{Sct.:_M4_Electrodynamics},
\ref{Sct.:_Conformal_higher_spins_in_even_dimensions} and
\ref{Sct.:_F_T_conformal_higher_spins_in_even_dimensions}.

To summarize, the construction is as follows. To write down all
$\alg_{\psubalg}$-invariant homogeneous equations on a finite
number of fields for a semi-simple Lie algebra $\alg$ one has to
classify all $\alg$-modules that are integrable with respect to
parabolic subalgebra $\psubalg\subset\alg$ with the Abelian
radical $\rad$. These consist of irreducible $\alg$-modules of
this class and all their extensions. The unfolded form of the
$\alg_{\psubalg}$-invariant homogeneous equations has the form of
the covariant constancy equation
(\ref{covariant_constancy_equation}) for the $0$-form section
$\bundsec{0}$ of the bundle $\bundle$. Dynamical fields form the
$0$-form section $\subbundsec{0}$ of $\subbundle$. Differential
field equations on the dynamical fields are characterized by the
cohomology $\cohomology{1}$, which is the linear space where the
nontrivial left hand sides of the equations
$\equations\subbundsec{0}=0$ take their values. Since equation
(\ref{covariant_constancy_equation}) is $\alg$-invariant, the
equation $\equations\subbundsec{0}=0$ is $\alg$-invariant as well,
i.e. $\alg$ maps its solutions to solutions. The construction is
universal because any differential equations can be ``unfolded" to
some covariant constancy equation by adding enough (usually
infinitely many) auxiliary fields expressed by virtue of the
unfolded equations through derivatives of the dynamical fields
$\subbundsec{0}$. If the original system of differential equations
is $\alg$-invariant, the corresponding unfolded equation is also
$\alg$-invariant, and auxiliary fields together with the dynamical
fields, span the space of sections of $\bundle$.

Now we are in a position to give the full list of conformally
invariant systems of differential equations in $\Minkowski$
($\dmn\geq 3$).

\section{Conformal Systems of Equations
\label{Sct.:_Conformal_Systems_of_equations}} We set
$\alg=\so(\dmn+2)$ with the commutation relations
(\ref{conformal_algebra_commutators})
($\so(\dmn+2)\sim\so(\dmn,2)$ for the complex case we focus on).
The structure of simple roots $\roots$ for $\so(\dmn+2)$ depends
on whether $\dmn$ is odd or even. For $\dmn=2q$,
$\so(\dmn+2)=D_{q+1}$ and $\roots$ is described by the Dynkin
diagram
\begin{equation}\label{even-dynkin}
   \unitlength=5pt
   \begin{picture}(20,10)
   \put(-3,3){
   \put(0,0){
    $
    \circ\kern2pt\oneline\circ\oneline\circ\oneline\,\,\,\dots\,\,\oneline\,
   \circ
    $
        }
   \put(26.5,0.4){\line(3,-2){4}}
   \put(26.5,0.7){\line(3, 2){4}}
   \put(30.5,-3.2){$\circ$}
   \put(30.5, 3.0){$\circ$}
   \put(0,-3){$\alpha_0$}
   \put(5.5,-3){$\alpha_1$}
   \put(11.5,-3){$\alpha_2$}
   \put(24.5,-3){$\alpha_{q-2}$}
   \put(32.5,-3){$\alpha_q\qquad .$}
   \put(32.5, 3){$\alpha_{q-1}$}
   }
\end{picture}
\end{equation}
For odd $\dmn=2q+1$, $\so(\dmn+2)=B_{q+1}$ and $\roots$ is
described by the Dynkin diagram
\begin{equation}\label{odd-dynkin}
   \unitlength=5pt
   \begin{picture}(20,10)
   \put(-3,3){
   \put(0,0){
    $
    \circ\kern2pt\oneline\circ\oneline\circ\oneline\,\,\,\dots\,\,\oneline
   \circ\doubleline\kern-12pt>\circ
    $
        }
   \put(0,-3){$\alpha_0$}
   \put(5.5,-3){$\alpha_1$}
   \put(11.5,-3){$\alpha_2$}
   \put(24.5,-3){$\alpha_{q-1}$}
   \put(32.5,-3){$\alpha_q\qquad .$}
   }
\end{picture}
\end{equation}
In both cases we choose $\subroots=(\ga_1,\ldots,\ga_q)$ and
hence $\psubalg=\iso(\dmn)\oplus\so(2)
=\so(\dmn)\oplus\so(2)\subplus\translations(\dmn)$ where
$\Levi=\so(\dmn)\oplus\so(2)$ is the direct sum of the Lorentz
algebra and the dilatation while $\rad=\translations(\dmn)$ is the
algebra of momenta. Since the algebra $\translations(\dmn)$ is
Abelian (cf. (\ref{conformal_algebra_commutators})), we can apply
results of section \ref{Sct.:_General_Construction}
to classify all linear conformally invariant systems of
differential equations in terms of the cohomology
$\socohomology{0}{\module}$ and  $\socohomology{1}{\module}$ of
$\translations(\dmn)$ with coefficients in various integrable
$\so(\dmn+2)$-modules $\module$.

For the conformal algebra $\so(\dmn+2)$ and its parabolic
subalgebra $\iso(\dmn+2)\oplus\so(2)$, we calculate the cohomology
$\socohomology{p}{\irrmodule}$
for any $p$ and any irreducible module $\irrmodule$ using the
information on the structure of the  generalized Verma
modules obtained by the methods developed
in~\cite{Vogan,[BB],eastw}. Once the cohomology
$\socohomology{p}{\irrmodule}$ for any irreducible
module~$\irrmodule$ is known, the cohomology
$\socohomology{p}{\extmodule}$ for any extension $\extmodule$ of
the irreducible modules can also be easily found.

\subsection{Irreducible Tensors and Spinor--tensors%
\label{Sct.:_Irreducible_Tensors_and_Spin-tensors}}
Consider an irreducible finite dimensional module
$\vacVerma{\gl}$ of $\iso(\dmn)\oplus\so(2)$ with some basis
elements $\vacbasis{\gl}{A}$  of the carrier space, labelled by
$A$,
\begin{equation}\label{vacuum_module}
(\cL^{nm}\vacbasis{\gl}{})^A=\cL_0^{nmA}{}_B\vacbasis{\gl}{B}\,,
\qquad\cD\vacbasis{\gl}{A}=\Gd\vacbasis{\gl}{A}\,,
\qquad \cP^n\vacbasis{\gl}{A}=0\,.
\end{equation}
We choose the highest weight of $\vacVerma{\gl}$ in the form
$(\gl)=(\gl_0,\gl_1,\ldots,\gl_q)$, where $\gl_0=-\Delta$ is the
highest weight of $\so(2)$ and $(\gl_1,\ldots,\gl_q)$ is the
highest weight of~$\so(\dmn)$. The condition that $\vacVerma{\gl}$
is finite dimensional demands
\begin{equation}\label{Lorentz_weights_finitedimesiality}
2\gl_1\equiv\cdots\equiv2\gl_q\; \mbox{ mod 2}\,,
\end{equation}
\begin{equation}\label{Lorentz_weights_even}
\gl_1\geq\gl_2\geq \ldots \geq|\gl_q|\geq0\,,\quad\dmn\quad \mbox{is even}\,,
\end{equation}
\begin{equation}\label{Lorentz weights_odd}
\gl_1\geq\gl_2\geq \ldots \geq\gl_q\geq0\,, \quad\dmn \quad \mbox{
is odd}\,.
\end{equation}

It is customary in physics to describe finite dimensional
representations of the Lorentz algebra as appropriate irreducible
spaces of traceless tensors or $\gga$-transversal spinor--tensors.
One possible realization is as follows. Let
$2\gl_1\equiv\cdots\equiv2\gl_q\equiv 0\; \mbox{ mod 2}$. Consider
the space of traceless tensors
\begin{equation}
T^{n^1(\gl_1),n^2(\gl_2),\ldots,n^q(|\gl_q|)}\,,
\qquad
\eta_{n^i n^j}
T^{n^1(\gl_1),n^2(\gl_2),\ldots,n^q(|\gl_q|)}=0\,,
\qquad 1\leq i,j\leq q \,,
\end{equation}
where, following  \cite{fort}, we write $n^i(\gl_i)$ instead of
writing a set of $\gl_i$ totally symmetrized indices
$n^i_1,n^i_2,\ldots,n^i_{\gl_i}$, i.e. we indicate in parentheses
how many indices are subject to total symmetrization. For example,
we write $T^{n(\gl)}$ instead of rank-$\gl$ symmetric tensor
$T^{n_1\ldots n_\gl }$. We use the convention that upper(lower)
indices denoted by the same latter inside parentheses are
symmetrized.  For example, $T^{(n_1}P^{n_2)}$ is equivalent to
$\half (T^{n_1}P^{n_2}+T^{n_2}P^{n_1})$. The tensor
$T^{n^1(\gl_1),n^2(\gl_2),\ldots,n^q(|\gl_q|)}$ is totally
symmetric within each group of $\gl_i$ indices $n^i$. We impose
the condition that the total symmetrization of indices
$n^i(\gl_i)$ with any index from some set $n^j(\gl_j)$ with $j>i$
gives zero. Such symmetry properties are described by the Young
tableau $\Gl$ composed of rows of length $\gl_1$, $\gl_2$, \ldots,
$|\gl_q|$. Such tensors span the irreducible representation
$\vacVerma{\gl}$ whenever $\dmn$ is odd or $\gl_q=0$. For even
$\dmn$ and $\gl_q\neq 0$ this space is
$\vacVerma{\gl_0,\gl_1,\ldots,\gl_q}\oplus
\vacVerma{\gl_0,\gl_1,\ldots,-\gl_q}$, where the direct summands
are the selfdual and antiselfdual parts of the tensors (see
below).

Let $\gs_1,\ldots,\gs_p$ be the heights of the columns of
$\Lambda$. Another basis in $\vacVerma{\gl}$ with explicit
antisymmetrizations consists of the traceless tensors
\begin{equation}
T^{m^1[\gs_1],m^2[\gs_2],\ldots,m^p[\gs_p]}\,, \qquad
\eta_{m^i m^j}
T^{m^1[\gs_1],m^2[\gs_2],\ldots,m^p[\gs_p]}=0\,, \qquad 1\leq i,j
\leq p\,,
\end{equation}
where $m^i[\gs_i]$ denotes a set of totally antisymmetrized
indices $m^i_1,m^i_2,\dots,m^i_{\gs_i}$. We use the convention
that upper(lower) indices denoted by the same latter inside square
brackets are antisymmetrized \cite{fort}.  For example,
$T^{[n_1}P^{n_2]}$ is equivalent to $\half(
T^{n_1}P^{n_2}-T^{n_2}P^{n_1})$. For a tensor associated with the
Young tableau $\Lambda$ the condition is imposed that the total
antisymmetrization of the indices $m^i[\gs_i]$ with any index from
some set $m^j[\gs_j]$ with $j>i$  gives zero.

{}From the formula
\begin{equation}
\gep_{{n_1}\ldots n_\dmn}\epsilon_{{m_1}\ldots m_\dmn}=
\sum_{p}(-1)^{\pi(p)}\eta_{n_1 m_{p (1)}}\ldots \eta_{n_\dmn m_{p
(\dmn)}}\,,
\end{equation}
where summation is over all permutations $p$ of indices
$m_i$, and ${\pi (p)}= 0$ or $1$ is the oddness of the permutation
$p$, it follows for traceless tensors  that
\begin{equation}\label{traceless_tensors_property}
T^{\ldots ,m^i[\gs_i],\ldots ,m^j[\gs_j],\ldots}=0
\end{equation}
if $\gs_i+\gs_j >\dmn$ for some $i\neq j$. {}From
(\ref{traceless_tensors_property}) along with the property that
\begin{equation}
T^{\ldots,m^i[\gs],\ldots,m^j[\gs],\ldots}=
T^{\ldots,m^j[\gs],\ldots,m^i[\gs],\ldots}\,,
\end{equation}
it follows that there is essentially one way to define the Hodge
conjugation operation ${}^*$ for such tensors,
\begin{equation}
({}^*T)^{k[\dmn-\gs_1],m^2[\gs_2],\ldots,m^p[\gs_p]}=
\frac{(i)^{\gs_1(\dmn-\gs_1)}}{\gs_1 !}
T^{m^1[\gs_1],m^2[\gs_2],\ldots,m^p[\gs_p]}
\gep_{m^1[\gs_1]}{}^{k[\dmn-\gs_1]}\,,
\end{equation}
where the normalization factor is fixed such that
\begin{equation}
({}^{**}T)^{m^1[\gs_1],\ldots,m^p[\gs_p]}=
T^{m^1[\gs_1],\ldots,m^p[\gs_p]}\,.
\end{equation}
For $\dmn=2q$ and $\gl_q\neq 0$, to single out the irreducible
part of the $\so(2q)$ tensor representation
$T^{m^1[q],m^2[\gs_2],\ldots, m^p[\gs_p]}$, we impose the
(anti)selfduality condition
\begin{equation}\label{anty_selfduality_conditions}
{}^* T^{m^1[q],m^2[\gs_2],\ldots,m^p[\gs_p]}=
\pm T^{m^1[q],m^2[\gs_2],\ldots,m^p[\gs_p]}\,.
\end{equation}

When $2\gl_1\equiv\cdots\equiv2\gl_q\equiv 1\; \mbox{ mod 2}$, the
basis $\vacbasis{\gl}{A}$ of the module $\vacVerma{\gl}$ can be
realized by spinor--tensors
\begin{equation}
T^{n^1(\gl_1-\half),n^2(\gl_2-\half),
\ldots,n^q(|\gl_q|-\half),\ga} \qquad\mbox{\rm or}\qquad
T^{m^1[\gs_1],m^2[\gs_2],\ldots,m^p[\gs_p],\ga}\,,
\end{equation}
where $\ga=1,\ldots,2^{[\dmn/2]}$ is the spinor index. They
satisfy analogous (anti)symmetry conditions and are
$\gga$-transversal, i.e.
\begin{align}\label{gamma_transversality}
&\gga_{n^i}{}^\gb{}_\ga T^{n^1(\gl_1-\half),n^2(\gl_2-\half),
\ldots,n^q(|\gl_q|-\half),\ga}=0\,,&& 1\leq i\leq q\,,\\
&\gamma_{m^j}{}^\gb{}_\ga T^{m^1[\gs_1],m^2[\gs_2],
\ldots,m^p[\gs_p],\ga}=0\,,&& 1\leq j\leq p\,,\nn
\end{align}
where $\gga$ matrices satisfy (\ref{gamma_matrices}).
{}From~(\ref{gamma_transversality}) it follows that
$T^{n^1(\gl_1-\half),n^2(\gl_2-\half),\ldots,
n^q(|\gl_q|-\half),\ga}$ and
$T^{m^1[\gs_1],m^2[\gs_2],\ldots,m^p[\gs_p],\ga}$ are traceless.
A counterpart of the identity (\ref{traceless_tensors_property})
for $\gga$-transversal spinor--tensors is
\begin{equation}\label{gga_transversal_spin_tensors_property}
T^{m^1[\gs_1],m^2[\gs_2],\ldots,m^p[\gs_p],\ga}=0
\end{equation}
if $2\gs_i>\dmn$ for some $i$.

For $\dmn=2q$, to single out the irreducible part of a
spinor--tensor $\so(2q)$ module, one imposes the additional
chirality condition
\begin{align}\label{chirality_condition}
&\Gga^\gb{}_\ga T^{n^1(\gl_1-\half),n^2(\gl_2-\half),\ldots,
n^q(|\gl_q|-\half),\ga}=
\pm T^{n^1(\gl_1-\half),n^2(\gl_2-\half),\ldots,
n^q(|\gl_q|-\half),\gb}\,,\\
&\Gga^\gb{}_\ga T^{m^1[\gs_1],m^2[\gs_2],\ldots,m^p[\gs_p],\ga}=
\pm T^{m^1[\gs_1],m^2[\gs_2],\ldots,m^p[\gs_p],\gb}\,,\nn
\end{align}
where
\begin{equation}
\Gga^\gb{}_\ga=(-i)^{q}(\gga^1\cdots
\gga^{2q})^\gb{}_\ga
\end{equation}
is normalized to have unit square
\begin{equation}
\Gga^\gb{}_\gga \Gamma^\ga{}_\gb=\gd^\ga_\gga\,.
\end{equation}
(Note that for odd $\dmn$,  $\Gga$ is the central element, which
is required to be $\pm\one$ in a chosen spinor  representation
and hence (\ref{chirality_condition}) is automatically
satisfied.) For even $\dmn$, a $\gga$-transversal chiral
spinor--tensor $T^{m^1[q],m^2[\gs_2],\ldots, m^p[\gs_p],\ga}$
that has definite Young properties, is automatically
(anti)selfdual because
\begin{equation}
{}^*T^{m^1[q],m^2[\gs_2],\ldots,m^p[\gs_p],\gb}=
\Gga^\gb{}_\ga T^{m^1[q],m^2[\gs_2],\ldots,m^p[\gs_p],\ga}\,.
\end{equation}

\subsection{Generalized Verma Modules%
\label{Sct.:_Generalized_Verma_Modules}} The generalized Verma
$\so(\dmn+2)$-module $\Verma{\gl}$ is freely generated from a
vacuum module $\vacVerma{\gl}$ (see section
\ref{Sct.:_Irreducible_Tensors_and_Spin-tensors}) by the operators
$\cK^n$. Recall that $(\gl)=(\gl_0,\ldots,\gl_q)$ satisfy
(\ref{Lorentz_weights_finitedimesiality}),
(\ref{Lorentz_weights_even}), (\ref{Lorentz weights_odd}). It is
convenient to represent the action of~$\cK^n$ as a multiplication
by an independent variable~$y^n$. Basis elements of $\Verma{\gl}$
are formed by homogeneous polynomials
\begin{equation}\label{Verma_monomial_l}
\Vermabasis{l}{n(l);A}=\underbrace{y^{(n}\cdots y^{n)}}_{l}\vacbasis{\gl}{A}
\,,\qquad {l=0,1,2\ldots}\,.
\end{equation}
A special universality property of generalized Verma modules that
makes them important for our analysis is that any irreducible
$\so(\dmn+2)$-module $\irrVerma{\gl}$ with the highest weight
$(\gl)$ integrable with respect to the parabolic subalgebra
$\iso(\dmn)\oplus\so(2)$ is a quotient of $\Verma{\gl}$.

The subspace $\Vermal{\gl}{l}\subset\Verma{\gl}$ spanned by
degree $l$ monomials (\ref{Verma_monomial_l})  is called the
$l$-th level of $\Verma{\gl}$. The associated
 grading in $\Verma{\gl}$ is
\begin{equation}\label{Verma_gradation}
\Verma{\gl}=\bigoplus_{l=0}^\infty\Vermal{\gl}{l}\,.
\end{equation}

The representation of the conformal algebra
in $\Verma{\gl}$ is
\begin{align}\label{so_action_on_Verma_module}
&\cL^{mk}\ketv{v}=
\left(y^k\d{y_m}-y^m\d{y_k}+\cL_0^{mk}\right)\ketv{v}\,,\\
&\cD\ketv{v}=\left(-\gl_0+y^j\d{y^j}\right)\ketv{v}\,,\\
&\cK^m\ketv{v}=y^m\ketv{v}\,,\\
&\cP^m\ketv{v}=\left(2\left(-\gl_0+y^j\d{y^j}\right)\d{y_m}-
y^m\d{y^j}\d{y_j}+2\cL_0^{mj}\d{y^j}\right)\ketv{v}\,,\label{P-in-verma}
\end{align}
where $\ketv{v}\in\Verma{\gl}$ and $\cL_0^{nm}$ acts in the vacuum
module (\ref{vacuum_module}). $\cL^{mk}$ and $\cD$ preserve level
$l$. $\cD$ is the grading operator, i.e. $\Vermal{\gl}{l}$ is the
eigenspace of $\cD$ with the eigenvalue $-\gl_0+l$. $\cK^m$ and
$\cP^m$ increase and decrease a level by one unit,  respectively.

Every level $\Vermal{\gl}{l}$ decomposes into a direct sum of
$\so(\dmn)\oplus\so(2)$ irreducible modules,
\begin{equation}\label{Verma_level_decomposition}
\Vermal{\gl}{l}=\bigoplus_{i=0}^{[l/2]}\vacVerma{\gl}\tensor
\vacVerma{-l,l-2i,0,\dots,0}=\bigoplus_{(\mu)\in\Gl_{(\gl),l}}\vacVerma{\mu}\,,
\end{equation}
where $\Gl_{(\gl),l}$ is the set of highest weights  in this
decomposition. A $\so(\dmn)\oplus\so(2)$-module $\singmodule{\mu}$
in decomposition (\ref{Verma_level_decomposition}) with $l\geq 1$
is called singular module if
\begin{equation}
\cP^n\singmodule{\mu}=0\,.
\end{equation}
Any vector from $\singmodule{\mu}$ is called singular vector. Let
singular vectors $\singbasis{A}$ form a basis of
$\singmodule{\mu}$. Any singular module
$\singmodule{\mu}\subset\Vermal{\gl}{l}$ induces the proper
submodule $\subVerma{\gl}{\mu}$ of $\Verma{\gl}$ with the
homogeneous elements of the form
\begin{equation}\label{submodule_monomial_m}
\ketv{m}^{n(m);A}=\underbrace{y^{(n}\cdots y^{n)}}_{m}
\singbasis{A}\,,\qquad m\geq 0\,.
\end{equation}
 Note that $\subVerma{\gl}{\mu}$ is not freely
generated from $\singmodule{\mu}$, i.e., the elements
$\ketv{m}^{n(m);A}$ are not necessarily linearly independent.
Also note that the grading (\ref{Verma_gradation}) defined for
generalized Verma modules differs from the grading
(\ref{gradation}) defined in section
\ref{Sct.:_General_Construction} for arbitrary
$\psubalg$-integrable modules. Namely, $\Verma{\gl}{}_{[0]}$
consists of~$\Vermal{\gl}{0}$ along with all  singular
subspaces of $\Verma{\gl}$. In what follows we use
the grading (\ref{Verma_gradation}).

If $\Verma{\gl}$ is irreducible it does not contain singular
modules.  For reducible $\Verma{\gl}$, let $\singmodule{\mu_1}$,
$\singmodule{\mu_2},\ldots$ list all singular modules of
$\Verma{\gl}$. Let $\msubVerma{\gl}$ be the image in $\Verma{\gl}$
of the module induced from
$\singmodule{\mu_1}\oplus\singmodule{\mu_2}\oplus\ldots$. Consider
the quotient $\quotVerma{\gl}=\Verma{\gl}/\msubVerma{\gl}$. A
singular module~$\singmodule{\mu}'$ of $\quotVerma{\gl}$ is called
a subsingular module of~$\Verma{\gl}$. Its elements are called
subsingular vectors. A singular module of the quotient
$\quotVerma{\gl}'=\quotVerma{\gl}/\msubVerma{\gl}'$ is called a
subsubsingular module $\singmodule{\mu}''$ of $\Verma{\gl}$ and so
on. For generalized Verma modules $\Verma{\gl}$ of the conformal
algebra the situation is relatively simple because~$\Verma{\gl}$
can have only singular and subsingular modules for~$\dmn$  even
and only singular modules for~$\dmn$  odd (see section
\ref{Sct.:_Structure of so(M+2) generalized Verma modules} and
Appendix A for more details).

\subsection{Contragredient Modules\label{Sct.:_Contragredient_Modules}}
Let $\module$ be an $\alg$-module. The module $\comodule$
contragredient to module $\module$ is the graded dual to
$\module$ vector space\footnote{Graded dual vector space to the
graded space $V=\oplus_i V_i$ with finite-dimensional homogeneous
components~$V_i$ is defined as $V^*=\oplus_i V^*_i$, where each
$V^*_i$ is dual to the corresponding~$V_i$.} with the action of
the algebra~$\alg$ defined as
\begin{equation}\label{contragredient_algebra_action}
f\ga(v)=\ga(\tau(f)v)\,,
\end{equation}
where $f\in\alg$, $v\in\module$, $\ga\in\comodule$ and $\tau$ is
the Chevalley involution (\ref{Chevalley_involution}). Note that
for any irreducible module~$\irrVerma{\gl}$ with the highest
weight $(\gl)$, the contragredient module $\coirrVerma{\gl}$ is
also irreducible with the same highest weight and, thus,
$\irrVerma{\gl}\sim\coirrVerma{\gl}$.

The  module $\coVerma{\gl}$ contragredient to the
generalized Verma module $\Verma{\gl}$ can be realized as follows.
Consider $\covacmodule{\gl}\sim\vacVerma{\gl}$ with the basis
$\covacbasis{\gl}{A}$  dual to $\vacbasis{\gl}{A}$
\begin{equation}\label{vacuum_scalar_product}
\covacbasis{\gl}{B}\vacbasis{\gl}{A}=\gd^A_B
\end{equation}
 and the following action of the $\iso(\dmn)\oplus\so(2)$ algebra
\begin{equation}\label{co-vacuum_module}
{}_A(\covacbasis{\gl}{}\cL^{nm})=
\covacbasis{\gl}{B}\cL_0^{nm}{}_A{}^B\,,
\qquad\covacbasis{\gl}{A}\cD=-\covacbasis{\gl}{A}\gl_0\,,
\qquad\covacbasis{\gl}{A}\cP^n=0\,.
\end{equation}
The vector space $\coVerma{\gl}$ can be realized as
the space of polynomials
of~$y^n$ with coefficients in~$\covacmodule{\gl}$.
It is convenient to extend the definition of the Chevalley
involution to this realization as follows:
\begin{equation}\label{Verma_scalar_product}
\tau (y^n)=\d{y_n}\,,\qquad\tau \left(\d{y_n}\right)=y^n\,.
\end{equation}
The $l$-th level $\coVermal{\gl}{l}$ of $\coVerma{\gl}$ is spanned
by the monomials
\begin{equation}\label{co-Verma_monomial_l}
\coVermabasis{l}{n(l);A}=\frac{1}{l!}\covacbasis{\gl}{A}
\underbrace{y_{(n}\cdots y_{n)}}_{l}\,.
\end{equation}
{}From
\begin{equation}\label{so_conjugation}
\tau(\cL^{nm})=-\cL^{nm}\,,\qquad
\tau(\cD)=\cD\,,\qquad
\tau(\cK^n)=\cP^n\,,\qquad
\tau(\cP^n)=\cK^n
\end{equation}
it follows that  the action
(\ref{contragredient_algebra_action}) of  $\so(\dmn+2)$
on $\coVerma{\gl}$ is
{\allowdisplaybreaks
\begin{align}\label{so_action_on_co-Verma_module}
&\brav{\ga}\cL^{mk}=
\brav{\ga}\Big(\ld{y_m}y^k-\ld{y_k}y^m+\cL_0^{mk}\Big)\,,\\
&\brav{\ga}\cD=\brav{\ga}\Big(-\gl_0+\ld{y_j}y^j\Big)\,,\\
&\brav{\ga}\cK^m=\brav{\ga}\Big(2(-\gl_0+\ld{y^j}y^j)y^m-
\ld{y_m}y^jy_j+2\cL_0^{mj}y_j\Big)\,,\\
&\brav{\ga}\cP^m=\brav{\ga}\ld{y_m}\label{eq:P-action-in-contragrad}\,,
\end{align}
} for  $\brav{\ga}\in\coVerma{\gl}$. Note that the elements
$\cP^n$ act co-freely in $\coVerma{\gl}$, i.e. any vector in
$\coVermal{\gl}{l}$ has a preimage under the action of~$\cP^n$
for every $n$.

\subsection{Structure of $\so(\dmn+2)$ Generalized Verma
Modules\label{Sct.:_Structure of so(M+2) generalized Verma
modules}} In this section we describe  the structure of
$\so(\dmn+2)$ generalized Verma modules. Singular modules in
$\so(\dmn+2)$ generalized Verma modules were completely
investigated in~\cite{[BCI],[BCII],eastw}.
To find subsingular modules we use general results
from~\cite{Vogan,[BB]}. This analysis is sketched in Appendix A.

\subsubsection{$\dmn=2q+1$\label{Sct.:_M=2q+1}}
It turns out that for odd $\dmn=2q+1$, $\dmn\geq 3$ any
$\so(\dmn+2)$ generalized Verma module $\Verma{\gl}$ does not have
subsingular modules\footnote{The fact that the homogeneous space
$SO(\dmn+2)/ISO(\dmn)\times SO(2)$ does not contain two cells of
the equal dimension forbids appearance of subsingular modules
\cite{Vogan}.}. This means that the maximal submodule
$\msubVerma{\gl}\subset\Verma{\gl}$ such that the quotient
$\quotVerma{\gl}=\Verma{\gl}/\msubVerma{\gl}$ is irreducible, is
induced from singular modules. For generic~($\lambda$),
$\Verma{\gl}$ is irreducible. There are two series of reducible
generalized Verma modules.

Let $(\gl)_0$ be an arbitrary dominant integral weight, i.e.{}
$\gl_0\ge\gl_1\geq\cdots\geq\gl_q$ and $2\gl_0\equiv\cdots\equiv
2\gl_q\mbox{ mod 2}$. The first series consists of the modules
with the following highest weights {\allowdisplaybreaks
\begin{align}\label{odd_dimmensional_sequence}
&(\gl)_0=(\gl_0,\gl_1,\ldots,\gl_q)\,,\nn\\
&(\gl)_1=(\gl_1-1,\gl_0+1,\gl_2,\ldots,\gl_q)\,,\nn\\[-0.2cm]
&\qquad\qquad\qquad\qquad\vdots\nn\\[-0.2cm]
&(\gl)_N=(\gl_N-N,\gl_0+1,\ldots,\gl_{N-1}+1,
    \gl_{N+1},\ldots,\gl_q)\,,&&\mbox{$N=0,\ldots,q$}\,,\nn\\[-0.2cm]
&\qquad\qquad\qquad\qquad\vdots\nn\\[-0.2cm]
&(\gl)_q=(\gl_q-q,\gl_0+1,\ldots,\gl_{q-1}+1)\,,\\
&(\gl)_{q+1}=(-\gl_q-q-1,\gl_0+1,\ldots,\gl_{q-1}+1)\,,\nn\\[-0.2cm]
&\qquad\qquad\qquad\qquad\vdots\nn\\[-0.2cm]
&(\gl)_{q+K}=(-\gl_{q+1-K}-q-K,\gl_0+1,\ldots,\gl_{q-K}+1,
    \gl_{q-K+2},\ldots,\gl_q)\,,&&\mbox{$K=1,\ldots,q$}\,,\nn\\[-0.2cm]
&\qquad\qquad\qquad\qquad\vdots\nn\\[-0.2cm]
&(\gl)_{2q-1}=(-\gl_2-2q+1,\gl_0+1,\gl_1+1,\gl_3\ldots,\gl_q)\,,\nn\\
&(\gl)_{2q}=(-\gl_1-2q,\gl_0+1,\gl_2,\ldots,\gl_q)\,.\nn
\end{align}
} The generalized Verma modules with highest weights from
(\ref{odd_dimmensional_sequence}) have the structure described by
the following short exact sequences
{\allowdisplaybreaks
\begin{align}
\label{odd_dimensional_exact_sequence_0}
&\exactsq{\sqirrVerma{\gl}{0}}{\sqVerma{\gl}{0}}{\sqirrVerma{\gl}{1}}\,;\\[-0.2cm]
&\qquad\qquad\qquad\qquad\vdots\nn\\[-0.2cm]
\label{odd_dimensional_exact_sequence_N}
&\exactsq{\sqirrVerma{\gl}{N}}{\sqVerma{\gl}{N}}{\sqirrVerma{\gl}{N+1}}\,,
&&N=0,\ldots,2q\,;\\[-0.2cm]
&\qquad\qquad\qquad\qquad\vdots\nn\\[-0.2cm]
\label{odd_dimensional_exact_sequence_2q}
&\exactsq{\sqirrVerma{\gl}{2q}}{\sqVerma{\gl}{2q}}{\sqirrVerma{\gl}{2q+1}}\,;\\[0.3cm]
\label{odd_dimensional_exact_sequence_2q+1}
&0\longrightarrow\sqVerma{\gl}{2q+1}\longrightarrow\sqirrVerma{\gl}{2q+1}
\longrightarrow 0\,,
\end{align}
} where $(\gl)_{2q+1}=(-\gl_0-2q-1,\gl_1,\ldots,\gl_q)$ and all
$\irrVerma{\gl}$ are irreducible. Equation
(\ref{odd_dimensional_exact_sequence_2q+1}) means that
$\sqVerma{\gl}{2q+1}=\sqirrVerma{\gl}{2q+1}$ is irreducible.
Equation
 (\ref{odd_dimensional_exact_sequence_2q}) means that
$\sqirrVerma{\gl}{2q+1}$ is the maximal submodule of $\sqVerma{\gl}{2q}$
and the quotient                $\sqirrVerma{\gl}{2q}=
\sqVerma{\gl}{2q}/\sqVerma{\gl}{2q+1}$ is irreducible.
The maximal submodule of $\sqVerma{\gl}{2q-1}$ is
$\sqirrVerma{\gl}{2q}$ and the quotient
$\sqirrVerma{\gl}{2q-1}=\sqVerma{\gl}{2q-1}/\sqirrVerma{\gl}{2q}$
is irreducible, and so on.

The second series consists of reducible generalized Verma modules
with non-integral highest weights. Let $\mu_1\geq\cdots\geq\mu_q$
and $2\mu_1\equiv\cdots\equiv 2\mu_q \mbox{ mod 2}$. Consider the
highest weight
\begin{equation}\label{odd_dimensional_series_two}
\begin{split}
(\mu)&=(\mu_0,\mu_1,\ldots,\mu_q)\,,\\
\mu_0&=-q+\half+\oN_0
\qquad\mbox{if $2\mu_1\equiv2\mu_q\equiv 0$ mod 2}\,,\\
\mu_0&=-q+\oN_0 \qquad\qquad\mbox{if $2\mu_1\equiv2\mu_q\equiv 1$
mod 2}\,.
\end{split}
\end{equation}
We have
\begin{equation}\label{odd_dimensional_exect_sequence_mu}
\exactsq{\irrVerma{\mu}}{\Verma{\mu}}{\Verma{\mu}{}_{'}}\,,
\end{equation}
where
\begin{equation}
(\mu)'=(-\mu_0-2q-1,\mu_1,\ldots,\mu_q)\,.
\end{equation}
The modules $\irrVerma{\mu}=\Verma{\mu}/\Verma{\mu}{}_{'}$
and $\Verma{\mu}{}_{'}$ are irreducible.

The described two series give the full list of
 reducible $\so(\dmn+2)$ generalized
Verma modules for odd $\dmn$.

\subsubsection{$\dmn=2q$\label{Sct.:_M=2q}}
The structure of $\so(\dmn+2)$ generalized Verma modules
$\Verma{\gl}$ for even $\dmn$ is more complicated because in the
even dimensional case some $\Verma{\gl}$ have subsingular modules
(no subsubsingular modules, however\footnote{The fact that the
homogeneous space $SO(\dmn+2)/ISO(\dmn)\otimes SO(2)$ does not
contain three cells of the equal dimension forbids appearance of
subsubsingular modules \cite{Vogan}.}). Again, there are  two
series of reducible generalized Verma modules.

Let $(\gl)_{-q}$ be an arbitrary dominant integral weight, i.e.{}
$\gl_0\ge\gl_1\geq\cdots\geq|\gl_q|$ and $2\gl_0\equiv
\cdots\equiv 2\gl_q\mbox{ mod 2}$. Consider the set of highest
weights {\allowdisplaybreaks
\begin{align}\label{even_dimmensional_sequence}
\kern20pt&(\gl)_{-q}=(\gl_0,\gl_1,\ldots,\gl_q)\,,\nn\\
&(\gl)_{-q+1}=(\gl_1-1,\gl_0+1,\gl_2,\ldots,\gl_q)\,,\nn\\[-0.2cm]
&\qquad\qquad\qquad\qquad\vdots\nn\\[-0.2cm]
&(\gl)_{-q+N}=(\gl_N-N,\gl_0+1,\ldots,\gl_{N-1}+1,
                       \gl_{N+1},\ldots,\gl_q)\,,
      \qquad\qquad\qquad\;\:{N=0,\ldots,q-1}\,,\nn\\[-0.2cm]
&\qquad\qquad\qquad\qquad\vdots\nn\\[-0.2cm]
&(\gl)_{-1}=(\gl_{q-1}-q+1,\gl_0+1,\ldots,\gl_{q-2}+1,\gl_q)\,,\nn\\[0.1cm]
&\kern-10pt(\gl)_0=(\gl_q-q,\gl_0+1,\ldots,\gl_{q-1}+1)\,,
\qquad
(\gl)_{0'}=(-\gl_q-q,\gl_0+1,\ldots,\gl_{q-2}+1,-\gl_{q-1}-1)\,,\nn\\[0.2cm]
&(\gl)_{1}=(-\gl_{q-1}-q-1,\gl_0+1,\ldots,\gl_{q-2}+1,-\gl_q)\,,\\[-0.2cm]
&\qquad\qquad\qquad\qquad\vdots\nn\\[-0.2cm]
&\kern-20pt(\gl)_{K}=(-\gl_{q-K}-q-K,\gl_0+1,\ldots,\gl_{q-K-1}+1,
                       \gl_{q-K+1},\ldots,\gl_{q-1},-\gl_q)\,,
                       \quad{K=1,\ldots,q-1}\,,\nn\\[-0.2cm]
&\qquad\qquad\qquad\qquad\vdots\nn\\[-0.2cm]
&(\gl)_{q-2}=(-\gl_2-2q+2,\gl_0+1,\gl_1+1,
       \gl_3 ,\ldots,\gl_{q-1},-\gl_q)\,,\nn\\
&(\gl)_{q-1}=(-\gl_1-2q+1,\gl_0+1,\gl_2,\ldots,\gl_{q-1},-\gl_q).\nn
\end{align}
}
The structure of the generalized Verma modules with the highest
weights~(\ref{even_dimmensional_sequence})
is described by the following short exact sequences
{\allowdisplaybreaks
\begin{align}
\label{even_dimensional_exact_sequence_0}
&\exactsq{\sqirrVerma{\gl}{-q}}{\sqVerma{\gl}{-q}}
{\sqquotVerma{\gl}{-q+1}}\,;\\[-0.2cm]
&\exactsq{\sqirrVerma{\gl}{-q+1}}{\sqquotVerma{\gl}{-q+1}}
{\sqVerma{\gl}{q}^\natural}\,;\\[0.5cm]
\label{even_dimensional_exact_sequence_1_quot}
&\exactsq{\sqirrVerma{\gl}{-q+1}}{\sqVerma{\gl}{-q+1}}
{\sqquotVerma{\gl}{-q+2}}\,;\\[-0.2cm]
\label{even_dimensional_exact_sequence_1_irr}
&\exactsq{\sqirrVerma{\gl}{-q+2}}{\sqquotVerma{\gl}{-q+2}}
{\sqVerma{\gl}{q-1}^\natural}\,;\\[0.2cm]
&\qquad\qquad\qquad\qquad\vdots\nn\\[0.2cm]
&\exactsq{\sqirrVerma{\gl}{N}}{\sqVerma{\gl}{N}}
{\sqquotVerma{\gl}{N+1}}\,,
                        &&N=-q,-q+1,\ldots,-2\,,\\[-0.2cm]
&\exactsq{\sqirrVerma{\gl}{N+1}}{\sqquotVerma{\gl}{N+1}}
{\sqVerma{\gl}{-N}^\natural}\,;\\[0.2cm]
&\qquad\qquad\qquad\qquad\vdots\nn\\[0.2cm]
\label{even_dimensional_exact_sequence_q-2_quot}
&\exactsq{\sqirrVerma{\gl}{-2}}{\sqVerma{\gl}{-2}}
{\sqquotVerma{\gl}{-1}}\,;\\[-0.2cm]
\label{even_dimensional_exact_sequence_q-2_irr}
&\exactsq{\sqirrVerma{\gl}{-1}}{\sqquotVerma{\gl}{-1}}
{\sqVerma{\gl}{2}^\natural}\,;\\[0.5cm]
\label{even_dimensional_exact_sequence_q-1_sub}
&\exactsq{\sqirrVerma{\gl}{-1}}
{\sqVerma{\gl}{-1}}{\sqquotVerma{\gl}{0}}\,,\\[-0.2cm]
\label{even_dimensional_exact_sequence_q-1_irr}
&\exactsq{\sqirrVerma{\gl}{0}\oplus\sqirrVerma{\gl}{0}{}_{'}}
{\sqquotVerma{\gl}{0}}
{\sqVerma{\gl}{1}^\natural}\,;\\[0.5cm]
\label{even_dimensional_exact_sequence_q}
&\exactsq{\sqirrVerma{\gl}{0}}{\sqVerma{\gl}{0}}
{\sqirrVerma{\gl}{1}}\,;\\[0.5cm]
\label{even_dimensional_exact_sequence_q'}
&\exactsq{\sqirrVerma{\gl}{0}{}_{'}}{\sqVerma{\gl}{0}{}_{'}}
{\sqirrVerma{\gl}{1}}\,;\\[0.5cm]
\label{even_dimensional_exact_sequence_q+1}
&\exactsq{\sqirrVerma{\gl}{1}}{\sqVerma{\gl}{1}}
{\sqirrVerma{\gl}{2}}\,;\\[0.2cm]
&\qquad\qquad\qquad\qquad\vdots\nn\\[0.2cm]
&\exactsq{\sqirrVerma{\gl}{N}}{\sqVerma{\gl}{N}}{\sqirrVerma{\gl}{N+1}}\,,
&&N=1,\ldots,q-1\,;\\[0.2cm]
&\qquad\qquad\qquad\qquad\vdots\nn\\[0.2cm]
\label{even_dimensional_exact_sequence_2q-1}
&\exactsq{\sqirrVerma{\gl}{q-1}}{\sqVerma{\gl}{q-1}}
{\sqirrVerma{\gl}{q}}\,;\\[0.5cm]
\label{even_dimensional_exact_sequence_2q}
&0\longrightarrow\sqVerma{\gl}{q}\longrightarrow\sqirrVerma{\gl}{q}
\longrightarrow 0\,.
\end{align}
} Here $(\gl)_{q}=(-\gl_0-2q,\gl_1,\ldots,\gl_{q-1},-\gl_q)$, and
all $\irrVerma{\gl}$ are irreducible. Analogously to the odd
dimensional case, (\ref{even_dimensional_exact_sequence_2q}) means
that $\sqVerma{\gl}{q}=\sqirrVerma{\gl}{q}$ is irreducible. From
(\ref{even_dimensional_exact_sequence_2q-1}) it follows that
$\sqirrVerma{\gl}{q}$ is the maximal submodule of $\sqVerma{\gl}{q-1}$
and the quotient                $\sqirrVerma{\gl}{q-1}$
is irreducible, which in
its turn is the maximal submodule of $\sqVerma{\gl}{q-2}$ and so on.
Continuing the same way one finally arrives at
$\sqirrVerma{\gl}{1}=\sqVerma{\gl}{1}/\sqirrVerma{\gl}{2}$
(\ref{even_dimensional_exact_sequence_q+1}). The structure of the
modules $\sqVerma{\gl}{1},\ldots,\sqVerma{\gl}{q-1}$ is analogous to
that of the odd dimensional case.

The modules $\sqVerma{\gl}{0}$ and $\sqVerma{\gl}{0}{}_{'}$ have the
common maximal submodule $\sqirrVerma{\gl}{1}$ (see
(\ref{even_dimensional_exact_sequence_q}),
(\ref{even_dimensional_exact_sequence_q'})) and the quotients
$\sqirrVerma{\gl}{0}=\sqVerma{\gl}{0}/\sqirrVerma{\gl}{1}$ and
$\sqirrVerma{\gl}{0}{}_{'}=\sqVerma{\gl}{0}{}_{'}/\sqirrVerma{\gl}{1}$
are irreducible. The module~$\sqVerma{\gl}{-1}$ has the most
complicated structure of submodules. Equation
(\ref{even_dimensional_exact_sequence_q-1_irr}) describes the
structure of the maximal submodule~$\sqquotVerma{\gl}{0}$
of~$\sqVerma{\gl}{-1}$. The appearance of the contragredient
module~$\sqVerma{\gl}{1}^\natural$
in~(\ref{even_dimensional_exact_sequence_q-1_irr}) means that the
maximal submodule of~$\sqVerma{\gl}{-1}$ cannot be generated from
singular modules because the module contragredient to a generalized Verma
module is not (unless it is irreducible) a highest-weight module
and therefore~$\sqVerma{\gl}{-1}$ contains a subsingular module.
Analogously the
modules~$\sqVerma{\gl}{-2}\ldots\sqVerma{\gl}{-q+1}$ contain
singular and subsingular modules as described by
(\ref{even_dimensional_exact_sequence_q-2_quot},
\ref{even_dimensional_exact_sequence_q-2_irr})-(\ref{even_dimensional_exact_sequence_1_quot},
\ref{even_dimensional_exact_sequence_1_irr}). Finally, the module
$\sqVerma{\gl}{-q}$ contains the
submodule~$\sqVerma{\gl}{q}^\natural$ but in this case subsingular
modules do not appear because~$\sqVerma{\gl}{q}^\natural$ is
isomorphic to~$\sqVerma{\gl}{q}=\sqirrVerma{\gl}{q}$, and
therefore the maximal submodule of~$\sqVerma{\gl}{-q}$ is
generated from singular modules.

Let $\mu_1\geq\cdots\geq\mu_{q-1}\geq |\mu_q|$ and
$2\mu_1\equiv\cdots\equiv 2\mu_q \hbox{ mod 2}$. The second series
of reducible generalized Verma $\so(\dmn+2)$ modules with even
$\dmn$ contains the modules with the singular highest
weights~$(\mu)=(\mu_0,\mu_1,\ldots,\mu_q)$ such that
{\allowdisplaybreaks
\begin{equation}\label{even_series_mu}
\begin{split}
&\mu_0=\mu_N-N \qquad\hbox{for some $N=1,\ldots,q$}\,,\\
&\mu_0\neq -q\,,\\
&\mu_0+\mu_q+q\in\oN_0\,,\\
&\mu_0-\mu_q+q\in\oN_0\,.
\end{split}
\end{equation}
}

The structure of $\Verma{\mu}$ is described by the short exact
sequence
\begin{equation}\label{even_dimensional_exact_sequence_mu}
\exactsq{\irrVerma{\mu}}{\Verma{\mu}}{\Verma{\mu}{}_{'}}\,,
\end{equation}
where $(\mu)'=(-\mu_0-2q,\mu_1,\ldots,\mu_{q-1},-\mu_q)$ and
$\irrVerma{\mu}=\Verma{\mu}/\Verma{\mu}{}_{'}$ is irreducible.

\subsection{Cohomology of Irreducible $\so(\dmn+2)$-Modules%
\label{Sct.:_Cohomology of irreducible o(M+2)-modules}} Any
irreducible $\so(\dmn+2)$-module $\irrVerma{\gl}$ with the highest
weight $(\gl)$ integrable with respect to the parabolic subalgebra
$\iso(\dmn)\oplus\so(2)$ is a quotient of an appropriate
generalized Verma $\so(\dmn+2)$-module $\Verma{\gl}$. (Recall that
$(\gl)$ is required to satisfy
(\ref{Lorentz_weights_finitedimesiality})-(\ref{Lorentz
weights_odd}).) In this section we show that once the structure of
all generalized Verma modules is known, one can calculate
$\socohomology{p}{\irrVerma{\gl}}$ (i.e. the cohomology of
$\translations(\dmn)$ with coefficients in $\irrVerma{\gl}$) for
any $p$ and irreducible~$\irrVerma{\gl}$. Recall that
$\translations(\dmn)$ is the subalgebra of $\so(\dmn+2)$ generated
by the momenta $\cP^n$.

Let us start with the following Lemma.
\begin{Lemma}\label{Lemma_4_1}
Let $\Verma{\gl}$ be the generalized Verma $\so(\dmn+2)$-module
induced from $\vacVerma{\gl}$. Then
\begin{align}
\label{Lemma:_coVerma_cohomology_0}
&\socohomology{0}{\coVerma{\gl}}=\vacVerma{\gl}\,,\\
\label{Lemma:_coVerma_cohomology_p}
&\socohomology{p}{\coVerma{\gl}}=0 \qquad\hbox{for $p=1,\ldots$.}
\end{align}
\end{Lemma}
\begin{prf}
{}From (\ref{eq:P-action-in-contragrad}) it follows that~$\sgminus
= \xi^n {\d{y^n}}$ (see (\ref{gs})) for any ${\coVerma{\gl}}$.
Equations (\ref{Lemma:_coVerma_cohomology_0}) and
(\ref{Lemma:_coVerma_cohomology_p}) follow from the standard
Poincar\'e Lemma.
\end{prf}
The following two Theorems describe the $\sgminus$
cohomology $\socohomology{p}{\irrVerma{\gl}}$ with coefficients
in $\irrVerma{\gl}$. Recall that any $\irrVerma{\gl}$ is a quotient
of the generalized Verma module $\Verma{\gl}$ induced from
$\vacVerma{\gl}$ as described in Sec.~\ref{Sct.:_Generalized_Verma_Modules}.
\begin{Thm}\label{Thm_cohomology_odd}\nopagebreak
Let  $\dmn$ be odd\nopagebreak
\begin{enumerate}
\item\label{Thm_cohomology_odd_1} If $\Verma{\gl}$ is irreducible then
\begin{align}
&\socohomology{0}{\irrVerma{\gl}}=\vacVerma{\gl}\,,\\
&\socohomology{p}{\irrVerma{\gl}}= 0\,,&&p=1,2,\ldots\,.
\end{align}
\item\label{Thm_cohomology_odd_2} If $\Verma{\gl}$ is reducible and
$(\gl)=(\gl)_N$ ($N=0,\ldots,2q$) belongs to the series
\eqref{odd_dimmensional_sequence} then
\begin{align}
&\socohomology{p}{\sqirrVerma{\gl}{N}}=\sqvacVerma{\gl}{p+N}\,,
&& p=0,\ldots,2q+1-N\,,\\
&\socohomology{p}{\sqirrVerma{\gl}{N}}=0\,,&& p=2q+2-N,\ldots\,.
\end{align}
\item\label{Thm_cohomology_odd_3} If $\Verma{\gl}$ is reducible and
$(\gl)=(\mu)$ belongs to the series
\eqref{odd_dimensional_series_two} then
\begin{align}
&\socohomology{0}{\irrVerma{\mu}}=\vacVerma{\mu}\,,\\
&\socohomology{1}{\irrVerma{\mu}}=\vacVerma{\mu}{}_{'}\,,\\
&\socohomology{p}{\irrVerma{\mu}}=0\,,&&p=2,\ldots\,.
\end{align}
\end{enumerate}
\end{Thm}
\begin{prf}
Item \ref{Thm_cohomology_odd_1} follows from Lemma \ref{Lemma_4_1}
and the observation that~$\Verma{\gl}$ is isomorphic
to~$\coVerma{\gl}$ whenever~$\Verma{\gl}$ is irreducible.

Items \ref{Thm_cohomology_odd_2} and \ref{Thm_cohomology_odd_3}
follow from Lemma \ref{Lemma_4_1} and long cohomological sequences
corresponding to short exact sequences contragredient to
(\ref{odd_dimensional_exact_sequence_0})-(\ref{odd_dimensional_exact_sequence_2q+1})
and~(\ref{odd_dimensional_exect_sequence_mu}).
\end{prf}
\begin{Thm}\label{Thm_cohomology_even}
Let $\dmn$ be even
\begin{enumerate}
\item If $\Verma{\gl}$ is irreducible then
\begin{align}
&\socohomology{0}{\irrVerma{\gl}}=\vacVerma{\gl}\,,\\
&\socohomology{p}{\irrVerma{\gl}}=0\,,&&p=1,2,\ldots\,.
\end{align}
\item If $\Verma{\gl}$ is reducible and
$(\gl)=(\gl)_N$ ($N=-q,-q+1,\ldots,-1,0,0',\ldots,q$) belongs to
the series \eqref{even_dimmensional_sequence} then
\begin{align}
&\socohomology{0}{\sqirrVerma{\gl}{N}}=\sqvacVerma{\gl}{N}\,,
&&N=-q,-q+1,\dots,-1,0,0',1,\ldots,q\,,
 \label{cohomology-0}\\
&\socohomology{p}{\sqirrVerma{\gl}{N}}=\sqtvacVerma{\gl}{p+N}
   \oplus\sqtvacVerma{\gl}{p-N}\,,
   &&p=1,\ldots,\quad N=-q+1,-q+2,\ldots -1\,,
    \label{cohomology-p-1}\\
&\socohomology{p}{\sqirrVerma{\gl}{N}}=\sqtvacVerma{\gl}{p+N}\,,
   &&p=1,\ldots,\quad N=-q,0,0',1,2,\ldots q\,,
  \label{cohomology-p-2}
\end{align}
where
\begin{align}
\sqtvacVerma{\gl}{N}&=\sqvacVerma{\gl}{N}&\mbox{\rm for
$N=-q,-q+1,\ldots,q$ and $N\neq 0$}\,,\nn\\
\sqtvacVerma{\gl}{0}&=\sqvacVerma{\gl}{0}
                   \oplus\sqvacVerma{\gl}{0}{}_{'}\,,&\label{n_s_tildoy}\\
\sqtvacVerma{\gl}{N}&=0&\mbox{\rm for
$N=q+1,\ldots$\,,\kern105pt}\nn
\end{align}
and $p+0=p+0'=p$.
\item If $\Verma{\gl}$ is reducible and
$(\gl)=(\mu)$ belongs to the series
\eqref{even_dimensional_exact_sequence_mu} then
\begin{align}
&\socohomology{0}{\irrVerma{\mu}}=\vacVerma{\mu}\,,\\
&\socohomology{1}{\irrVerma{\mu}}=\vacVerma{\mu}{}_{'}\,,\\
&\socohomology{p}{\irrVerma{\mu}}=0\,,&&p=2,\ldots\,.
\end{align}
\end{enumerate}
\end{Thm}
\begin{prf}
Item 1 is analogous to that of Theorem \ref{Thm_cohomology_odd}.

Let us prove item 2. For the module $\sqirrVerma{\gl}{-q}$ there
exists the BGG resolution~\cite{eastw}
  \begin{multline}
    0\longrightarrow\sqirrVerma{\gl}{-q}\longrightarrow
    \sqVerma{\gl}{-q}^\natural\longrightarrow
    \sqVerma{\gl}{-q+1}^\natural\longrightarrow\dots\longrightarrow
    \sqVerma{\gl}{-1}^\natural\longrightarrow
    \sqVerma{\gl}{0}^\natural\oplus\sqVerma{\gl}{0'}^\natural
   \longrightarrow\\
    \longrightarrow\sqVerma{\gl}{1}^\natural\longrightarrow\dots
    \longrightarrow\sqVerma{\gl}{q}^\natural\longrightarrow0
  \end{multline}
and for the modules $\sqirrVerma{\gl}{N}$ for $N=0,0',1,\dots q$
there exist the resolutions
  \begin{equation}
    0\longrightarrow\sqirrVerma{\gl}{N}\longrightarrow
    \sqVerma{\gl}{N}^\natural\longrightarrow\sqVerma{\gl}{N+1}^\natural
    \longrightarrow\dots\longrightarrow
    \sqVerma{\gl}{q}^\natural\longrightarrow0\qquad
     N=0,0',1,2,\dots q\,.
  \end{equation}
The standard spectral sequence technique
together with the definition of~$\socohomology{0}{\bullet}$ as
invariants of~$\translations(\dmn)$
allows us to calculate the
cohomology of the irreducible modules for~$N=-q,0,0',1,2,\dots,q$.
Using these we have
  \begin{align}
    \socohomology{0}{\sqirrVerma{\gl}{N}}&=\sqvacVerma{\gl}{N}
   &\mbox{\rm for $N=-q,-q+1,\dots,-1,0,0',1,2,\dots,q$}\,,
   \label{cohomol-0}\\
    \socohomology{p}{\sqirrVerma{\gl}{N}}&=\sqtvacVerma{\gl}{N+p}
   &\qquad\mbox{\rm for $p=1,\ldots,$ $N=-q,0,0',1,2,\dots,q\,.$}\label{cohomol-p}
  \end{align}
This proves~(\ref{cohomology-0}) and~(\ref{cohomology-p-2}). In
order to prove~(\ref{cohomology-p-1}) we consider the short exact
sequences contragredient
to~(\ref{even_dimensional_exact_sequence_0})--(\ref{even_dimensional_exact_sequence_q-1_irr})
  \begin{align}\label{exact-1}
    \exactsq{\sqquotVerma{\gl}{N+1}^\natural}{\sqVerma{\gl}{N}^\natural}
          {\sqtirrVerma{\gl}{N}}\,,\\
\exactsq{\sqVerma{\gl}{-N}}{\sqquotVerma{\gl}{N+1}^\natural}
          {\sqtirrVerma{\gl}{N+1}}\label{exact-2}\,,
  \end{align}
where $N=-q,-q+1,\dots,-1$ and
$\sqtirrVerma{\gl}{N}=\sqirrVerma{\gl}{N}$ for $N\neq0$
and~$\sqtirrVerma{\gl}{0}=\sqirrVerma{\gl}{0}\oplus\sqirrVerma{\gl}{0'}$.
The long cohomological exact sequence corresponding
to~(\ref{exact-1}) gives
  \begin{equation}
    \socohomology{p}{\sqquotVerma{\gl}{N+1}^\natural}=
    \socohomology{p+1}{\sqtirrVerma{\gl}{N}}.
  \end{equation}
Then substituting this into the long cohomological exact sequence
corresponding to~(\ref{exact-2})
  \begin{align}
    \dots&
    \stackrel{g_N^{p-1}}{\longrightarrow}
    \socohomology{p}{\sqtirrVerma{\gl}{N+1}}
    \longrightarrow\socohomology{p}{\sqquotVerma{\gl}{N+1}^\natural}
    \stackrel{f_N^{p}}{\longrightarrow}
    \socohomology{p}{\sqVerma{\gl}{-N}}
    \stackrel{g_N^{p}}{\longrightarrow}&\\
    &\stackrel{g_N^{p}}{\longrightarrow}
    \socohomology{p+1}{\sqtirrVerma{\gl}{N+1}}
    \longrightarrow\socohomology{p+1}{\sqquotVerma{\gl}{N+1}^\natural}
    \stackrel{f_N^{p+1}}{\longrightarrow}
    \socohomology{p+1}{\sqVerma{\gl}{-N}}
    \stackrel{g_N^{p+1}}{\longrightarrow}&\dots\nn
  \end{align}
we obtain the long exact sequence
  \begin{align}\label{main-long-1}
    \dots&
    \stackrel{g_N^{p-1}}{\longrightarrow}
    \socohomology{p}{\sqtirrVerma{\gl}{N+1}}
    \longrightarrow\socohomology{p+1}{\sqtirrVerma{\gl}{N}}
    \stackrel{f_N^{p}}{\longrightarrow}
    \socohomology{p}{\sqVerma{\gl}{-N}}
    \stackrel{g_N^{p}}{\longrightarrow}&\\
    &\stackrel{g_N^{p}}{\longrightarrow}
    \socohomology{p+1}{\sqtirrVerma{\gl}{N+1}}
    \longrightarrow\socohomology{p+2}{\sqtirrVerma{\gl}{N}}
    \stackrel{f_N^{p+1}}{\longrightarrow}
    \socohomology{p+1}{\sqVerma{\gl}{-N}}
    \stackrel{g_N^{p+1}}{\longrightarrow}&\dots\nn
  \end{align}
Using~(\ref{cohomol-p}), (\ref{cohomol-0}) and short exact
sequences~(\ref{even_dimensional_exact_sequence_q})--(\ref{even_dimensional_exact_sequence_2q})
we calculate the cohomology of the generalized Verma
modules~$\sqVerma{\gl}{N}$ for $N=0,0',1,2,\dots,q$
  \begin{align}
    \socohomology{0}{\sqVerma{\gl}{N}}=&\sqvacVerma{\gl}{N}
     \oplus\sqtvacVerma{\gl}{N+1}
   \qquad\mbox{\rm for $N=0,0',1,2,\dots,q$\,,}\\
    \socohomology{p}{\sqVerma{\gl}{N}}=&\sqtvacVerma{\gl}{N+p}
     \oplus\sqtvacVerma{\gl}{N+p+1}
   \qquad\mbox{\rm for $p=1,\ldots$, $N=0,0',1,2,\dots,q$.}
  \end{align}
Substituting this into~(\ref{main-long-1}) we have
  \begin{align}
    \dots&
    \stackrel{g_N^{p-1}}{\longrightarrow}
    \socohomology{p}{\sqtirrVerma{\gl}{N+1}}
    \longrightarrow\socohomology{p+1}{\sqtirrVerma{\gl}{N}}
    \stackrel{f_N^{p}}{\longrightarrow}
    \sqtvacVerma{\gl}{-N+p}\oplus\sqtvacVerma{\gl}{-N+p+1}
    \stackrel{g_N^{p}}{\longrightarrow}&\\
    &\stackrel{g_N^{p}}{\longrightarrow}
     \socohomology{p+1}{\sqtirrVerma{\gl}{N+1}}
    \longrightarrow\socohomology{p+2}{\sqtirrVerma{\gl}{N}}
    \stackrel{f_N^{p+1}}{\longrightarrow}
     \sqtvacVerma{\gl}{-N+p+1}\oplus\sqtvacVerma{\gl}{-N+p+2}\nn
    \stackrel{g_N^{p+1}}{\longrightarrow}&\dots
  \end{align}
whence we can obtain the following recurrent relation between
cohomology
  \begin{align}
    \socohomology{1}{\sqtirrVerma{\gl}{N}}&=
    \socohomology{0}{\sqtirrVerma{\gl}{N+1}}\oplus\Img f_N^0&
     \mbox{\rm for $N=-q,-q+1,\dots,-1$}\\[2pt]
    \socohomology{p}{\sqtirrVerma{\gl}{N}}&=
    \bigl(\socohomology{p-1}{\sqtirrVerma{\gl}{N+1}}/\Img
    g_N^{p-2}\bigr)
    \oplus\Img f_N^{p-1}& \\[-2pt]
     &&\kern-90pt\mbox{\rm for $N=-q,-q+1,\dots,-1$ and $p\geq2$.}\nn
  \end{align}
These  relations interpolate
between~$\socohomology{p}{\sqirrVerma{\gl}{-q}}$
and~$\socohomology{p}{\sqtirrVerma{\gl}{0}}$ calculated above.
This allows us to calculate
  \begin{align}
    \Img f_N^{p}&=\sqtvacVerma{\gl}{-N+p+1}\\
    \Img g_N^{p}&=\sqtvacVerma{\gl}{-N+p}.
  \end{align}
Then we have
  \begin{align}
    \socohomology{1}{\sqtirrVerma{\gl}{N}}&=
    \socohomology{0}{\sqtirrVerma{\gl}{N+1}}\oplus\sqtvacVerma{\gl}{-N+1}&
     \mbox{\rm for $N=-q,-q+1,\dots,-1$}\\[2pt]
    \socohomology{p}{\sqtirrVerma{\gl}{N}}&=
    \bigl(\socohomology{p-1}{\sqtirrVerma{\gl}{N+1}}/\sqtvacVerma{\gl}{-N+p-2}\bigr)
    \oplus\sqtvacVerma{\gl}{-N+p}& \\[-2pt]
     &&\kern-90pt\mbox{\rm for $N=-q,-q+1,\dots,-1$ and $p\geq2$.}\nn
  \end{align}
Finally these recurrent relations give~(\ref{cohomology-p-1}).

Item 3 is analogous to that of Theorem \ref{Thm_cohomology_odd}
\end{prf}
According to Sec.~\ref{Sct.:_M=2q}, items 2 and 3 in Theorems
\ref{Thm_cohomology_odd} and \ref{Thm_cohomology_even} describe
all reducible $\Verma{\gl}$.

Let us summarize  the results for
$\socohomology{0}{\irrVerma{\gl}}$ and
$\socohomology{1}{\irrVerma{\gl}}$, which are most
important for this paper:
\begin{equation}\label{cohomology_0}
\socohomology{0}{\irrVerma{\gl}}=\vacVerma{\gl}\,,
\end{equation}
\begin{align}
&\socohomology{1}{\irrVerma{\gl}}= 0&&\hbox{if
$\irrVerma{\gl}\sim\Verma{\gl}$}\,,\\
\label{cohomology_1_Imu}
&\socohomology{1}{\irrVerma{\gl}}=\vacVerma{\mu}{}_{'}&&\hbox{if
$(\gl)=(\mu)$ from (\ref{odd_dimensional_series_two}) or
(\ref{even_series_mu})}\,,\\
&\socohomology{1}{\sqirrVerma{\gl}{N}}=\sqvacVerma{\gl}{N+1}&
&\mbox{if $\dmn=2q+1$, $N=0,\ldots,2q$ and $(\gl)_N$ belongs to
(\ref{odd_dimmensional_sequence}),}\\
&&&\mbox{or if $\dmn=2q$, $N=-q,1,\ldots,q-1$ and $(\gl)_N$
belongs to (\ref{even_dimmensional_sequence})\,. }\nn
\end{align}
In addition, for $\dmn=2q$
\begin{align}
\label{cohomology_1_I0_nonsubsingular}
&\socohomology{1}{\sqirrVerma{\gl}{0}}=\socohomology{1}{\sqirrVerma{\gl}{0}{}_{'}}=
\sqvacVerma{\gl}{1}\,,\\
&\socohomology{1}{\sqirrVerma{\gl}{-1}}=\sqvacVerma{\gl}{0}\oplus\sqvacVerma{\gl}{0}{}_{'}
\oplus\sqvacVerma{\gl}{2}\,,\\
&\socohomology{1}{\sqirrVerma{\gl}{N}}=\sqvacVerma{\gl}{N+1}\oplus
\sqvacVerma{\gl}{-N+1}&&\mbox{if $(\gl)_N$ with $N=-2,\ldots,-q+1$ }\\
&&&\mbox{ belongs to (\ref{even_dimmensional_sequence})}\,.\nn
\end{align}

\begin{Rem}\label{Thm_Cohomology_Sing}\rm
For any irreducible module ${\irrVerma{\gl}}$,
$\socohomology{1}{\irrVerma{\gl}}$ is equal to the direct sum of those
singular and subsingular modules of the
generalized Verma module ${\Verma{\gl}}$,
that are not descendants of some other singular module in ${\Verma{\gl}}$.
This property is
expected  because, as one can see
from the examples in sections \ref{Sct.:_Examples of the Simplest
Conformal Systems}, \ref{Sct.:_Conformal_equations},
both $\socohomology{1}{\irrVerma{\gl}}$ and
singular and subsingular modules determine the structure of
differential equations on the dynamical fields.
\end{Rem}

\subsection{Examples of Calculating Cohomology of Reducible $\so(\dmn+2)$-Modules%
\label{Sct.:_Examples of calculating cohomology for reducible
conformal modules}} Using Theorems \ref{Thm_cohomology_odd} and
\ref{Thm_cohomology_even} one can easily calculate
$\socohomology{p}{\module}$ for any integrable module $\module$.
Let $\extmodule_{\irrmodule_1,\irrmodule_2}$ be the first extension of
the irreducible modules $\irrmodule_1$, $\irrmodule_2$ given by
the nonsplittable short exact sequence
\begin{equation}
0\longrightarrow\irrmodule_1\longrightarrow
\extmodule_{\irrmodule_1,\irrmodule_2}\longrightarrow
\irrmodule_2\longrightarrow 0\,.
\end{equation}
From the long exact sequence for cohomology
\begin{equation}\label{long}
0\longrightarrow\socohomology{0}{\irrmodule_1}\longrightarrow
\socohomology{0}{\extmodule_{\irrmodule_1,\irrmodule_2}}\longrightarrow
\socohomology{0}{\irrmodule_2}\longrightarrow\socohomology{1}{\irrmodule_1}
\longrightarrow\cdots\,,
\end{equation}
where $\socohomology{p}{\irrmodule_1}$ and
$\socohomology{p}{\irrmodule_2}$ are given by Theorems
\ref{Thm_cohomology_odd} and \ref{Thm_cohomology_even}, one
obtains $\socohomology{p}{\extmodule_{\irrmodule_1,\irrmodule_2}}$.

Using Theorem \ref{Thm_cohomology_odd} it is not hard to
see that in the case
$\dmn=2q+1$ any extension of an irreducible conformal module is
isomorphic to a contragredient generalized Verma module. This
means that any odd dimensional conformal system of equations is
either primitive or decomposes into independent primitive subsystems.
We therefore focus on the even dimensional case.

As an example, let us calculate cohomology of the module
$\extmodule_{A,F}$ which corresponds to the case of $\dmn=4$ electrodynamics
considered in section~\ref{Sct.:_M4_Electrodynamics}.
The module $\extmodule_{A,F}$ is defined by the short exact
sequence
\begin{equation}\label{E_A_F}
  0\to\irrVermaex{A}\to\extmodule_{A,F}\to\mathfrak{K}_F\to0\,,
\end{equation}
where $\irrVermaex{A}=\sqirrVerma{\gl}{-1}$ and
$\mathfrak{K}_F=\sqirrVerma{\gl}{0}\oplus\sqirrVerma{\gl}{0'}$
belong to the series (\ref{even_dimmensional_sequence}) that starts
from the dominant highest weight $(\gl)_{-2}=(0,0,0)$,
$\dmn=2q=4$. {}From Theorem \ref{Thm_cohomology_even} we obtain
the long exact cohomology sequence
\begin{multline}
  0\to\sqvacVerma{\gl}{-1}\to\socohomology{0}{\extmodule_{A,F}}\to
   \sqvacVerma{\gl}{0}\oplus\sqvacVerma{\gl}{0'}\to\\
  \to\sqvacVerma{\gl}{0}\oplus\sqvacVerma{\gl}{0'}\oplus\sqvacVerma{\gl}{2}\to
\socohomology{1}{\extmodule_{A,F}}\to\sqvacVerma{\gl}{1}\oplus\sqvacVerma{\gl}{1}
   \to\sqvacVerma{\gl}{1}\oplus\sqvacVerma{\gl}{3}\to\dots
\end{multline}
whence
\begin{equation}
  \socohomology{0}{\extmodule_{A,F}}=\sqvacVerma{\gl}{-1}\,,\qquad
  \socohomology{1}{\extmodule_{A,F}}=
  \sqvacVerma{\gl}{1}\oplus\sqvacVerma{\gl}{2}\,.
\end{equation}

As a generalization of (\ref{E_A_F}) let us consider the module
$\extmodule_{\irrmodule_{(\gl)_{-N}},
  \irrmodule_{(\gl)_{-N+1}}}$ defined by the short exact
sequence
\begin{equation}
\label{act}
0\longrightarrow\irrmodule_{(\gl)_{-N}}\longrightarrow
\extmodule_{\irrmodule_{(\gl)_{-N}},
  \irrmodule_{(\gl)_{-N+1}}}\longrightarrow
\tilde\irrmodule_{(\gl)_{-N+1}}\longrightarrow 0\,,
\end{equation}
where $(\gl)_{-N}$ and $(\gl)_{-N+1}$ with $N=1,\ldots,q-1$
belong to (\ref{even_dimmensional_sequence}),
$\tilde\irrmodule_{(\gl)_{N}}=\irrmodule_{(\gl)_{N}}$  for $N\neq 0$ and
$\tilde\irrmodule_{(\gl)_{0}}
=\irrmodule_{(\gl)_{0}}\oplus\irrmodule_{(\gl)_{0'}}$.
 Cohomology
of $\extmodule_{\irrmodule_{(\gl)_{-N}},
  \irrmodule_{(\gl)_{-N+1}}}$ is calculated from
\begin{multline}\label{qwerty}
  0\to\sqvacVerma{\gl}{-N}\to\socohomology{0}{\extmodule_{\irrmodule_{(\gl)_{-N}},
  \irrmodule_{(\gl)_{-N+1}}}}\to
   \sqtvacVerma{\gl}{-N+1}\to\sqtvacVerma{\gl}{-N+1}\oplus\sqvacVerma{\gl}{N+1}\to\\
\to\socohomology{1}{\extmodule_{\irrmodule_{(\gl)_{-N}},
  \irrmodule_{(\gl)_{-N+1}}}}\to\sqtvacVerma{\gl}{-N+2}\oplus\sqvacVerma{\gl}{N}
   \to\sqtvacVerma{\gl}{-N+2}\oplus\sqvacVerma{\gl}{N+2}\to\dots\,,
\end{multline}
where $\sqtvacVerma{\gl}{N}$ is defined in (\ref{n_s_tildoy}). {}From
(\ref{qwerty}) we have that
\begin{equation}\label{cohomology_example}
  \socohomology{0}{\extmodule_{\irrmodule_{(\gl)_{-N}},
  \irrmodule_{(\gl)_{-N+1}}}}=\sqvacVerma{\gl}{-N}\,,\qquad
  \socohomology{1}{\extmodule_{\irrmodule_{(\gl)_{-N}},
  \irrmodule_{(\gl)_{-N+1}}}}=
  \sqvacVerma{\gl}{N}\oplus\sqvacVerma{\gl}{N+1}\,.
\end{equation}
Equations corresponding to
$\extmodule_{\irrmodule_{(\gl)_{-N}}, \irrmodule_{(\gl)_{-N+1}}}$
are considered for $N=1$ in subsection
\ref{Sct.:_Conformal_higher_spins_in_even_dimensions} and for
$N=q-1$ in subsection
\ref{Sct.:_F_T_conformal_higher_spins_in_even_dimensions}.
An important general property of the dynamical systems associated with
the module
$\extmodule_{\irrmodule_{(\gl)_{-N}}, \irrmodule_{(\gl)_{-N+1}}}$
in (\ref{act}) is that the Lorentz algebra representations of the
dynamical fields and dynamical equations are isomorphic while the
sum of their conformal dimensions is $2q$ which is the canonical
dimension of a Lagrangian density.
We therefore expect that all these dynamical systems are Lagrangian.

\subsection{Conformal Equations\label{Sct.:_Conformal_equations}}
\label{Sct.:_More_examples_of_conformal_equations} Now it is
straightforward to write down conformal equations
$\equations_\module\subbundsec{0}=0$ corresponding to any
conformal module $\module$. First,  one represents $\module$ as
an extension of irreducible conformal modules. Then (as explained
in section \ref{Sct.:_Examples of calculating cohomology for
reducible conformal modules}) the results of Theorem
\ref{Thm_cohomology_odd} (for odd $\dmn$) and Theorem
\ref{Thm_cohomology_even} (for even $\dmn$) are used to calculate
$\socohomology{0}{\module}$ and $\socohomology{1}{\module}$.
Finally, along the lines of the proof of Theorem \ref{Theorem_1}
one expresses auxiliary fields contained in $\bundsec{0}$ (see
Remark \ref{Thm.:_Remark_3_4}) in terms of derivatives of the
dynamical field $\subbundsec{0}$ and reconstructs the nontrivial
equations $\equations_\module\subbundsec{0}=0$ on the latter.
These equations are associated with $\socohomology{1}{\module}$.
In practice, it is most useful to use Remark
\ref{Thm_Cohomology_Sing}, which identifies the left hand sides
of the field equations with the singular and subsingular modules
of $\Verma{\gl}$. In those cases where $\Verma{\gl}$ does not
contain modules of the Levi factor $\Levi$ equivalent to (but
different from) the singular and subsingular modules, the
explicit form of conformal equations corresponding to the
irreducible conformal module $\irrVerma{\gl}$ can be obtained  by
replacing $\cK_n$ by $\d{x^n}$ in the expressions for a basis of
the singular and subsingular modules.

The examples given in section \ref{Sct.:_Examples of the Simplest
Conformal Systems} and  in the rest of this section result from
the application of this general scheme to the following modules
(here $\irrmodule$ denotes an irreducible module and $\extmodule$
denotes an extension).
\begin{enumerate}
\item $\irrVermaex{((2-\dmn)/2,0,\ldots,0)}$ corresponds
to Klein--Gordon equation (\ref{Klein_Gordon_equation})
(primitive)
\item $\irrVermaex{((1-\dmn)/2,1/2,\ldots,1/2)}$ corresponds
to Dirac equation (\ref{Dirac_equation}) (primitive)
\item
$\irrVermaex{(-p,\underbrace{\scriptstyle
1,\ldots,1}_p,0,\ldots,0)}$ for odd $\dmn$ or for even $\dmn$ and
$p=0$ corresponds to closedness equation (\ref{closery_condition})
on a $p$-form or equivalent conservation equation
(\ref{conservation_condition}) on a $(\dmn-p)$-polyvector
(primitive);\\
for even $\dmn$ and $p>0$, $\irrVermaex{(-p,\underbrace{\scriptstyle
1,\ldots,1}_p,0,\ldots,0)}$
corresponds to the  system
(\ref{closery_condition}), (\ref{subsingular_forms_closeness}) on
a $p$-form or the equivalent system (\ref{conservation_condition}),
(\ref{subsingular_forms_conservation}) on a $(\dmn-p)$-polyvector
(primitive)
\item $\irrVermaex{(p-\dmn,\underbrace{\scriptstyle
1,\ldots,1}_p,0,\ldots,0)}$ $p>0$ corresponds to conservation
equation (\ref{conservation_condition}) on a $p$-polyvector or the
equivalent closedness equation (\ref{closery_condition}) on a
$(\dmn-p)$-form (primitive)
\item $\irrVermaex{(\dmn/2,1,\ldots,1,\pm 1)}$ for even $\dmn$
corresponds to (anti)selfduality equation
(\ref{antiselfduality_equation_1}),
(\ref{antiselfduality_equation_2}) (primitive)
\item $\irrVermaex{(-2,1,1)}\oplus \irrVermaex{(-2,1,-1)}$
corresponds to the field strength form of Maxwell equations
(\ref{Maxwell_1}), (\ref{Maxwell_2}) (non-primitive)
\item $\extmodule_{A,F}$ corresponds to the potential form of
Maxwell equations (\ref{Maxwell_1}), (\ref{Maxwell_2}),
(\ref{Maxwell_potential}) in conformal gauge (\ref{Maxwell_gauge})
(non-primitive)
\item $\extmodule_{A,F,J}$ corresponds to the off-mass-shell version
of Maxwell electrodynamics (\ref{Maxwell_1}),
(\ref{Maxwell_potential}), (\ref{Maxwell_current}),
(\ref{Maxwell_current_conservation}) in conformal gauge
(\ref{Maxwell_gauge}) (non-primitive)
\item $\extmodule_{A,F,J,G}$ corresponds to the off-mass-shell gauge
invariant version of Maxwell electrodynamics (\ref{Maxwell_1}),
(\ref{Maxwell_potential}), (\ref{Maxwell_current}),
(\ref{Maxwell_current_conservation}),
(\ref{Maxwell_gauge_invariant}) (non-primitive)
\item \label{flat_limits}$\irrVermaex{((2-\dmn)/2,\underbrace{\scriptstyle
\gl,\ldots,\gl}_\nu,0,\ldots,0)}$ for odd $\dmn$ or for even
$\dmn$ with either $\nu\leq q-2$ or $\nu=q$, $\gl=1$ corresponds
to Klein--Gordon--like equation
(\ref{Klein_Gordon_equation_for_blok}) on a tensor field
described by the $\gl\times\nu$-rectangular Young tableau
(primitive)
\item $\irrVermaex{((1-\dmn)/2,\underbrace{\scriptstyle
\gl+1/2,\ldots,\gl+1/2}_\nu,1/2,\ldots,\pm1/2)}$ for odd $\dmn$
or for even $\dmn$ with $\nu\leq q-1$ corresponds to Dirac--like
equation (\ref{Dirac_equation_for_block}) on a spinor--tensor
field described by the $\gl\times\nu$-rectangular Young tableau
(primitive)
\item $\mathfrak{K}_{(\gl)_F}=\irrVermaex{(\gl)_+}\oplus\irrVermaex{(\gl)_-}$ for even $\dmn$
corresponds to the field strength form of conformal higher spin
equations (\ref{generalized_electrodynamic_1}),
(\ref{generalized_electrodynamic_2}) (non-primitive)
\item $\extmodule_{\irrVermaex{(\gl)_A},\mathfrak{K}_{(\gl)_F}}$
for even $\dmn$ corresponds to the gauge fixed potential form of
conformal higher spin equations
(\ref{generalized_electrodynamic_1}),
(\ref{generalized_electrodynamic_2}),
(\ref{generalized_potetial}), (\ref{generalized_gauge})
(non-primitive)
\item $\extmodule_{\irrVermaex{(\gl)_A},\mathfrak{K}_{(\gl)_F},
\irrVermaex{(\gl)_J}}$ for even $\dmn$ corresponds to the gauge
fixed off-mass-shell version of conformal higher spin equations
(\ref{generalized_electrodynamic_1}),
(\ref{generalized_potetial}), (\ref{generalized_gauge}),
(\ref{generalized_electrodynamic_2_current}),
(\ref{generalized_current}) (non-primitive)
\item $\extmodule_{\irrVermaex{(\gl)_A},\mathfrak{K}_{(\gl)_F},
\irrVermaex{(\gl)_J},\irrVermaex{(\gl)_{G}}}$ for even $\dmn$
corresponds to the gauge invariant off-mass-shell version of
conformal higher spin equations
(\ref{generalized_electrodynamic_1}),
(\ref{generalized_potetial}),
(\ref{generalized_electrodynamic_2_current}),
(\ref{generalized_current}), (\ref{generalized_gauge_inv_1}),
(\ref{generalized_gauge_1}) (non-primitive)
\item
$\irrVermaex{C}$ for even $\dmn$ corresponds to the condition that
the generalized Weyl tensor for spin $\gl\geq 1/2$ symmetric
tensor field equals to zero
(\ref{F_T_S_zero_Weyl_tensor_equation}) supplemented with the
gauge fixing condition (\ref{F_T_S_gauge_fixing}) (primitive)
\item $\extmodule_{\irrVermaex{C},\irrVermaex{W}}$ for even $\dmn$
corresponds to gauge fixed spin $\gl\geq 1/2$ Fradkin--Tseytlin
conformal higher spin equation (\ref{F_T_S_gauge_fixing}),
(\ref{F_T_S_Weyl_tensor_definition}),
(\ref{F_T_S_Weyl_tensor_equation}) (non-primitive)
\item $\extmodule_{\irrVermaex{C},\irrVermaex{W},
\irrVermaex{G}}$ for even $\dmn$ corresponds to gauge invariant
spin $\gl\geq 1/2$ Fradkin--Tseytlin conformal higher spin
equation (\ref{F_T_S_Weyl_tensor_definition}),
(\ref{F_T_S_Weyl_tensor_equation}) (\ref{F_T_S_gauge_invariant})
(non-primitive)
\end{enumerate}

Note that flat limits of the most non-flat conformal equations
considered in \cite{ER},\cite{nfl_1}-\cite{pnfl_3} belong to the
case 10. The system of conformal equations considered in
\cite{P_S_T} corresponds to the case 8.

\subsubsection{Conformal Klein--Gordon and Dirac--like
equations for a block%
\label{Sct.:_Conformal_Klein-Gordon_and_Dirac_-like_equations_for_a_block}}
Let
$(\gl)=(-(\dmn-2)/2,\underbrace{\gl,\ldots,\gl}_{\nu},0,\ldots,0)$,
$\gl\in\oN$, and $\irrVerma{\gl}$ be the irreducible conformal
module with the highest weight $(\gl)$. It is represented by the
short exact sequence~(\ref{odd_dimensional_exect_sequence_mu}) for
odd $\dmn$ and by~(\ref{even_dimensional_exact_sequence_mu}) for
even $\dmn$. Let us consider the bundle
$\bundle_{(\gl)}=\Minkowski\times\irrVerma{\gl}$ and its subbundle
$\bundle_{(\gl)}\supset\subbundle_{(\gl)}
=\Minkowski\times\vacVerma{\gl}$. Consider a section
\begin{equation}
\subbundsec{}=C_{n^1(\gl),n^2(\gl),\ldots,n^\nu(\gl)}(x)
\ketv{(\gl)}^{n^1(\gl),n^2(\gl),\ldots,n^\nu(\gl)}
\end{equation}
of $\subbundle_{(\gl)}$ and a section $\bundsec{}$ of
 $\bundle_{(\gl)}$ such that,
$\bundsec{}\Big|_{\subbundle_{(\gl)}}=\subbundsec{}$,
\begin{equation}
\bundsec{}=\sum_{l=0}\frac{1}{l!}C_{n^1(\gl),n^2(\gl),\ldots,
n^\nu(\gl);m(l)}(x)\underbrace{y^{(m}\cdots y^{m)}}_{l}
\ketv{(\gl)}^{n^1(\gl),n^2(\gl),\ldots,n^\nu(\gl)}\,.
\end{equation}
Here $\ketv{(\gl)}^{n^1(\gl),n^2(\gl),\ldots,n^\nu(\gl)}$ form a
basis of $\vacVerma{\gl}$. The symmetry properties of
$\ketv{(\gl)}^{n^1(\gl),n^2(\gl),\ldots,n^\nu(\gl)}$ imply that
symmetrization over any $\lambda +1$ indices gives zero. The
corresponding Young tableau is a rectangle of length $\lambda$
and height $\nu$ and is referred to as a block. Note that fields
that appear in most of physical applications belong to this class.

As shown in section \ref{Sct.:_General_Construction} the covariant
constancy equation (\ref{Con.:_lift}) encodes the differential
equations on the dynamical variables that take values in
$\socohomology{0}{\irrVerma{\gl}}$. The form of these
differential equations is determined by
$\socohomology{1}{\irrVerma{\gl}}$. These cohomology groups are
determined in (\ref{cohomology_0}) and (\ref{cohomology_1_Imu}).
Using the symmetry properties of the block Young tableau it can be
easily seen that $\socohomology{1}{\irrVerma{\gl}}$ corresponds
to the singular module $\singmodule{\gl}$ of $\Verma{\gl}$
described by the block tableau with the conformal weight
$\dmn/2+1$, i.e. it has the weights
$(\gl)=(-(\dmn+2)/2,\underbrace{\gl,\ldots,\gl}_{\nu},0,\ldots,0)$.
It is easy to see that $\ketv{s}\in \singmodule{\gl}$ has the form
\begin{equation}\label{K_G_singular_vector}
\ketv{s}=\psi^{n^1(\gl),\ldots,n^\nu(\gl)} \Big (y^m
y_m\delta_{n^\nu}^k- \frac{4\gl\nu}{2\gl-2\nu+\dmn} y_{n^\nu}
y^k\Big) \ketv{\lambda}_{n^1(\gl),\ldots,n^\nu(\gl-1)k}\,,
\end{equation}
where $\psi^{n^1(\gl),\ldots,n^\nu(\gl)}$ is an arbitrary
parameter taking values in the $\gl\times\nu$ traceless block
tableau. In fact, $\psi^{n^1(\gl),\ldots,n^\nu(\gl)}$ can be
thought of as an arbitrary element of the dual space of
$\socohomology{1}{\irrVerma{\gl}}$. The conformal equation
associated with $\socohomology{1}{\irrVerma{\gl}}$ is
\begin{equation}\label{Klein_Gordon_equation_for_blok}
\psi^{n^1(\gl),\ldots,n^\nu(\gl)} \Big (\square
C_{n^1(\gl),\ldots,n^\nu(\gl)}(x)- \frac{4\gl\nu}{2\gl-2\nu+\dmn}
\ptl_{n^\nu} \ptl^m C_{n^1(\gl),\ldots,n^\nu(\gl-1)m}(x)\Big
)=0\,.
\end{equation}
This is the Klein--Gordon type conformal equation for a field with
the block symmetry properties and conformal weight $\dmn/2-1$.

Note that for even $\dmn$, (\ref{even_series_mu}) requires either
$\nu\leq q-2$ or $\nu=q$, $\gl=1$. This is in accordance with our
analysis because, although being conformally invariant, the
equations (\ref{Klein_Gordon_equation_for_blok}) with $\nu=q-1$,
$\dmn=2q$ are non-primitive (see section
\ref{Sct.:_Conformal_higher_spins_in_even_dimensions}). Also one
can see for even $\dmn$ and $\nu=q$, $\gl\geq2$ that the singular
vector (\ref{K_G_singular_vector}) is zero\footnote{One way to
see this is to observe that for the case of $\nu=q$ the tensor
contracted with $\psi^{n^1(\gl),\ldots,n^q(\gl)}$ on the left
hand side of (\ref{K_G_singular_vector}) has opposite
(anti)selfduality properties for the first and last columns of
the corresponding rectangular Young tableau, that is only
possible when it is zero.} and equation
(\ref{Klein_Gordon_equation_for_blok}) becomes the identity $0=0$.

For the particular cases of $\nu=1$, $\lambda=0,1,2$ equation
(\ref{Klein_Gordon_equation_for_blok}) reads
\begin{align}
\label{Klein_Gordon_equation_s0}
&\square C(x)=0\,,\\
\label{Klein_Gordon_equation_s1} &\psi^n\Big(\square
C_n(x)-\frac{4}{\dmn}\ptl_n\ptl^m C_m(x)\Big)=0\,,
\\
\label{Klein_Gordon_equation_s2} &\psi^{n_1n_2}\Big(\square C_{n_1
n_2}(x)-\frac{8}{2+\dmn}\ptl_{n_1}\ptl^m C_{n_2m}(x)\Big)=0\,.
\end{align}
Equation (\ref{Klein_Gordon_equation_s0}) is the usual
Klein--Gordon equation. Equation (\ref{Klein_Gordon_equation_s1})
for $\dmn=4$ corresponds to Maxwell electrodynamics formulated in
terms of potential. Equation (\ref{Klein_Gordon_equation_s1}) for
$\dmn\neq 4$ and equation (\ref{Klein_Gordon_equation_s2})
correspond to non-unitary field-theoretical models.

The Dirac--like equations are associated with the bundles
$\subbundle_{(\gl)}$ and $\bundle_{(\gl)}$ with
$$(\gl)=(-(\dmn-1)/2,\underbrace{\gl+\half,\ldots,\gl+\half}_{
\nu},\half,\ldots,\pm\half),\qquad\gl\in\oN$$ and their sections
\begin{equation}
\subbundsec{}=C_{n^1(\gl),n^2(\gl),\ldots,n^\nu(\gl),\ga}(x)
\ketv{(\gl)}^{n^1(\gl),n^2(\gl),\ldots,n^\nu(\gl),\ga}
\end{equation}
and
\begin{equation}
\bundsec{}=\sum_{l=0}\frac{1}{l!}C_{n^1(\gl),n^2(\gl),\ldots,
n^\nu(\gl),\ga;m(l)}(x)\underbrace{y^{(m}\cdots y^{m)}}_l
\ketv{(\gl)}^{n^1(\gl),n^2(\gl),\ldots,n^\nu(\gl),\ga}\,,
\end{equation}
where $\bundsec{}\Big|_{\subbundle}=\subbundsec{}$. Here
$\ga=1,\ldots, 2^{\left[ \frac{\dmn}{2} \right]}$ is a spinorial
index. $C_{n^1(\gl),\ldots,n^\nu(\gl),\ga}(x)$ is a
$\gamma$-transversal block spinor--tensor with definite
chirality. The cohomology groups
$\socohomology{0}{\irrVerma{\gl}}$ and
$\socohomology{1}{\irrVerma{\gl}}$ are given in
(\ref{cohomology_0}) and (\ref{cohomology_1_Imu}), respectively.
$\socohomology{1}{\irrVerma{\gl}}$ corresponds to the singular
module $\singmodule{\gl}$ in~$\Verma{\gl}$ with the general
element
\begin{equation}\label{D_singular_vector}
\ketv{s} =\psi^{n^1(\gl),\ldots,n^\nu(\gl),}{}_\ga \Big (y^m
\gga_m{}^\ga{}_\gb \delta_{n^\nu}^k-\frac{2\gl\nu}{2\gl-2\nu+\dmn}
\gga_{n^\nu}{}^\ga{}_\gb y^k \Big
)\ketv{\lambda}_{n^1(\gl),\ldots,n^\nu(\gl-1)k,}{}^\gb\,.
\end{equation}
Here $\psi^{n^1(\gl),\ldots,n^\nu(\gl),}{}_\ga$ is an arbitrary
$\gamma$-transversal chiral spinor--tensor parameter taking values
in the $\gl\times\nu$-block tableau. The conformal equation
encoded by the covariant constancy equation~(\ref{Con.:_lift}) is
\begin{equation}\label{Dirac_equation_for_block}
\psi^{n^1(\gl),\ldots,n^\nu(\gl),}{}_\ga \Big (\ptl^m
\gga_m{}^\ga{}_\gb C_{n^1(\gl),\ldots,n^\nu(\gl),}{}^\gb(x)
{}-\frac{2\gl\nu}{2\gl-2\nu+\dmn} \gga_{n^\nu}{}^\ga{}_\gb \ptl^m
C_{n^1(\gl),\ldots,n^\nu(\gl-1)m,}{}^\gb(x)\Big )=0.
\end{equation}
This is the conformally invariant generalization of the Dirac
equation to a block spinor--tensor with conformal weight
$(\dmn-1)/2$. For the particular cases of $\nu=1$, $\lambda=0,1$
we get
\begin{align}
\label{Dirac_equation_s1/2} &\ptl^m \gamma_m{}^\alpha{}_\beta
C_,{}^\beta(x)=0
\,,\\
\label{Dirac_equation_s3/2} &\psi^{n,}{}_\ga\Big(\ptl^m
\gamma_m{}^\alpha{}_\beta C_{n,}{}^\beta(x)-
\frac{2}{\dmn}\gga_n{}^\ga{}_\gb \ptl^m C_{m,}{}^\gb(x)\Big)=0\,.
\end{align}
Equation (\ref{Dirac_equation_s1/2}) is the usual Dirac equation.
Note that conditions (\ref{even_series_mu}) require $\nu\leq q-1$
for even $\dmn$. Analogously to the case of Klein--Gordon type
equations one can prove that singular vector
(\ref{D_singular_vector}) is zero for even $\dmn$, $\nu=q$, and
corresponding equation (\ref{Dirac_equation_for_block}) becomes
identity $0=0$.

Analogous conformally invariant generalizations of the
Klein--Gordon and Dirac equations exist for tensor fields of
other symmetry types. They correspond to other irreducible
modules $\irrVerma{\gl}$ from the
series~(\ref{odd_dimensional_exect_sequence_mu}) for odd $\dmn$
and~(\ref{even_dimensional_exact_sequence_mu}) for even $\dmn$.
All these systems however are not expected to correspond to
unitary field-theoretical models in accordance with the general
fact \cite{sieg,metsaev,FF} that conformal field equations
compatible with unitarity
are exhausted by the massless equations
for a scalar, a spinor and blocks of the height $[(\dmn-1)/2]$.

\subsubsection{Conformal higher spins in even
dimensions\label{Sct.:_Conformal_higher_spins_in_even_dimensions}}

Here we describe a generalization of the equations for $\dmn=4$
massless higher spin fields to a broad class of conformal field
equations for tensor fields in $\dmn=2q$ dimensions (the
following construction can be easily formulated also for
spinor--tensor fields). Let
$(\gl)_\pm=(-q,\lambda_1,\lambda_2,\dots,\lambda_{q-1},\pm1)$,
where $\gl_i\in\oN$ and
$\gl_1\geq\gl_2\geq\cdots\geq\gl_{q-1}\geq1$. Let $q=\mu_1
>\mu_2\geq\mu_3\geq\cdots\geq\mu_p$ be the heights of the columns
in the Young tableau corresponding to $\vacVerma{\gl}{}_{{}_\pm}$.
(Note that the first column is required to have the maximal height
$q$, while the second one is required to be smaller.) Let us
denote
$\mathfrak{K}_{(\gl)_F}=\irrVermaex{(\gl)_+}\oplus\irrVermaex{(\gl)_-}$.
Consider the bundle
$\bundle_F=\Minkowski\times\mathfrak{K}_{(\gl)_F}$ and its
subbundle
$\subbundle_F=\Minkowski\times(\vacVerma{\gl}{}_{{}_+}\oplus
\vacVerma{\gl}{}_{{}_-})$. Irreducible modules
$\irrVerma{\gl}{}_{{}_+}$ and $\irrVerma{\gl}{}_{{}_-}$ are
defined by the short exact sequences
(\ref{even_dimensional_exact_sequence_q}) and
(\ref{even_dimensional_exact_sequence_q'}), respectively. Choose a
section of $\subbundle_F$
\begin{equation}
\ketv{\phi_F(x)}=F_{n^1[q],n^2[\mu_2],\ldots,n^p[\mu_p]}(x)
\ket{(\gl)_F}^{n^1[q],n^2[\mu_2],\ldots,n^p[\mu_p]}\,,
\end{equation}
where $\ket{(\gl)_F}^{n^1[q],n^2[\mu_2],\ldots,n^p[\mu_p]}$ is a
basis in $\vacVerma{\gl}{}_{{}_+}\oplus\vacVerma{\gl}{}_{{}_+}$,
i.e. it contains both selfdual and antiselfdual parts. We treat
$\ketv{\phi_F(x)}$ as a higher spin field strength. Let
$\ketv{\Phi_F(x)}$ be a section of $\bundle_F$ such that
$\ketv{\Phi_F(x)}\big|_{\subbundle_F}=\ketv{\phi_F(x)}$
\begin{equation}
\ketv{\Phi_F(x)}=\sum_{l=0}\frac{1}{l!}F_{n^1[q],n^2[\mu_2],\ldots,
n^p[\mu_p];m(l)}(x)\underbrace{y^{(m}\cdots y^{m)}}_l
\ket{(\gl)_F}^{n^1[q],n^2[\mu_2],\ldots,n^p[\mu_p]}\,.
\end{equation}
As follows from (\ref{cohomology_0}) and
(\ref{cohomology_1_I0_nonsubsingular}), the condition
\begin{equation}
\diff \ketv{\Phi_F(x)}=0
\end{equation}
implies the equations
\begin{align}
\label{generalized_electrodynamic_1}
\psi^{n^1[q-1],n^2[\mu_2],\ldots,n^p[\mu_p]}
\ptl^m({}^*F)_{mn^1[q-1],n^2[\mu_2],\dots,n^p[\mu_p]}(x)&=0\,,\\
\label{generalized_electrodynamic_2}
\psi^{n^1[q-1],n^2[\mu_2],\ldots,n^p[\mu_p]}
\ptl^mF_{mn^1[q-1],n^2[\mu_2],\dots,n^p[\mu_p]}(x)&=0\,,
\end{align}
where an arbitrary element
$\psi^{n^1[q-1],n^2[\mu_2],\ldots,n^p[\mu_p]}$ of the irreducible
$\so(M)$-module associated with the Young tableau with columns of
heights $q-1,\mu_2, \ldots, \mu_p$ is introduced to avoid
complicated projection operators. For the particular case of the
block with $\mu_2 = \mu_3 =\ldots = q-1$ these are equations of
motion (formulated in terms of field strengths)
for the conformal fields that respect unitarity
\cite{sieg,metsaev,FF}. For $q=2$ one recovers the usual
equations of motion for massless fields in four dimensions
formulated in terms of field strengths. For~$q=3$ the conformal
massless higher spins of this type were discussed in~\cite{Hull}.

The system (\ref{generalized_electrodynamic_1}),
(\ref{generalized_electrodynamic_2}) admits extensions analogous
to that of the system (\ref{Maxwell_1}), (\ref{Maxwell_2}). In
particular, one can introduce potentials to the field strength
$F_{n^1[q],n^2[\mu_2],\dots,n^p[\mu_p]}(x)$ in both gauge
invariant and conformal gauge fixed forms. To this end we consider
the nontrivial extension
$\extmodule_{\irrmodule_{(\gl)_A},\mathfrak{K}_{(\gl)_F}}$ of the
module $\mathfrak{K}_{(\gl)_F}$ by the module
$\irrVerma{\gl}{}_{{}_A}$ where
$(\gl)_A=(-q+1,\gl_1,\ldots,\gl_{q-1},0)$.
$\extmodule_{\irrmodule_{(\gl)_A},\mathfrak{K}_{(\gl)_F}}$ is
defined by the short exact sequence
\begin{equation}\label{ext_A_F}
0\longrightarrow\irrmodule_{(\gl)_A}\longrightarrow
\extmodule_{\irrmodule_{(\gl)_A},\mathfrak{K}_{(\gl)_F}}
\longrightarrow\mathfrak{K}_{(\gl)_F} \longrightarrow0\,.
\end{equation}
The module
$\extmodule_{\irrmodule_{(\gl)_A},\mathfrak{K}_{(\gl)_F}}$ can be
described as follows. Let
$\ket{(\gl)_A}^{n^1[q-1],n^2[\mu_2],\dots,n^p[\mu_p]}$ be the
basis in $\vacVerma{\gl}{}_{{}_A}$. Impose the following relations
\begin{align}
\label{relation_IF} &\psi_{n^1[q-1],n^2[\mu_2],\ldots,n^p[\mu_p]}
y_{m}\ket{(\gl)_F}^{mn^1[q-1],n^2[\mu_2],\dots,n^p[\mu_p]}=0\,,\\
&\psi_{n^1[q-1],n^2[\mu_2],\ldots,n^p[\mu_p]}
y_{m}({}^*\ket{(\gl)_F}^{mn^1[q-1],n^2[\mu_2],\dots,n^p[\mu_p]}=0\,,
\nonumber\\
\label{relation_IA_1} &\psi_{n^1[q],n^2[\mu_2],\ldots,n^p[\mu_p]}
y^{n^1}\ket{(\gl)_A}^{n^1[q-1],n^2[\mu_2],\dots,n^p[\mu_p]}=0\,,\\
\label{relation_IA_2}
&\psi_{n^1[q-2],n^2[\mu_2-1],\ldots,n^{\gl_{q-1}}[\mu_{\gl_{q-1}}-1],
n^{\gl_{q-1}+1}[\mu_{\gl_{q-1}+1}],\ldots,n^p[\mu_p]}y^n y_n
y_{n^1}\cdots y_{n^{\gl_{q-1}}}\\
&\kern220pt\ket{(\gl)_A}^{n^1[q-1],n^2[\mu_2],\dots,n^p[\mu_p]}=0\,,
\end{align}
which single out the modules $\mathfrak{K}_{(\gl)_F}$ and
$\irrmodule_{(\gl)_A}$, respectively. The nontrivial extension is
defined by the condition
\begin{equation}\label{relation_ext_IA_IF}
\psi_{n^1[q],n^2[\mu_2],\ldots,n^p[\mu_p]}
\cP^m\ketv{(\gl)_F}^{n^1[q],n^2[\mu_2],\ldots,n^p[\mu_p]}=
-\psi_{n^1[q],n^2[\mu_2],\ldots,n^p[\mu_p]}
\eta^{mn^1}\ketv{(\gl)_A}^{n^1[q-1],n^2[\mu_2],\ldots,n^p[\mu_p]}\,.
\end{equation}
The module
$\extmodule_{\irrmodule_{(\gl)_A},\mathfrak{K}_{(\gl)_F}}$ is
generated by $y^n$ from
$\ket{(\gl)_F}^{n^1[q],n^2[\mu_2],\dots,n^p[\mu_p]}$ and
$\ket{(\gl)_A}^{n^1[q-1],n^2[\mu_2],\dots,n^p[\mu_p]}$.

Consider the bundle $\bundle_{A,F}=\Minkowski\times
\extmodule_{\irrmodule_{(\gl)_A},\mathfrak{K}_{(\gl)_F}}$.
$\subbundle_F$ and
$\subbundle_A=\Minkowski\times\vacVerma{\gl}{}_{{}_A}$ are its
subbundles.  Consider a section $\ketv{\Phi_{A,F}(x)}$ of
$\bundle_{A,F}$
\begin{multline}
\ketv{\Phi_{A,F}(x)}=\sum_{l=0}\frac{1}{l!}F_{n^1[q],n^2[\mu_2],\ldots,
n^p[\mu_p];m(l)}(x)\underbrace{y^{(m}\cdots y^{m)}}_l
\ketv{(\gl)_F}^{n^1[q],n^2[\mu_2],\ldots,n^p[\mu_p]}\\
+\sum_{l=0}\frac{1}{l!}A_{n^1[q-1],n^2[\mu_2],\ldots,n^p[\mu_p];m(l)}(x)
\underbrace{y^{(m}\cdots y^{m)}}_l
\ketv{(\gl)_A}^{n^1[q-1],n^2[\mu_2],\ldots,n^p[\mu_p]}\,.
\end{multline}
Cohomology
$\socohomology{0}{\extmodule_{\irrmodule_{(\gl)_A},\mathfrak{K}_{(\gl)_F}}}$,
$\socohomology{1}{\extmodule_{\irrmodule_{(\gl)_A},\mathfrak{K}_{(\gl)_F}}}$
is given in (\ref{cohomology_example}) for $N=1$. Condition
$\diff\ketv{\Phi_{F,A}(x)}=0$ implies
\begin{equation}\label{generalized_potetial}
\psi^{mn^1[q-1],n^2[\mu_2],\ldots,n^p[\mu_p]} \Big(\ptl_m
A_{n^1[q-1],n^2[\mu_2],\dots,n^p[\mu_p]}(x)
-F_{mn^1[q-1],n^2[\mu_2],\dots,n^p[\mu_p]}(x) \Big)=0\,,
\end{equation}
\begin{equation}\label{generalized_gauge}
\psi^{n^1[q-2],n^2[\mu_2-1],\ldots,n^{\gl_{q-1}}[\mu_{\gl_q-1}-1],
n^{\gl_{q-1}+1}[\mu_{\gl_{q-1}+1}],\ldots,n^p[\mu_p]}\square
\ptl^{n^1}\cdots \ptl^{n^{\gl_{q-1}}}
A_{n^1[q-1],n^2[\mu_2],\dots,n^p[\mu_p]}(x)=0
\end{equation}
together with (\ref{generalized_electrodynamic_1}) and
(\ref{generalized_electrodynamic_2}). This extension introduces
gauge potentials $A_{n^1[q-1],n^2[\mu_2],\ldots,n^p[\mu_p]}(x)$
to the field strength, along  with the conformally invariant gauge
condition (\ref{generalized_gauge}).

Now we introduce the module
$\extmodule_{\irrmodule_{(\gl)_A},\mathfrak{K}_{(\gl)_F},
\irrmodule_{(\gl)_J}}$ that extends
$\extmodule_{\irrmodule_{(\gl)_A},\mathfrak{K}_{(\gl)_F}}$ by the
module $\irrmodule_{(\gl)_J}$, where $(\gl)_J=(-q-1,\gl_1,\ldots
,\gl_{q-1},0)$.
$\extmodule_{\irrmodule_{(\gl)_A},\mathfrak{K}_{(\gl)_F},
\irrmodule_{(\gl)_J}}$ is described by the short exact sequence
\begin{equation}\label{ext_J_A_F}
0\longrightarrow\extmodule_{\irrmodule_{(\gl)_A},
\mathfrak{K}_{(\gl)_F}}\longrightarrow
\extmodule_{\irrmodule_{(\gl)_A},\mathfrak{K}_{(\gl)_F},
\irrmodule_{(\gl)_J}}
\longrightarrow\irrmodule_{(\gl)_J}\longrightarrow0\,.
\end{equation}
The module
$\extmodule_{\irrmodule_{(\gl)_A},\mathfrak{K}_{(\gl)_F},
\irrmodule_{(\gl)_J}}$ is generated by $y^n$ from
$\ket{(\gl)_F}^{n^1[q],n^2[\mu_2],\dots,n^p[\mu_p]}$,
$\ket{(\gl)_A}^{n^1[q-1],n^2[\mu_2],\dots,n^p[\mu_p]}$ and
$\ket{(\gl)_J}^{n^1[q-1],n^2[\mu_2],\dots,n^p[\mu_p]}$ satisfying
conditions (\ref{relation_IF})--(\ref{relation_ext_IA_IF}) along
with
\begin{equation}\label{relation_IJ}
\psi_{n^1[q-2],n^2[\mu_2-1],\ldots,n^{\gl_{q-1}}[\mu_{\gl_{q-1}}-1],
n^{\gl_{q-1}+1}[\mu_{\gl_{q-1}+1}],\ldots,n^p[\mu_p]}
y_{n^1}\cdots y_{n^{\gl_{q-1}}}
\ket{(\gl)_J}^{n^1[q-1],n^2[\mu_2],\dots,n^p[\mu_p]}=0
\end{equation}
and
\begin{equation}\label{relation_ext_IJ_IA_IF}
\psi_{mn^1[q-1],n^2[\mu_2],\ldots,n^p[\mu_p]}
\cP^{m}\ketv{(\gl)_J}^{n^1[q-1],n^2[\mu_2],
\ldots,n^p[\mu_p]}=-\frac{q}{3}
\psi_{mn^1[q-1],n^2[\mu_2],\ldots,n^p[\mu_p]}
\ketv{(\gl)_F}^{mn^1[q-1],n^2[\mu_2],\ldots,n^p[\mu_p]}\,.
\end{equation}
Consider a section $\ketv{\Phi_{A,F,J}(x)}$ of the bundle
$\Minkowski\times\extmodule_{\irrmodule_{(\gl)_A},\mathfrak{K}_{(\gl)_F},
\irrmodule_{(\gl)_J}}$
\begin{multline}
\ketv{\Phi_{A,F,J}(x)}=\sum_{l=0}\frac{1}{l!}F_{n^1[q],n^2[\mu_2],\ldots,
n^p[\mu_p];m(l)}(x)\underbrace{y^{(m}\cdots y^{m)}}_l
\ketv{(\gl)_F}^{n^1[q],n^2[\mu_2],\ldots,n^p[\mu_p]}\\
+\sum_{l=0}\frac{1}{l!}A_{n^1[q-1],n^2[\mu_2],\ldots,n^p[\mu_p];m(l)}(x)
\underbrace{y^{(m}\cdots y^{m)}}_l
\ketv{(\gl)_A}^{n^1[q-1],n^2[\mu_2],\ldots,n^p[\mu_p]}\\
+\sum_{l=0}\frac{1}{l!}J_{n^1[q-1],n^2[\mu_2],\ldots,n^p[\mu_p];m(l)}(x)
\underbrace{y^{(m}\cdots y^{m)}}_l
\ket{(\gl)_J}^{n^1[q-1],n^2[\mu_2],\ldots,n^p[\mu_p]}\,.
\end{multline}
Calculating the cohomology
$\socohomology{p}{\extmodule_{\irrmodule_{(\gl)_A},\mathfrak{K}_{(\gl)_F},
\irrmodule_{(\gl)_J}}}$ from (\ref{ext_J_A_F}) one obtains that
the condition $\diff\ketv{\Phi_{A,F,J}(x)}=0$ implies equations
(\ref{generalized_electrodynamic_1}),
(\ref{generalized_potetial}), (\ref{generalized_gauge}) along with
equations
\begin{equation}\label{generalized_electrodynamic_2_current}
\psi^{n^1[q-1],n^2[\mu_2],\ldots,n^p[\mu_p]} \Big(
\ptl^{n^1}F_{n^1[q],n^2[\mu_2],\dots,n^p[\mu_p]}(x)-
J_{n^1[q-1],n^2[\mu_2],\dots,n^p[\mu_p]}(x)\Big)=0\,,
\end{equation}
\begin{equation}\label{generalized_current}
\psi^{n^1[q-2],n^2[\mu_2-1],\ldots,n^{\gl_{q-1}}[\mu_{\gl_{q-1}}-1],
n^{\gl_{q-1}+1}[\mu_{\gl_{q-1}+1}],\ldots,n^p[\mu_p]}
\ptl^{n^1}\cdots \ptl^{n^{\gl_{q-1}}}
J_{n^1[q-1],n^2[\mu_2],\dots,n^p[\mu_p]}(x)=0 \,.
\end{equation}

For $\gl_{q-1}=1$  (equivalently $\mu_2\leq q-2$) the system
(\ref{generalized_electrodynamic_1}),
(\ref{generalized_potetial}),
(\ref{generalized_electrodynamic_2_current}),
(\ref{generalized_current}) generalizes the ordinary $\dmn=4$
electrodynamics to any even space--time dimension and arbitrary
tensor structure of fields. Here (\ref{generalized_potetial})
defines the generalized field strength
$F_{mn^1[q-1],n^2[\mu_2],\dots,n^p[\mu_p]}(x)$ via the generalized
potential $A_{n^1[q-1],n^2[\mu_2],\dots,n^p[\mu_p]}(x)$. Equation
(\ref{generalized_electrodynamic_1}) is the Bianchi identity for
generalized field strength. Equation
(\ref{generalized_electrodynamic_2_current}) describes
``interaction" with the ``current" (see
section~\ref{Sct.:_M4_Electrodynamics})
$J_{n^1[q-1],n^2[\mu_2],\dots,n^p[\mu_p]}(x)$, which conserves due
to equation (\ref{generalized_current}). The system
(\ref{generalized_electrodynamic_1}),
(\ref{generalized_potetial}),
(\ref{generalized_electrodynamic_2_current}),
(\ref{generalized_current}) is gauge invariant under the
generalized gradient transformations
\begin{equation}\label{generalized_gradient_transformations}
\psi^{n^1[q-1],n^2[\mu_2],\ldots,n^p[\mu_p]}\gd
A_{n^1[q-1],n^2[\mu_2],\dots,n^p[\mu_p]}(x)=
\psi^{n^1[q-1],n^2[\mu_2],\ldots,n^p[\mu_p]}\ptl_{n^1}
\gep_{n^1[q-2],n^2[\mu_2],\dots,n^p[\mu_p]}(x)
\end{equation}
with an arbitrary parameter
$\gep_{n^1[q-2],n^2[\mu_2],\dots,n^p[\mu_p]}(x)$. Equation
(\ref{generalized_gauge}) fixes conformal gauge,
generalizing equation
(\ref{Maxwell_gauge}).

Analogously to the example in section
\ref{Sct.:_M4_Electrodynamics}, one can relax the gauge fixing
condition (\ref{generalized_gauge}) by considering the module
$\extmodule_{\irrmodule_{(\gl)_A},\mathfrak{K}_{(\gl)_F},
\irrmodule_{(\gl)_J},\irrmodule_{(\gl)_{G}}}$ defined by the short
exact sequence
\begin{equation}\label{ext_J_A_F_gen}
0\longrightarrow\extmodule_{\irrmodule_{(\gl)_A},
\mathfrak{K}_{(\gl)_F},\irrmodule_{(\gl)_J}}
\longrightarrow\extmodule_{\irrmodule_{(\gl)_A},
\mathfrak{K}_{(\gl)_F},\irrmodule_{(\gl)_J},
\irrmodule_{(\gl)_{G}}}
\longrightarrow\irrmodule_{(\gl)_{G}}\longrightarrow0\,,
\end{equation}
where $(\gl)_{G}=(-\gl_{q-1}-q-1,\gl_1,\ldots,\gl_{q-2},0,0)$. The
covariant constancy condition for the
section~$\ketv{\Phi_{A,F,J,G}}$ implies equations
(\ref{generalized_electrodynamic_1}),
(\ref{generalized_potetial}),
(\ref{generalized_electrodynamic_2_current}),
(\ref{generalized_current}) along with the equation
\begin{align}\label{generalized_gauge_inv_1}
\psi^{n^1[q-2],n^2[\mu_2],\ldots,n^p[\mu_p]}\Big(\square
\ptl^{n^1}A_{n^1[q-1],n^2[\mu_2],\dots,n^p[\mu_p]}(x)-
G{}_{n^1[q-2],n^2[\mu_2],\dots,n^p[\mu_p]}(x)\Big)=0
\end{align}
instead of (\ref{generalized_gauge}). The field
$G{}_{n^1[q-1],n^2[\mu_2],\dots,n^p[\mu_p]}(x)$ satisfies the
equation
\begin{equation}\label{generalized_gauge_1}
\psi^{n^1[q-3],n^2[\mu_2-1],\ldots,n^{\gl_{q-2}}[\mu_{\gl_{q-2}}-1],n^{\gl_{q-2}+1}
[\mu_{\gl_{q-2}+1}],\ldots,n^p[\mu_p]}
\ptl^{n^1}\cdots\ptl^{n^{\gl_{q-2}}}G{}_{n^1[q-2],n^2[\mu_2],\dots,n^p[\mu_p]}(x)=0\,.
\end{equation}


\subsubsection{Fradkin--Tseytlin conformal higher spins in even
dimensions\label{Sct.:_F_T_conformal_higher_spins_in_even_dimensions}}
Consider highest weight $(\gl)_{C}=(\gl-2,\gl,0\ldots,0)$,
$\gl_i\in\oN$ (the case of half-integer $\gl_i$ can be considered
analogously). Let
$\irrmodule_{(\gl)_{C}}$ be irreducible conformal module with
the highest weight $(\gl)_{C}$. Using Theorems~\ref{Theorem_1}
and \ref{Thm_cohomology_even} we obtain primitive
conformal system corresponding to the module
$\irrmodule_{(\gl)_{C}}$. It has the form
\begin{align}
\label{F_T_S_zero_Weyl_tensor_equation}
&\psi^{n(\gl),m(\gl)}\underbrace{\ptl_m\cdots\ptl_m}_{\gl}C_{n(\gl)}(x)=0\,,\\
\label{F_T_S_gauge_fixing}
&\psi^{n(\gl-1)}(\ptl\cdot\ptl)^{\gl+q-1}\ptl^nC_{n(\gl)}(x)=0\,.
\end{align}
Here $C_{n(\gl)}(x)$ is a symmetric traceless tensor field,
$\psi^{n(\gl),m(\gl)}$ is an arbitrary traceless tensor parameter
corresponding to the $\gl\times 2$-block Young tableaux.
$(\ptl\cdot\ptl)^{\gl+q-1}$ is an order $2(\gl+q-1)$
differential operator
\begin{equation}
(\ptl\cdot\ptl)^{\gl+q-1}C_{n(\gl)}(x) =  \sum_{p+r=\gl+q-1}
a(p,r)\square^p \underbrace{\ptl_{(n}\cdots\ptl_n}_r\underbrace{\ptl^m
\cdots\ptl^m}_rC_{n(\gl-r))m(r)}
\end{equation}
for some $a(p,r)$.

The left hand side of the
equation (\ref{F_T_S_zero_Weyl_tensor_equation}) can be
interpreted as the generalized Weyl tensor for the field $C_{n(\gl)}(x)$
\begin{equation}\label{F_T_S_Weyl_tensor_definition}
\psi^{n(\gl),m(\gl)}\underbrace{\ptl_m\cdots\ptl_m}_{\gl}C_{n(\gl)}(x)=
\psi^{n(\gl),m(\gl)}W_{n(\gl),m(\gl)}(x)\,.
\end{equation}
It is gauge invariant under the gauge transformations
\begin{equation}\label{F_T_S_gauge_transformations}
\psi^{n(\gl)}\gd
C_{n(\gl)}(x)=\psi^{n(\gl)}\ptl_n\gep_{n(\gl-1)}(x)\,,
\end{equation}
where $\gep_{n(\gl-1)}(x)$ is a gauge parameter.
Equation (\ref{F_T_S_zero_Weyl_tensor_equation}) sets
$W_{n(\gl),m(\gl)}(x)$ to zero and is dynamically trivial
(i.e. describes pure gauge degrees of freedom).
Equation (\ref{F_T_S_gauge_fixing}) is the conformal gauge
condition for $C_{n(\gl)}(x)$. (Note that, as any covariant gauge condition,
it is incomplete.)

A non-trivial dynamical system with nonzero Weyl tensor is
non-primitive and results from the reducible module
$\extmodule_{\irrmodule_{(\gl)_{C}},\irrmodule_{(\gl)_{W}}}$
defined by the short exact sequence
\begin{equation}
\label{ext_F,T_W} 0\longrightarrow\irrmodule_{(\gl)_{C}}
\longrightarrow
\extmodule_{\irrmodule_{(\gl)_{C}},\irrmodule_{(\gl)_{W}}}
\longrightarrow\irrmodule_{(\gl)_{W}}\longrightarrow0\,,
\end{equation}
where $\irrmodule_{(\gl)_{W}}$ is the irreducible conformal module
with the highest weight $(\gl)_{W}=(-2,\gl,\gl,0,\ldots,0)$
corresponding to the Weyl tensor $W_{n(\gl),m(\gl)}(x)$.
Cohomology of
$\extmodule_{\irrmodule_{(\gl)_{C}},\irrmodule_{(\gl)_{W}}}$ is
given in (\ref{cohomology_example}) for $N=q-1$. The module
$\extmodule_{\irrmodule_{(\gl)_{C}},\irrmodule_{(\gl)_{W}}}$
gives rise to the gauge fixing equation (\ref{F_T_S_gauge_fixing})
along with the definition of the Weyl tensor
(\ref{F_T_S_Weyl_tensor_definition}) and the equation
\begin{equation}\label{F_T_S_Weyl_tensor_equation}
\psi^{n(\gl)}\square^{2q-4}\underbrace{\ptl^m\cdots\ptl^m}_{\gl}
W_{n(\gl),m(\gl)}=0\,.
\end{equation}
This class of conformal equations was found by Fradkin and Tseytlin
in \cite{FrT} along with the analogous equations for spinor--tensors
for $\dmn=4$ and generalized to arbitrary even $\dmn=2q$
in \cite{Segal}.

Gauge invariant form of the same system (i.e., without equation
(\ref{F_T_S_gauge_fixing})) results from our
construction applied to the module
$\extmodule_{\irrmodule_{(\gl)_{C}},\irrmodule_{(\gl)_{W}},
\irrmodule_{(\gl)_{G}}}$ defined by the short exact sequence
\begin{equation}
\label{ext_F,T_W_G}0\longrightarrow
\extmodule_{\irrmodule_{(\gl)_{C}},\irrmodule_{(\gl)_{W}}}
\longrightarrow
\extmodule_{\irrmodule_{(\gl)_{C}},\irrmodule_{(\gl)_{W}},
\irrmodule_{(\gl)_{G}}}
\longrightarrow\irrmodule_{(\gl)_{G}}\longrightarrow0\,.
\end{equation}
Here $\irrmodule_{(\gl)_{G}}$ is the irreducible conformal module with
the highest weight $(\gl)_{G}=(-\gl-2q+1,\gl-1,0,\ldots,0)$.
Module
$\extmodule_{\irrmodule_{(\gl)_{C}},\irrmodule_{(\gl)_{W}},
\irrmodule_{(\gl)_{G}}}$ gives rise to the system containing equations
(\ref{F_T_S_Weyl_tensor_definition}),
(\ref{F_T_S_Weyl_tensor_equation}) and the equation
\begin{equation}\label{F_T_S_gauge_invariant}
\psi^{n(\gl-1)}(\ptl\cdot\ptl)^{\gl+q-1}\ptl^nC_{n(\gl)}(x)=
\psi^{n(\gl-1)}G_{n(\gl-1)}\,,
\end{equation}
which relaxes the gauge fixing equation (\ref{F_T_S_gauge_fixing}).


\section{Conclusions\label{Sct.:_Conclusions}}
In this paper we study a general framework, which allows us to
classify and obtain the explicit form of all linear homogeneous
$\alg_{\psubalg}$-invariant $\dmn$-dimensional equations for an arbitrary
semi-simple Lie algebra $\alg$ which has a parabolic subalgebra
$\psubalg$ with an $\dmn$-dimensional Abelian radical $\rad$.
These equations are written in the form of the covariant constancy
conditions
\begin{equation}\label{concl_1}
\diff\bundsec{p}=(\extdiff+\go_0(x))\bundsec{p}=0\,.
\end{equation}
Here the connection 1-form $\go_0(x)$ takes values in $\alg$ and
is flat, i.e. $(\extdiff+\go_0(x))^2$=0. A particularly useful
choice of the connection is  $\go_0(x)=\gs_-$, where $\gs_-$ takes
values in radical $\rad$ and is $x$-independent, i.e., $d \gs_- +
\gs_- d =0$, $\gs_-^2=0$. The $p$-forms $\bundsec{p}$ take values
in an $\alg$-module $\module$ that is required to be
$\psubalg$-integrable. We prove that (\ref{concl_1}) leads to a
linear homogeneous $\alg$-invariant equation
\begin{equation}\label{concl_2}
\equations_\module\subbundsec{p}=0
\end{equation}
on the set of dynamical fields $\subbundsec{p}$ that are elements
of the $p$-th cohomology of $\gs_-$ (see Remark
\ref{Thm.:_Remark_3_4}). All other fields from the set
$\bundsec{p}$ are either pure gauge or auxiliary fields expressed
in terms of derivatives of the dynamical fields. The form of
equations (\ref{concl_2}) is determined by the $p+1$-th cohomology
of $\gs_-$. $\alg_{\psubalg}$-invariant equations (\ref{concl_2})
are classified by the modules $\module$. This classification is
complete because any equation can be unfolded to the form
(\ref{concl_1}) by introducing  auxiliary fields. A constructive
procedure is described, which allows one to obtain the explicit
form of the $\alg_{\psubalg}$-invariant equation associated with
$\module$. In this paper, the proposed general construction is
applied to obtain the complete classification of conformally
invariant differential equations in terms of singular and
subsingular modules of generalized Verma modules of the conformal
algebra in $\dmn$ dimensions.

The approach proposed in this paper can be further applied to
several problems. The most straightforward application is to study
free (i.e., linear) equations invariant under symmetries different
from the usual conformal symmetry. A particularly interesting
example is that of the symplectic algebra $sp(m)$ which was shown
\cite{BHS} to be a proper extension of the usual conformal
algebra, acting on the infinite systems of fields of higher spins.
More examples of $sp(m)$-invariant equations were obtained
recently in~\cite{GV}. It is also tempting to apply our approach
to the study of $\dmn=2$ conformal systems starting with the
related infinite-dimensional symmetries.

Another interesting generalization to be studied consists in
relaxing the requirement that the radical $\rad$ is Abelian. In
this case one can still formulate invariant equations in the form
(\ref{concl_1}). The resulting equations are not translationally
invariant because $\go_0 (x)$ is necessarily $x$-dependent. Also
it is not clear how to implement the analysis of the dynamical
content of the invariant equations in terms of cohomology. Let us
note that this case is not of a purely ``academic" interest. An
important class of equations of this type is provided by
superfield equations for supersymmetric systems, which are known
to contain an explicit dependence on anticommuting variables
through the supercovariant derivatives. It is well known that it
is sometimes difficult to distinguish between constraints and
``true" field equations in superspace. As mentioned in \cite{BHS},
the origin of this difficulty can be traced back to the absence of
a distinct $\gs_-$ cohomology description.

One of the most important problems is to go beyond the class of
linear equations. A suggestive feature of our approach mentioned
in section \ref{Sct.:_M4_Electrodynamics} is that it allows a
natural definition of current modules. As a result, the
interaction problem admits a reformulation in terms of the realization
of current modules as tensor products (i.e., nonlinear
combinations) of modules associated with matter fields.

By analogy with higher spin theory, to put interacting theory in
the framework of gravity with the gravitational field being one of
the dynamical fields (i.e., not just a background one as in this
paper) it is important to extend the formalism to (extensions of)
field equations formulated in terms of differential $p$-forms with
$p>0$. Among other things, this requires clarifying the
relationship between the dynamical equations formulated in terms
of 0-forms as in this paper and those formulated in terms of
higher differential forms (in particular, 1-forms) as in higher
spin gauge theory \cite{VD,gol}. In this respect Theorems 4.2, 4.3
in this paper and their generalizations to other Lie algebras to
be worked out are likely to play the key role because they link
together cohomology groups which determine dynamical fields and
field equations in terms of various   differential forms.

Finally, it would be very instructive to make contact with other
cohomological approaches such as developed e.g. in
\cite{A-inf,Henneaux,D-VH}.

\section*{Acknowledgments}
We are grateful to A.~Semikhatov for
useful discussions and numerous useful comments on the manuscript.
We are grateful to R.~Metsaev, B.~Feigin and
M.~Finkelberg for valuable discussions. The work was supported by
INTAS, Grant No.00-01-254, the RFBR, Grant No.02-02-17067 and
Russian Federation President Grant No.LSS-1578.2003.2. TIY is
partially supported by the RFBR Grant No.02-02-16944, RFBR Grant
No.03-01-06135 and the Russian Science Support Formation. SOV is
partially supported by the RFBR Grant No.03-02-06465 and the
Landau Scholarship Foundation, Forschungszentrum J$\ddot{\rm
u}$lich.

\appendix
\section{Relevant facts from representation theory\label{Sct.:_Appendix}}
The structure of generalized Verma modules can be investigated
using methods developed
in~\cite{Vogan,[BB],[BCI],[BCII],[KKz],Zhelob}. Let us first
recall some notations. Let~$\Cartan$ be the Cartan subalgebra
and~$\Cartan^*$ is its
 dual space. Let simple roots be denoted $\alpha_0$,
$\alpha_1$, \dots, $\alpha_q$ and  $\subroots$
consists of~$\alpha_1$, \dots, $\alpha_q$
{(see section~\ref{Sct.:_General_Construction}). The Weyl group $W^{q+1}$ is
generated by reflections
$r_{\ga_i}\equiv r_i$ ($0 \leq i \leq q$)
 of $\Cartan^*$ over the hyperplane orthogonal to the
simple root $\ga_i$
\begin{equation}
   r_i\lambda=\lambda-2\frac{(\lambda,\alpha_i)}{(\alpha_i,\alpha_i)}
    \alpha_i\,,
\end{equation}
$\gl\in\Cartan^*$. The action $r_\ga\cdot\lambda$ (nonlinear
representation)
 of $W^{q+1}$ in $\Cartan^*$ is defined by the formula
\begin{equation}\label{shiftact}
   r_\ga\cdot\lambda=
   \lambda-2\frac{(\lambda+\rho,\alpha)}{(\alpha,\alpha)}\alpha
\end{equation}
for any $\ga$, $\gl\in\Cartan^*$. Here $\rho$ is half of the sum
of positive roots\footnote{Note that this formula is universal:
given linear representation of a group $G$ in a linear space
$V$ and a fixed vector $\rho \in V$, the transformations $
A\cdot\lambda=A \lambda+(A-\one)\rho $ for $A\in G$ and $\lambda
\in V$ define the (nonlinear) action of $G$ in $V$.}.         Let
$W^q$ be the subgroup of the Weyl group generated by simple
reflections~$r_i$ with~$1\leq i\leq q$. Denote by~$\lattice$ the
root lattice $\{\oZ\alpha_0+\oZ\alpha_1+\dots+\oZ\alpha_q\}$. For
any highest weight~$\lambda$, let ~$W^{q+1}_\lambda$ be the
subgroup constituted by such elements $w\in W^{q+1}$ that
\begin{equation}
   w\cdot\lambda\in\lambda+\lattice\,.
\end{equation}
Let $S_\lambda\subset W^{q+1}_\lambda$ be the stability subgroup
of $\lambda$
\begin{equation}
   s\cdot\lambda=\lambda\,,\qquad s\in S_\lambda\,.
\end{equation}
Consider the quotient
\begin{equation}
   T_\lambda=(W^q\cap W^{q+1}_\lambda)\backslash
      W^{q+1}_\lambda/(W^q\cap S_\lambda)\,.
\end{equation}
Denote by~$\weights$ the set of highest weights of the
form~$\lambda=(\lambda_0,\lambda_1,\lambda_2,\dots\lambda_q)$
where~$(\lambda_1,\lambda_2,\dots\lambda_q)$ is a dominant
integral highest weight of $B_q$ ($D_q$)(i.e.
$\gl_1\geq\gl_2\geq\cdots\geq\gl_q$ ($\gl_1\geq\gl_2\geq\cdots\geq
|\gl_q|$) and $2\gl_i$ are all even or odd simultaneously). For
any equivalence class from $T_\lambda$ one can choose a
representative $t$ such that~$t\cdot\lambda\in\weights$
whenever~$\lambda\in\weights$. Let $\cT_\gl\subset T_\gl$ denote
the set of all such representatives. For any
weight~$\nu\in\weights$, the set of elements $\cT_\gl$ generates
the set of highest weights~$\{t\cdot\nu\}_{t\in \cT_\gl}$.

Elements~$t\in \cT_\lambda$ are ordered with respect to their
${\mathsf{length}}(t)$, where the ${\mathsf{length}}(t)$ is the
number of the multipliers in the reduced (i.e. minimal)
decomposition of~$t$ into a product of the elementary reflections
generated by the simple roots. The reduced decomposition is
unique. We write~$t_1\prec t_2$
whenever~${\mathsf{length}}(t_1)<{\mathsf{length}}(t_2)$. Note
that such defined order is partial because any two elements with
the same length can not be compared. The main point is that the
generalized Verma module~$\soW_{t_2\cdot\nu}$ admits a nontrivial
homomorphism into the generalized Verma
module~$\soW_{t_1\cdot\nu}$ whenever~$t_1\prec t_2$~\cite{Zhelob}.
Applying this general method to the conformal algebra one obtains
the structure of singular modules in~$\soW_\lambda$ in the
cases~$B_{q+1}$ and~$D_{q+1}$, which was completely studied
in~\cite{[BCI],[BCII]} (see also~\cite{eastw} for a textbook).
This exhausts the case of $B_{q+1}$. In the case of~$D_{q+1}$
subsingular modules exist and their structure should be
investigated separately. Let us sketch the final results
separately for the cases $B_{q+1}$ (i.e. $\dmn=2q+1$) and
$D_{q+1}$ (i.e. $\dmn=2q$).

Let $\dmn=2q+1$.
The Dynkin diagram of the algebra $B_{q+1}$ is~(\ref{odd-dynkin}). Choose an
orthogonal basis~$\epsilon_i$ $0\leq i \leq q$ in~$\Cartan^*$.
Then
\begin{equation}
   \alpha_i=\epsilon_i-\epsilon_{i+1}\,,\ 0\leq i\leq q-1\,,\qquad
   \alpha_q=\epsilon_q\,.
\end{equation}
Introduce the basis in~$\Cartan$ dual to~$\epsilon_i$ (i.e.
$\epsilon_i(\epsilon^j)=\delta_i^j$)
\begin{equation}
   \epsilon^0=-\cD\,,\qquad\epsilon^1=\cL_{12}\,,\qquad
   \epsilon^i=\sqrt{-1}\cL_{2i-1,2i}\,,\ 1<i\leq q-1\,,\qquad
    \epsilon^q=\sqrt{-1}\cL_{2q,2q+1}\,.
\end{equation}
Then
\begin{equation}
   \cH_i=\epsilon^i-\epsilon^{i+1}\,,\ 0\leq i\leq q-1\,,\qquad
   \cH_q=2\epsilon^q\,.
\end{equation}
Half the sum of all positive roots is in this case
\begin{equation}
   \rho=\sum_{i=0}^q(q-i+\half)\epsilon_i\,.
\end{equation}

Recall that $r_i$ denote the simple reflections
$r_i=r_{\ga_i}=r_{\epsilon_i-\epsilon_{i+1}}$ for $0\leq i \leq
q-1$ and $r_q=r_{\ga_q}=r_{\epsilon_q}$.

In the case of dominant integral~$\lambda$ the stability subgroup
is trivial and the set~$\cT_\lambda$ consists of the following
elements~\cite{eastw}
\begin{multline}
e\prec r_0\prec r_1r_{\epsilon_0-\epsilon_2}\prec
r_1r_2r_{\epsilon_0-\epsilon_3}\prec\dots\prec r_1r_2\dots
r_{q-1}r_{\epsilon_0-\epsilon_q}\prec r_1r_2\dots r_{q-1}r_q
r_{\epsilon_0+\epsilon_q}\prec\\ \prec r_1r_2\dots
r_{q-2}r_{\epsilon_{q-1}}r_{\epsilon_0+\epsilon_{q-1}}\prec \dots
\prec r_1r_{\epsilon_2}r_{\epsilon_0+\epsilon_2}\prec
r_{\epsilon_1}r_{\epsilon_0+\epsilon_1}\prec r_{\epsilon_0}\,.
\end{multline}
Note that these elements are written in the non-reduced form,
which, however, is more convenient for calculations. This gives
rise to the diagram~(\ref{oddembeddings}) (see the end of the
paper) of homomorphisms of modules~$\soW_\lambda$,  where
$\lambda_0\geq\lambda_1\geq\ldots\geq\lambda_q\geq0$ and
$2\lambda_i$ are either all even or all odd $0\leq i \leq q$.
Composition of any two homomorphisms (arrows) in the diagram is
zero.

For non-integral $\lambda$, homomorphisms are associated
with~$\cT_\lambda=\{e \prec r_{\epsilon_0}\}$. In this case the
parameters of the highest weight should satisfy
   \begin{equation}\label{eq:Klein-Gordon}
     \lambda_0=-q-\frac{1}{2}+n\,,\quad n\in\oN\,,
     \qquad\lambda_i\in\oN\,,\quad1\leq i\leq q\,,
   \end{equation}
   or
   \begin{equation}\label{eq:Dirac}
     \lambda_0=-q+n\,,\ n\in\oN_0\,,\qquad
     \lambda_i\in\half+\oN_0\,,\quad1\leq i\leq q\,.
   \end{equation}
This leads to the following diagram of homomorphisms
   \begin{equation}\label{non-standart_homomorphism_odd}
     \soW_{(\lambda_0,\lambda_1,\dots,\lambda_q)}
     \stackrel{(r_{\epsilon_0},l)}{\longleftarrow}
     \soW_{(-\lambda_0-2q-1,\lambda_1,\dots,\lambda_q)}\,.
   \end{equation}
   For the case~\eqref{eq:Klein-Gordon}~$l=2n$. For the
case~\eqref{eq:Dirac}~$l=2n+1$.

Let $\dmn=2q$.
The Dynkin diagram of the algebra~$D_{q+1}$
is~(\ref{even-dynkin}). Choose an orthogonal basis~$\epsilon_i$
in~$\Cartan^*$. Then
\begin{equation}
   \alpha_i=\epsilon_i-\epsilon_{i+1}\,,\ 0\leq i\leq q-1\,,\qquad
   \alpha_q=\epsilon_{q-1}+\epsilon_q\,.
\end{equation}
The half of the sum of all positive roots is
\begin{equation}
   \rho=\sum_{i=0}^{q-1}(q-i)\epsilon_i \,.
\end{equation}

The analysis analogous to that of the odd dimensional case gives
that~$\cT_\lambda$ with a dominant integral~$\lambda$ consists of
the following elements~\cite{eastw}
\begin{multline}
e\prec r_0\prec r_1r_{\epsilon_0-\epsilon_2}\prec
r_1r_2r_{\epsilon_0-\epsilon_3}\prec\dots\prec
r_1r_2\dots r_{q-2}r_{\epsilon_0-\epsilon_{q-1}}\prec\\
\prec\atop{r_1r_2\dots r_{q-1}r_{\epsilon_0-\epsilon_q}}
     {r_1r_2\dots r_{q-2}r_{q}r_{\epsilon_0+\epsilon_q}}
\prec r_1r_2\dots r_{q-1}r_{q}r_{\epsilon_0+\epsilon_{q-1}}
\prec\\ \prec r_1r_2\dots r_{q-3}r_{\epsilon_{q-2}-\epsilon_q}
r_{\epsilon_{q-2}+\epsilon_q}r_{\epsilon_0+\epsilon_{q-2}}\prec
r_1r_2\dots r_{q-4}r_{\epsilon_{q-3}-\epsilon_q}
r_{\epsilon_{q-3}+\epsilon_q}r_{\epsilon_0+\epsilon_{q-3}}\prec
\dots \prec\\ \prec r_1r_{\epsilon_2-\epsilon_q}
r_{\epsilon_2+\epsilon_q}r_{\epsilon_0+\epsilon_2}\prec
r_{\epsilon_1-\epsilon_q}r_{\epsilon_1+\epsilon_q}
r_{\epsilon_0+\epsilon_1}\prec
r_{\epsilon_0-\epsilon_q}r_{\epsilon_0+\epsilon_q}
\end{multline}
The diagram of $\soW_\gl$-homomorphisms is~(\ref{2q-diagram}) (see
the end of the paper), where $\gl_0\geq\gl_1\geq\cdots\geq
|\gl_q|$ and $2\gl_i$ are either all even or all odd $0\leq i \leq
q$. Here, the composition of any two homomorphisms, except for
those in the central rhombus and those that are labeled by NS, is
zero. There exist also~$q-1$ \textit{nonstandard}
homomorphisms~\cite{[BCI]} (they are labeled by the symbol NS in
the diagram (\ref{2q-diagram})) between modules in this diagram
that correspond to the
element~$r_{\epsilon_0-\epsilon_q}r_{\epsilon_0+\epsilon_q}$
from~$\cT_\lambda$
\begin{equation}\label{more}
  \soW_{(\lambda_N-N,\lambda_0+1,\lambda_1+1,\dots,
           \lambda_{N-1}+1,\lambda_{N+1},\dots,\lambda_q)}
\stackrel{(r_{\epsilon_0-\epsilon_q}r_{\epsilon_0+\epsilon_q},
2\lambda_N-2N+2q)}
{\longleftarrow\kern-6pt-\kern-6pt-\kern-6pt-\kern-6pt-
\kern-6pt-\kern-6pt-\kern-6pt-\kern-6pt-}
\soW_{(-\lambda_N+N-2q,\lambda_0+1,\lambda_1+1,\dots,
           \lambda_{N-1}+1,\lambda_{N+1},\dots,-\lambda_q)}
\end{equation}
for\footnote{\label{homomorphisms}For $N=q-1$, this  homomorphism
amounts to the composition of the homomorphisms that constitute the
rhombus.} $0\leq N < q-1$.

There are also nonstandard homomorphisms in the case
when~$\lambda$ is singular i.e.~$\lambda+\rho$ lies on a wall of
the Weyl chamber. Then~$\cT_\lambda=\{e \prec
r_{\epsilon_0-\epsilon_q}r_{\epsilon_0+\epsilon_q}\}$ and the
parameters of the highest weight satisfy the following relations
  \begin{alignat}{2}
    \lambda_0-\lambda_N+N&=0\quad\mbox{\rm for some $N=1,2,\dots,q$}\,,
     \label{eq:wall}\\
    \lambda_0+\lambda_q+q&=n\in\oN_0\,,\label{eq:int1}\\
    \lambda_0-\lambda_q+q&=m\in\oN_0\quad\mbox{\rm and $m+n\neq0$}\,.
    \label{eq:int2}
  \end{alignat}
Here \eqref{eq:wall} is the condition that the highest weight is
singular and~\eqref{eq:int1}, \eqref{eq:int2} are conditions
that~$r_{\epsilon_0-\epsilon_q}r_{\epsilon_0+\epsilon_q}\lambda$
belongs to the weight lattice. These homomorphisms are
\begin{equation}\label{weylchamber}
  \soW_{(\lambda_0,\lambda_1,\lambda_2,\dots,\lambda_q)}
\stackrel{(r_{\epsilon_0-\epsilon_q}r_{\epsilon_0+\epsilon_q},
2\lambda_0+2q)}
{\longleftarrow\kern-6pt-\kern-6pt-\kern-6pt-\kern-6pt-
\kern-6pt-\kern-6pt-\kern-6pt-\kern-6pt-}
\soW_{(-\lambda_0-2q,\lambda_1,\lambda_2,\dots,-\lambda_q)}\,.
\end{equation}

The quotient of an arbitrary generalized Verma module $\soW_\gl$
over the submodule $\msubVerma{\gl}$ generated from  all
singular submodules of $\soW_\gl$ is not necessarily irreducible.
In fact, the module $\soW_\gl$ can have subsingular submodules
(those that are singular in $\soW_\gl/\msubVerma{\gl}$),
subsubsingular submodules etc.\dots\ In the conformal algebra case
subsubsingular submodules do not appear.

To describe the structure of~$\soW_{\lambda}$ for the highest
weight~$(\lambda)$ belonging to
series~(\ref{even_dimmensional_sequence}) we start with the case
of~$(\gl)_{-q}=(0,0,\dots,0)$. All other cases can be obtained
from this one by application of the shift functor~\cite{Vogan} to
modules belonging to the case~$(\gl)_{-q}=(0,0,\ldots,0)$. So let
us consider the case {\allowdisplaybreaks
\begin{align}\label{even_dimmensional_sequence-trivial}
\kern10pt&(\gl)_{-q}=(0,0,\ldots,0)\,,\nn\\
&(\gl)_{-q+1}=(-1,1,0,\ldots,0)\,,\nn\\[-0.2cm]
&\qquad\qquad\qquad\qquad\vdots\nn\\[-0.2cm]
&(\gl)_{-q+N}=(-N,\underbrace{1,\ldots,1}_{N},
                       0,\ldots,0)\,,
      \qquad\qquad\qquad\qquad\;\:{N=0,\ldots,q-1}\,,\nn\\[-0.2cm]
&\qquad\qquad\qquad\qquad\vdots\nn\\[-0.2cm]
&(\gl)_{-1}=(-q+1,1,\ldots,1,0)\,,\nn\\[0.1cm]
&(\gl)_0=(-q,1,\ldots,1)\,, \qquad
(\gl)_{0'}=(-q,1,\ldots,1,-1)\,,\nn\\[0.2cm]
&(\gl)_{1}=(-q-1,1,\ldots,1,0)\,,\\[-0.2cm]
&\qquad\qquad\qquad\qquad\vdots\nn\\[-0.2cm]
&(\gl)_{K}=(-q-K,\underbrace{1,\ldots,1}_{q-K}, 0,\ldots,0)\,,
\qquad\qquad\qquad\qquad\;\:{K=1,\ldots,q-1}\,,\nn\\[-0.2cm]
&\qquad\qquad\qquad\qquad\vdots\nn\\[-0.2cm]
&(\gl)_{q-2}=(-2q+2,1,1,0,\ldots,0)\,,\nn\\
&(\gl)_{q-1}=(-2q+1,1,0,\ldots,0)\,,\nn\\
&(\gl)_{q}=(-2q,0,\ldots,0)\,.\nn
\end{align}
} The structure of generalized Verma modules with these highest
weights can be elaborated by the direct calculation. Solving
explicitly the system of equations
\begin{equation}
\cP_nF_A(y^m)\ketv{(\gl)_N}^A=0
\end{equation}
for the polynomials~$F_A(y^m)$, where~$\cP_n$ are differential
operators~(\ref{P-in-verma}) we obtain that the
module~$\soW_{(\gl)_{-q}}$ contains singular vectors
\begin{align}
&\ketv{s^1_{(\gl)_{-q}}}^m=y^m\ketv{(\gl)_{-q}}\,,\\
&\ketv{s^2_{(\gl)_{-q}}}=(y^2)^q\ketv{(\gl)_{-q}}\,.
\end{align}
The modules~$\soW_{(\gl)_{N}}$ for~$N=-q+1, \ldots,-1$ contain
singular vectors
\begin{align}
&\ketv{s^1_{(\gl)_N}}^{m[N+q+1]}=y^{[m}\ketv{(\gl)_{N}}^{m[N+q]]}\,,\\
&\ketv{s^2_{(\gl)_N}}^{m[N+q]}=(y^2)^{-N}\ketv{(\gl)_{N}}^{m[N+q]}-(N+q)
(y^2)^{-N-1}y_n y^{[m}\ketv{(\gl)_{N}}^{nm[N+q-1]]}
\end{align}
and subsingular vectors
\begin{equation}
\ketv{subs_{(\gl)_N}}^{m[N+q-1]}=(y^2)^{-N}y_n\ketv{(\gl)_{N}}^{nm[n+q-1]}\,.
\end{equation}
The modules~$\soW_{(\gl)_{N}}$ for~$N=0,0',1,\dots, q-1$ contain
singular vectors
\begin{equation}
\ketv{s_{(\gl)_N}}^{m[q-N-1]}=y_{m}\ketv{(\gl)_{N}}^{m[q-N]}\,.
\end{equation}

The completeness of this list of singular and subsingular modules
follows from the theory intersection cohomology
sheaves~\cite{[BB]}.

\newpage

\section{Homomorphism diagrams\label{app:hom_diag}}
{\small
\begin{equation}\label{oddembeddings}
\begin{CD}
     \soW_{(\gl)_{0}}=\soW_{(\lambda_0,\lambda_1,\lambda_2,\dots,\lambda_q)}\\
     @AA(r_{\epsilon_0-\epsilon_1},\,\lambda_0-\lambda_1+1)A\\
     \soW_{(\gl)_{1}}=\soW_{(\lambda_1-1,\lambda_0+1,\lambda_2,\dots,\lambda_q)}\\
     @AA(r_{\epsilon_0-\epsilon_2},\,\lambda_1-\lambda_2+1)A\\
     \dots\\
     @AA(r_{\epsilon_0-\epsilon_N},\,\lambda_{N-1}-\lambda_N+1)A\\
     \soW_{(\gl)_{N}}=\soW_{(\lambda_N-N,\lambda_0+1,\lambda_1+1,\dots,
           \lambda_{N-1}+1,\lambda_{N+1},\dots,\lambda_q)}\\
     @AA(r_{\epsilon_0-\epsilon_{N+1}},\,\lambda_N-\lambda_{N+1}+1)A\\
     \soW_{(\gl)_{N+1}}=\soW_{(\lambda_{N+1}-N-1,\lambda_0+1,\lambda_1+1,\dots,
           \lambda_N+1,\lambda_{N+2},\dots,\lambda_q)}\\
     @AA(r_{\epsilon_0-\epsilon_{N+2}},\,\lambda_{N+1}-\lambda_{N+2}+1)A\\
     \dots\\
     @AA(r_{\epsilon_0-\epsilon_q},\,\lambda_{q-1}-\lambda_q+1)A\\
     \soW_{(\gl)_{q}}=\soW_{(\lambda_q-q,\lambda_0+1,\lambda_1+1,\dots,\lambda_{q-1}+1)}\\
     @AA(r_{\epsilon_0},\,2\lambda_q+1)A\\
     \soW_{(\gl)_{q+1}}=\soW_{(-\lambda_q-q-1,\lambda_0+1,\lambda_1+1,\dots,\lambda_{q-1}+1)}\\
     @AA(r_{\epsilon_0+\epsilon_q},\,\lambda_{q-1}-\lambda_q+1)A\\
     \dots\\
     \\
     @AA(r_{\epsilon_0+\epsilon_{N+1}},\,\lambda_{N}-\lambda_{N+1}+1)A\\
     \soW_{(\gl)_{2q+1-N}}=\soW_{(-\lambda_N+N-2q-1,\lambda_0+1,\lambda_1+1,\dots,
           \lambda_{N-1}+1,
           \lambda_{N+1},\dots,\lambda_q)}\\
     @AA(r_{\epsilon_0+\epsilon_N},\,\lambda_{N-1}-\lambda_N+1)A\\
     \soW_{(\gl)_{2q+2-N}}=\soW_{(-\lambda_{N-1}+N-2q-2,\lambda_0+1,\lambda_1+1,\dots,
           \lambda_{N-2}+1,\lambda_N,\dots,\lambda_q)}\\
     @AA(r_{\epsilon_0+\epsilon_{N-1}},\,\lambda_{N-2}-\lambda_{N-1}+1)A\\
     \dots\\
     @AA(r_{\epsilon_0+\epsilon_1},\,\lambda_0-\lambda_1+1)A\\
     \soW_{(\gl)_{2q+1}}=\soW_{(-\lambda_0-2q-1,\lambda_1,\lambda_2,\dots,\lambda_q)}
\end{CD}
\end{equation}
} The label~$(r,l)$ at a homomorphism arrow has the following
meaning. $r$ denotes the reflection that connects highest weights
of the two modules. $l$ is the level  at which a singular module
resulting from the arrow homomorphism is situated.
\newpage
\begin{picture}(40,50)
  \put(40.5,25){NS}
  \qbezier[500](13.7,1.6)(40,2)(40,15)
  \put(40,15){\line(0,1){20}}
  \qbezier[500](40,35)(40,48)(15,47)
  \put(15,47){\vector(-3,1){1}}
  \put(30.5,17){NS}
  \qbezier[500](13.7,9)(30,9)(30,14)
  \put(30,14){\line(0,1){6}}
  \qbezier[500](30,20)(30,25.9)(15,25.4)
  \put(15,25.4){\vector(-3,1){1}}
  {\linethickness{1.2pt}
  \qbezier[10](33,19)(33,22)(33,22.9)
  }
  \put(33,23){\line(0,1){3}}
  \qbezier[500](33,26)(33,33.5)(15,32.4)
  \put(15,32.4){\vector(-3,1){1}}
\put(0,11){
  {\linethickness{1.2pt}
  \qbezier[10](37,19)(37,22)(37,22.9)
  }
  \put(37,23){\line(0,1){3}}
  \qbezier[500](37,26)(37,33.5)(15,32.4)
  \put(15,32.4){\vector(-3,1){1}}
  }
\put(0,5.5){
  {\linethickness{1.2pt}
  \qbezier[10](35,19)(35,22)(35,22.9)
  }
  \put(35,23){\line(0,1){3}}
  \qbezier[500](35,26)(35,34.5)(15,34.4)
  \put(15,34.4){\vector(-3,1){1}}
  }
\put(-3,0){ \vbox{{\small
\begin{equation}\label{2q-diagram}
\kern-90pt\begin{array}{c}
\begin{CD}
\soW_{(\gl)_{-q}}=\soW_{(\lambda_0,\lambda_1,\lambda_2,\dots,\lambda_q)}\\
     @AA(r_{\epsilon_0-\epsilon_1},\,\lambda_0-\lambda_1+1)A\\
\soW_{(\gl)_{-q+1}}=\soW_{(\lambda_1-1,\lambda_0+1,\lambda_2,\dots,\lambda
_q)}\\
     @AA(r_{\epsilon_0-\epsilon_2},\,\lambda_1-\lambda_2+1)A\\
     \dots\\
     @AA(r_{\epsilon_0-\epsilon_N},\,\lambda_{N-1}-\lambda_N+1)A\\
     \soW_{(\gl)_{-q+N}}=\soW_{(\lambda_N-N,\lambda_0+1,\lambda_1+1,\dots,
           \lambda_{N-1}+1,\lambda_{N+1},\dots,\lambda_q)}\\
     @AA(r_{\epsilon_0-\epsilon_{N+1}},\,\lambda_N-\lambda_{N+1}+1)A\\
\soW_{(\gl)_{-q+N+1}}=\soW_{(\lambda_{N+1}-N-1,\lambda_0+1,\lambda_1+1,\dots,
           \lambda_{N-1}+1,\lambda_N+1,\lambda_{N+2},\dots,\lambda_q)}\\
     @AA(r_{\epsilon_0-\epsilon_{N+2}},\,\lambda_{N+1}-\lambda_{N+2}+1)A\\
     \dots\\
     @AA(r_{\epsilon_0-\epsilon_{q-2}},\,\lambda_{q-3}-\lambda_{q-2}+1)A\\
     \soW_{(\gl)_{-2}}=\soW_{(\lambda_{q-2}-q+2,\lambda_0+1,\lambda_1+1,\dots,
             \lambda_{q-3}+1,\lambda_{q-1},\lambda_q)}\\
     @AA(r_{\epsilon_0-\epsilon_{q-1}},\,\lambda_{q-2}-\lambda_{q-1}+1)A\\
     \soW_{(\gl)_{-1}}=\soW_{(\lambda_{q-1}-q+1,\lambda_0+1,\lambda_1+1,\dots,
             \lambda_{q-2}+1,\lambda_q)}\\
\end{CD}\\
\begin{picture}(5,3)
\put(-2,-1){\vector(2,3){2}}
\put(-8.5,0){${}_{(r_{\epsilon_0-\epsilon_q},\,\lambda_{q-1}-\lambda_q+1)}$}
\put(8,-1){\vector(-2,3){2}}
\put(8,0){${}_{(r_{\epsilon_0+\epsilon_q},\,\lambda_{q-1}+\lambda_q+1)}$}
\end{picture}\\ \\ \kern-15pt\soW_{(\gl)_{0}}\!\!\!=\!\soW_{(\lambda_q-q,
       \lambda_0+1,\lambda_1+1,\dots,\lambda_{q-2}+1, \lambda_{q-1}+1)}
\kern10pt
\soW_{(\gl)_{0'}}\!\!\!=\!\soW_{(-\lambda_q-q,\lambda_0+1,\lambda_1+1,\dots,
\lambda_{q-2}+1,-\lambda_{q-1}-1)}\\
\begin{picture}(5,3)
\put(0.3,-1){\vector(-2,3){2}}
\put(-8.5,0){${}_{(r_{\epsilon_0+\epsilon_q},\,\lambda_{q-1}+\lambda_q+1)}$}
\put(6,-1){\vector(2,3){2}}
\put(8,0){${}_{(r_{\epsilon_0-\epsilon_q},\,\lambda_{q-1}-\lambda_q+1)}$}
\end{picture}\\ \\
\begin{CD}
     \soW_{(\gl)_{1}}=\soW_{(-\lambda_{q-1}-q-1,\lambda_0+1,\lambda_1+1,\dots,
                                \lambda_{q-2}+1,-\lambda_q)}\\
     @AA(r_{\epsilon_0+\epsilon_{q-1}},\,\lambda_{q-2}-\lambda_{q-1}+1)A\\
     \soW_{(\gl)_{2}}=\soW_{(-\lambda_{q-2}-q-2,\lambda_0+1,\lambda_1+1,\dots,
      \lambda_{q-3}+1,\lambda_{q-1},-\lambda_q)}\\
     @AA(r_{\epsilon_0+\epsilon_{q-2}},\,\lambda_{q-3}-\lambda_{q-2}+1)A\\
     \dots\\ @AA(r_{\epsilon_0+\epsilon_1},\,\lambda_0-\lambda_1+1)A\\
\soW_{(\gl)_{q}}=\soW_{(-\lambda_0-2q,\lambda_1,\dots,\lambda_{q-1},-\lambda_q)}
\end{CD} \end{array} \end{equation} }}} \end{picture}

The label~$(r,l)$ at a homomorphism arrow has the following
meaning. $r$ denotes the reflection that connects highest weights
of the two modules. $l$ is the level  at which a singular module
resulting from the arrow homomorphism is situated.


\begin{thebibliography}{33}
\bibitem{Ann}   M.A. Vasiliev, {\it Ann.Phys.} (N.Y.) {\bf 190} (1989)
59.
\bibitem{more} M.A. Vasiliev,  {\it Phys.Lett.} {\bf B 243} (1990) 378;
{\it Class.Quant.Grav.} {\bf 8} (1991) 1387; {\it Phys.Lett.}  {
\bf B 285} (1992) 225.
\bibitem{gol} M.A. Vasiliev,
Contributed article to Golfand's Memorial Volume, ed. by
M.~Shifman; ({\tt hep-th/9910096}).
\bibitem{SV}O.V. Shaynkman and M.A. Vasiliev,
{\it Theor.Math.Phys.} {\bf 123} (2000); ({\tt hep-th/0003123}).
\bibitem{BHS} M.A. Vasiliev, {\it Phys.Rev.} {\bf D 66} 066006 (2002);
({\tt hep-th/0106149}).
\bibitem{VD} M.A. Vasiliev,
{\it Phys.Lett.} {\bf B 567} (2003) 139--151; ({\tt
hep-th/0304049}).
\bibitem{Frrr}
E.S.~Fradkin and V.Ya.~Linetsky,
{\it Nucl.Phys.} {\bf B 431} (1994) 569.
\bibitem{Fed}
B.V. Fedosov, {\it Deformation Quantization and Index Theory}, Berlin,
Germany:Akademie-Verl. 1996 (Mathematical Topics:9).
\bibitem{Sakh}
M.A.~Vasiliev,
{\it Higher-spin theories and Sp(2M) invariant space-time},
  ({\tt hep-th/0301235}).
\bibitem{HSalg}E.S. Fradkin and M.A. Vasiliev,
{\it Ann.of Phys.} {\bf 177} (1987) 63.
\bibitem{KV1} S.E. Konstein and M.A. Vasiliev, {\it Nucl.Phys.\/}
{\bf B 331} (1990) 475.
\bibitem{FLA} E.S. Fradkin and V.Ya. Linetsky, {\it
Ann.of Phys.}  { \bf 198} (1990) 252.
\bibitem{conf3} O.V. Shaynkman, M.A. Vasiliev, {\it Theor.Math.Phys.} {\bf 128}
(2001) 1155-1168; ({\tt hep-th/0103208}).
\bibitem{SS}E. Sezgin and P. Sundell, {\it Nucl.Phys.} {\bf B 634} (2002)
120-140; ({\tt hep-th/0112100}).
\bibitem{east}  M.G.~Eastwood, {\it Higher symmetries of the laplacian}, ({\tt
hep-th/0206233}).
\bibitem{unfolded} M.A. Vasiliev, {\it Class.Quant.Grav.} {\bf
11} (1994) 649.
\bibitem{Vogan} D.A.~Vogan, {\it Representations of real reductive Lie
groups}, Prog. Math. vol. 15. Birkhauser 1981.
\bibitem{[BB]} A.~Beilinson and J.~Bernstein,
Advances in soviet mathematics, v. 16, 1 (1993).
\bibitem{Kos} B. Kostant, {\it Lect.Notes.Math.} {\bf 466} (1975) 101--128.
\bibitem{ER} M.G. Eastwood and J.W. Rice, {\it Comm.Math.Phys.}
{\bf 109} (1987) 207--228; Erratum, {\it Comm.Math.Phys.} {\bf
144} (1992) 213.
\bibitem{A-inf} T.~Lada and J.~Stasheff, {\it Internat.J.Theoret.Phys.}
{\bf 32} (1993) 1087; ({\tt hep-th/9209099}).
\bibitem{A-inf_}H. Kajiura,
{\it Nucl.Phys.} {\bf B 630} (2002) 361; ({\tt hep-th/0112228}).
\bibitem{tri}M.A. Vasiliev,
{\it Nucl.Phys.} {\bf B 324} (1989) 503.
\bibitem{Mayer}D.H. Mayer,
{\it J.Math.Phys.} {\bf 16} 4 (1975) 884.
\bibitem{Bayen_Flato}F. Bayen and M. Flato,
{\it J.Math.Phys.} {\bf 17} 7 (1976) 1112.
\bibitem{P_S_T}V.B. Petkova, G.M. Sotkov and I.T. Todorov,
{\it Comm.Math.Phys.} {\bf 97} (1985) 227.
\bibitem{Fr_P}E.S. Fradkin and M.Ya. Palchik,
{\it Conformal quantum field theory in D-dimensions}, Kluwer
Academic Publishers, v. 376, 1996.
\bibitem{eastw}  R.J.~Baston and M.G.~Eastwood, {\it
The Penrose transform. Its interaction with representation
theory}, Clarendon Press, Oxford 1989.
\bibitem{fort}   M.A. Vasiliev, {\it Fortschr.Phys.} {\bf 36}
(1988) 1.
\bibitem{[BCI]} B.D.~Boe and D.H.~Collingwood,
{\it J.Algebra} {\bf 54} (1985) 511--545.
\bibitem{[BCII]} B.D.~Boe and D.H.~Collingwood,
{\it Math.Z.} {\bf 190} (1985) 1--11.
\bibitem{nfl_1} T.P. Branson, {\it Comm.Math.Phys.} {\bf 178}
(1996) 301.
\bibitem{nfl_2} T. Parker and S. Rosenberg, {\it J.Diff.Geom.} {\bf 25}
(1987) 199.
\bibitem{nfl_3} R.J. Riegert, {\it Phys.Lett.} {\bf B 134}
(1984) 56.
\bibitem{nfl_4} S. Paneitz, {\it A quartic conformally
covariant differential operators for arbitrary pseudo-Riemannian
manifolds}, MIT preprint (1983).
\bibitem{nfl_5} A. Iorio, L. O'Raifeartaigh, I. Sachs and C. Wiesendanger,
{\it Nucl.Phys.} {\bf B 495} (1997) 433--450; ({\tt
hep-th/9607110}).
\bibitem{pnfl_1}J.~Erdmenger,
{\it Class.Quant.Grav.} {\bf 14} (1997) 2061; ({\tt
hep-th/9704108}).
\bibitem{pnfl_2}J.~Erdmenger and H.~Osborn,
{\it Class.Quant.Grav.} {\bf 15}, (1998) 273; ({\tt
gr-qc/9708040}).
\bibitem{pnfl_3}L.~Dolan, C.~R.~Nappi and E.~Witten,
JHEP {\bf 0110}, (2001) 016; ({\tt hep-th/0109096}).
\bibitem{sieg}W. Siegel,
{\it Int.J.Mod.Phys.} {\bf A 4} (1989) 2015-2020.
\bibitem{metsaev}R.~R.~Metsaev,
{\it Mod.Phys.Lett.} {\bf A 10} (1995) 1719.
\bibitem{FF} S. Ferrara and C. Fronsdal,
{\it Conformal fields in higher dimensions}, ({\tt
hep-th/0006009}).
\bibitem{Hull} C.M. Hull, {\it JHEP 0012:007} (2000);
({\tt hep-th/0011215}).
\bibitem{FrT} E.S. Fradkin and A.A. Tseytlin,
{\it Phys.Rep.} {\bf 119} (1985) 233--362.
\bibitem{Segal} A.Y. Segal, {\it Nucl.Phys.} {\bf B 664} (2003) 59--130; ({\tt
hep-th/0207212}).
\bibitem{GV}O.A. Gelfond and M.A. Vasiliev,
{\it Higher rank conformal fields in the sp(2m) symmetric
generalized space--time}, ({\tt hep-th/0304020}).
\bibitem{Henneaux}G. Barnich, F. Brandt, and M. Henneaux,
{\it Commun.Math.Phys.}  {\bf 174} (1995) 57; ({\tt hep-th/9405109}).
\bibitem{D-VH} M. Dubois-Violette and M. Henneaux, {\it Commun.Math.Phys.}
{\bf 226} (2002) 393-418; ({\tt math.QA/0110088}).
\bibitem{[KKz]} V.G.~Kac and D.A.~Kazhdan,
{\it Adv.Math.}, {\bf 34} (1984) 97-103.
\bibitem{Zhelob} D.P.~Zhelobenko, {\it Representations of reductive Lie
algebras,} Nauka, Fizmatlit Publishing Company, Moscow 1993.
\end{thebibliography}
\end{document}